\documentclass[a4paper,11pt]{article}
\pdfoutput=1

\usepackage{amssymb,amsmath,bm}
\usepackage{a4wide}
\usepackage{color}
\usepackage{slashed}
\usepackage{graphicx}
\usepackage{amsfonts}
\usepackage{lscape}
\usepackage{hyperref}
\usepackage{amsthm}
\hypersetup{
    colorlinks=true, 
    linktoc=all,     
    linkcolor=blue,  
    citecolor= blue
}
\usepackage{booktabs}
\usepackage{array}
\usepackage{rotating}
\usepackage[numbers, sort&compress]{natbib}
\usepackage{float}
\usepackage[utf8]{inputenc}
\usepackage[T1]{fontenc}
\usepackage{bm}
\newcommand{\GeV}{{\, \rm GeV}}
\newcommand{\TeV}{{\, \rm TeV}}

\newcommand{\ord}[1]{\mathcal{O}\left( #1 \right)}

\newcommand{\be}{\begin{equation}}
\newcommand{\ee}{\end{equation}}

\newcommand{\partialLR}{\overset{\leftrightarrow}{\partial^\mu}}
\newcommand{ \mysmall}[1]{\scriptscriptstyle #1} 

\newcommand{\cref}[1]{Chapter~\ref{ch:.#1}}

\newcommand{\beq}{\begin{equation}} 
\newcommand{\eeq}{\end{equation}} 
\newcommand{\ba}{\begin{array}}  
\newcommand{\ea}{\end{array}} 
\newcommand{\bea}{\begin{eqnarray}}  
\newcommand{\eea}{\end{eqnarray} }  
\newcommand{\bal}{\begin{align}}
\newcommand{\eal}{\end{align}}   
\newcommand{\bi}{\begin{itemize}}  
\newcommand{\ei}{\end{itemize}}  
\newcommand{\ben}{\begin{enumerate}}  
\newcommand{\een}{\end{enumerate}}  
\newcommand{\bc}{\begin{center}}
\newcommand{\ec}{\end{center}} 
\newcommand{\bt}{\begin{table}}
\newcommand{\et}{\end{table}}  
\newcommand{\btb}{\begin{tabular}}
\newcommand{\etb}{\end{tabular}}

\definecolor{Grn}{rgb}{0.,0.85,0.1}
\definecolor{Ylw}{rgb}{0.98,0.85,0.2}
\newcommand{\Grn}[1]{\begin{boldmath}#1\end{boldmath}}

\newcommand{\Red}[1]{#1}
\newcommand{\Ylw}[1]{\begin{boldmath}#1\end{boldmath}}

\begin{document}

\vspace{1cm}
\begin{titlepage}
\vspace*{-1.0truecm}
\begin{flushright}
CERN-TH-2018-068 \\
 \vspace*{2mm}
 \end{flushright}
\vspace{0.8truecm}

\begin{center}
\boldmath

{\Large\textbf{
Minimal Models for Dark Matter and the Muon $g-2$ Anomaly
}}
\unboldmath
\end{center}

\vspace{0.4truecm}

\begin{center}
{\bf Lorenzo  Calibbi$^a$,  Robert Ziegler$^{b,c}$, and Jure Zupan$^d$}
\vspace{0.4truecm}

{\footnotesize

$^a${\sl CAS Key Laboratory of Theoretical Physics, Institute of Theoretical Physics, \\ Chinese Academy of Sciences, Beijing 100190, P. R. China \vspace{0.2truecm}}

$^b${\sl Institute for Theoretical Particle Physics, Karlsruhe Institute of Technology, \\ Engesserstrasse 7, D-76128 Karlsruhe, Germany \vspace{0.2truecm}}

$^c${\sl CERN, Theory Division, CH-1211 Geneva 23, Switzerland \vspace{0.2truecm}}

$^d${\sl Department of Physics, University of Cincinnati, Cincinnati, Ohio 45221, USA \vspace{0.2truecm}}

}
\end{center}

\begin{abstract}
\noindent We construct models with minimal field content that can simultaneously explain the muon $g-2$ anomaly and give the correct dark matter relic abundance. These models fall into two general classes, whether or not the new fields couple to the Higgs. For the general structure of models without new Higgs couplings, we provide analytical expressions  that only depend on the $SU(2)_L$ representation. These results allow to demonstrate that only few models in this class can simultaneously explain $(g-2)_\mu$ and account for the relic abundance. The experimental constraints and perturbativity considerations exclude all such models, apart from a few fine-tuned regions in the parameter space, with new states in the few 100 GeV range. In the models with new Higgs couplings, the new states can be parametrically heavier by a factor $\sqrt{1/y_\mu}$, with $y_\mu$ the muon Yukawa coupling, resulting in masses for the new states in the TeV regime. At present these models are not well constrained experimentally, which we illustrate on two representative examples.

\end{abstract}

\end{titlepage}
\tableofcontents

\newpage

\renewcommand{\theequation}{\arabic{section}.\arabic{equation}}


\section{Introduction}
\setcounter{equation}{0}
The explanation of Dark Matter (DM) requires  physics beyond the Standard Model (SM).
A plausible possibility is that DM is  a new stable neutral particle with electroweak scale mass that is a thermal relic. To test concrete realizations of this scenario  it often suffices  to use simplified models. 
These keep only the minimal set of phenomenologically relevant fields out of the full set contained in complete new physics models. Only the lightest new physics states are important for the freeze-out, so that simplified models already describe all the relevant physics. For several examples of such an approach, see e.g.~Refs.~\cite{Abdallah:2015ter,Freitas:2015hsa}.

Similarly, simplified models can capture most of the features of the SM extensions that address the longstanding muon $g-2$ anomaly, i.e., the $\approx 3.5\,\sigma$ discrepancy between theoretical predictions and experimental determination of the muon anomalous magnetic moment~\cite{Bennett:2006fi,Bennett:2004pv,Bennett:2002jb, Brown:2001mga, Jegerlehner:2009ry, Hagiwara:2011af, Davier:2010nc,Blum:2013xva}.  
For instance, it is possible to build minimal extensions of the SM 
addressing the muon $g-2$ anomaly with a single new field\,---\,as systematically discussed in \cite{Freitas:2014pua,Queiroz:2014zfa,Biggio:2014ela,Biggio:2016wyy}\,---\,including leptoquarks \cite{Chakraverty:2001yg,ColuccioLeskow:2016dox}, a second Higgs doublet \cite{Broggio:2014mna,Cherchiglia:2016eui}, and axion-like particles \cite{Marciano:2016yhf}.

In this work we build the simplest extensions of the SM that {\it i)} have a stable DM candidate, and {\it ii)} can
simultaneously explain the muon $g-2$ anomaly. None of the single-field extensions, mentioned above, provide a DM candidate, since in these cases the new particle necessarily couples to two SM fields with sizeable couplings and thus decays quickly.
For this reason, the minimal models we construct require at least two extra fields, assumed to be odd under a $Z_2$ symmetry.
If neutral under color and electromagnetic interactions, the lightest state is then a stable DM candidate. 
We only introduce the fields that can enter the loop diagrams contributing to the muon $g-2$. Moreover, the new fields need to be part of an $SU(2)_L \times U(1)_Y$ multiplet, for gauge invariance, and be color neutral, to avoid colored stable particles. 

The above requirements severely restrict the space of possible models. We can divide them into two general classes, according to the new contributions to the muon $g-2$. In {\em Class I }
the required chirality flip is  provided by a Higgs vev insertion on the external fermion leg, while in {\em Class II} the chirality flip is due to the Higgs vev insertion in the loop. In Class I models the new physics (NP) contribution to $(g-2)_\mu$ is proportional to the small muon Yukawa coupling. This means that the new states need to be relatively light, with masses of the order of a few 100 GeV or below. The muon Yukawa suppression is avoided in Class II models, 
where an additional large coupling to the Higgs can provide a parametric enhancement and allow for NP masses to be as large as a few TeV. 

Non-supersymmetric scenarios addressing both DM and the muon $g-2$ have been discussed before, for instance in Refs.~\cite{Chen:2009ata,Agrawal:2014ufa,Baek:2015fea,Belanger:2015nma,Kowalska:2017iqv}. 
The recent analysis in Ref.~\cite{Kowalska:2017iqv} follows an approach similar to the one outlined above. However, the discussion was limited to only two possible DM candidates: a scalar $SU(2)_L \times U(1)_Y$ singlet, and the neutral component of a scalar doublet with the quantum numbers of the SM Higgs. In the present manuscript we instead systematically build all possible models that have a stable DM candidate and  can
simultaneously explain the muon $g-2$ anomaly, including examples with fermion DM candidates.

The paper is organized as follows.
In Section \ref{Setup} we introduce the general setup, listing models that have NP fields  coupling to muons and contain a DM candidate. Section \ref{sec:pheno} contains general results pertaining to the phenomenology of the introduced models, the contributions to the muon $g-2$, the relic density, constraints from direct detection, and LHC contraints. We apply these results to the Class I models in Section \ref{sec:noHiggs} and to Class II models in Section \ref{sec:Higgs}. 
Section \ref{ref:conclusions} contains our conclusions, while a number of more technical results are collected in appendices.

\section{General Setup}
\label{Setup}

We are interested in models that extend the SM with new fields that  both (i) contain a DM candidate and (ii) contribute to $(g-2)_\mu$ at one-loop. The two requirements significantly constrain possible models with minimal field content. The new fields need to contain a DM candidate, i.e., an electromagnetically neutral color singlet state that is stable on cosmological time-scales. To ensure DM stability we impose a $Z_2$ symmetry under which the new fields are odd, while the SM fields are $Z_2$ even. In order to contribute to $(g-2)_\mu$ the new fields also need to couple (pairwise) to the muon. Therefore, the $Z_2$-odd sector contains at least one new fermion and one new scalar, both of which are color neutral,  and have EW quantum numbers compatible with a DM candidate.  The minimal new field content consists of a 2-component Weyl fermion, $F$, and a heavy complex scalar, $S$, which couples to the LH muon $\mu $ and/or the RH muon $\mu^c$. For $F$ in a complex representation of the SM gauge group we need to add also a field in the conjugate representation, $F^c$, to allow for a fermion mass term. For the $SU(2)_L\times U(1)_Y$ quantum numbers we use the convention
\begin{align}
F & \sim \left( n_F \right)_{Y_F}\, ,  & S & \sim \left(n_S\right)_{Y_S} \, , & \mu & \sim 2_{-1/2} \, , & \mu^c & \sim 1_{1} \, .
\end{align} 
If $Y_{F} = 0$ ($Y_S = 0$) we can take $F$ ($S$) to be a Majorana fermion (real scalar), i.e., for $Y_{F} = 0$ fermion there is no need to add $F^c$ to the model. 

To make the notation more readable we denote by $F_R$ and $S_R$ the fermions and scalars that couple to the LH muon, and by $F_L$ and $S_L$ the fields that couple to the RH muon. The resulting models can then be divided into two classes:
\begin{itemize}
\item
{\bf {\em \underline {``Class I'' models:}}} The Higgs does not couple to the new fields
\begin{itemize}
\item {\em \bf ``LL" Models}: Couplings only to the LH muon
\begin{align}\label{eq:LHmuon}
{\cal L}_{\rm LL} & \supset \left( \lambda_{L} \mu F_R S_R  - M_{F_R} F_R F_R^c  + {\rm h.c.} \right) - M_{S_R}^2 S_R^* S_R \, . \end{align}
\item {\em \bf ``RR" Models}: Couplings only to the RH muon
\begin{align}\label{eq:RHmuon}
{\cal L}_{\rm RR} & \supset \left(  \lambda_{R} \mu^c F_L S_L - M_{F_L} F_L F_L^c +{\rm h.c.} \right) - M_{S_L}^2 S_L^* S_L  \, . 
\end{align}
\end{itemize}
\item
{\bf \underline {{\em ``Class II''} models:}} The new fields couple to the Higgs and both LH and RH muons 
\begin{itemize}
\item {\em \bf ``FLR" Models}: The Higgs couples to new fermions
\begin{align}\label{eq:FLR}
{\cal L}_{\rm FLR} & \supset \left( y_F H F_L F_R + \lambda_{L} \mu F_R S_R + \lambda_{R} \mu^c F_L S_R^*  +{\rm h.c.} \right) \nonumber \\
& - \left( M_{F_L} F_L F_L^c + M_{F_R} F_R F_R^c +{\rm h.c.} \right)  - M_{S_R}^2 S_R^* S_R \, .
\end{align}
Note that the field content is the sum of the LL and RR fields upon identifying $S_L$ with $S_R^*$. 
\item {\em \bf ``SLR" Models}: The Higgs couples to new scalars 
\begin{align}\label{eq:SLR}
{\cal L}_{\rm SLR} & \supset \left( a H S_L S_R + \lambda_{L} \mu F_R S_R + \lambda_{R} \mu^c F_R^c S_L  + {\rm h.c.} \right)  \nonumber \\
& - \left( M_{F_R} F_R F_R^c + {\rm h.c.} \right) - M_{S_L}^2 S_L^* S_L - M_{S_R}^2 S_R^* S_R \, .
\end{align}
Note that the field content is the sum of the LL and RR fields upon identifying $F_L$ with $F_R^c$. 
\end{itemize}
\end{itemize}
In writing the above Lagrangians we used a two-component spinor notation, and all the fermion fields are left-handed Weyl fermions (including $F_R$).

\begin{figure}[t]
\centering
\includegraphics[scale=0.65]{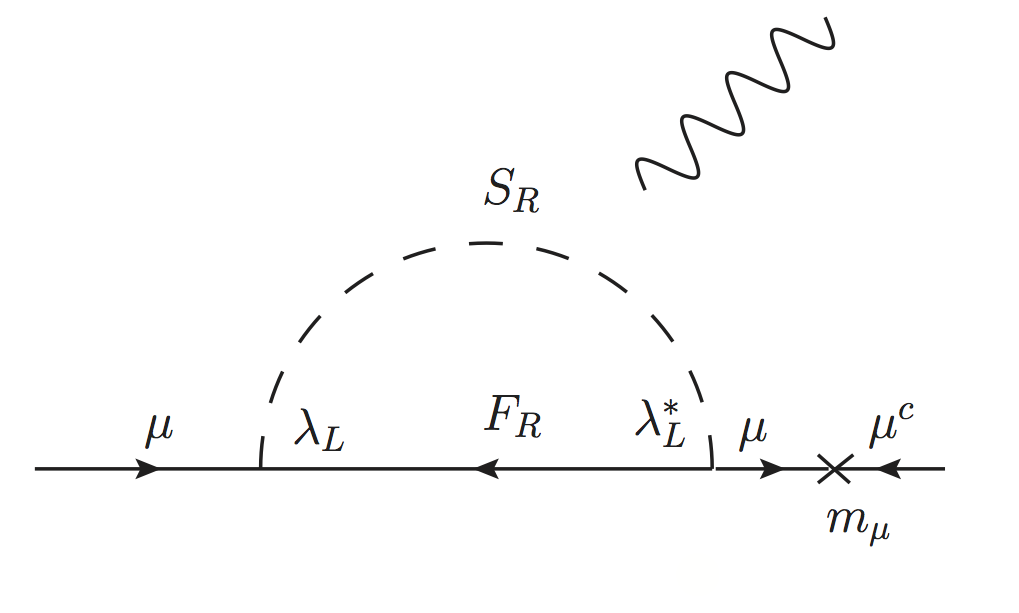}
~~~~
\includegraphics[scale=0.65]{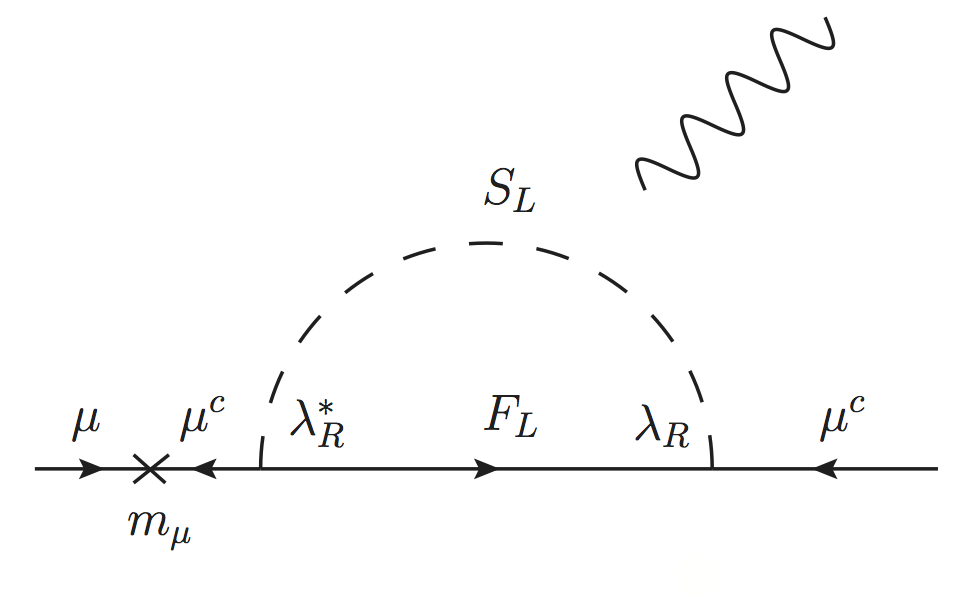}
\includegraphics[scale=0.65]{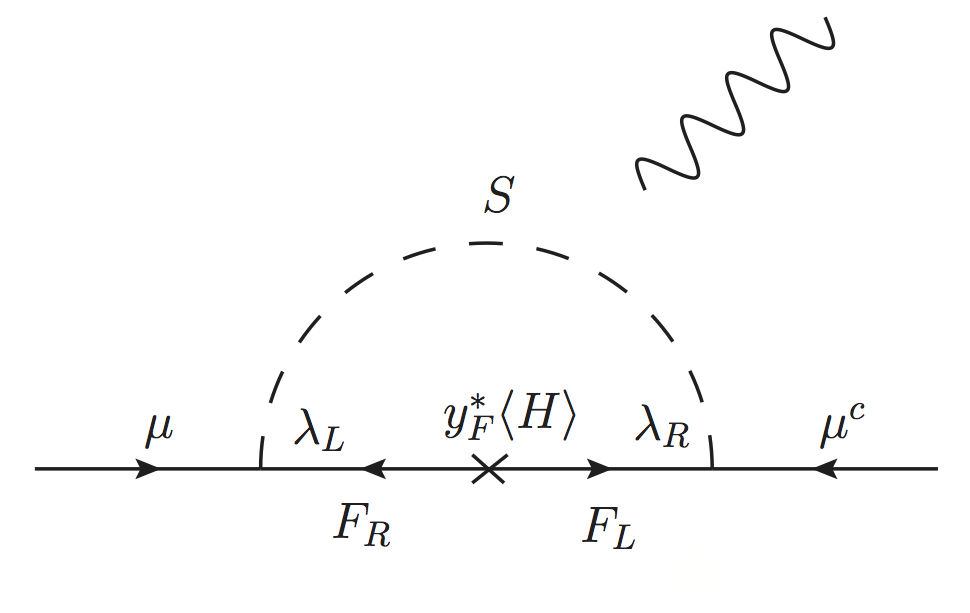}
~~~~
\includegraphics[scale=0.65]{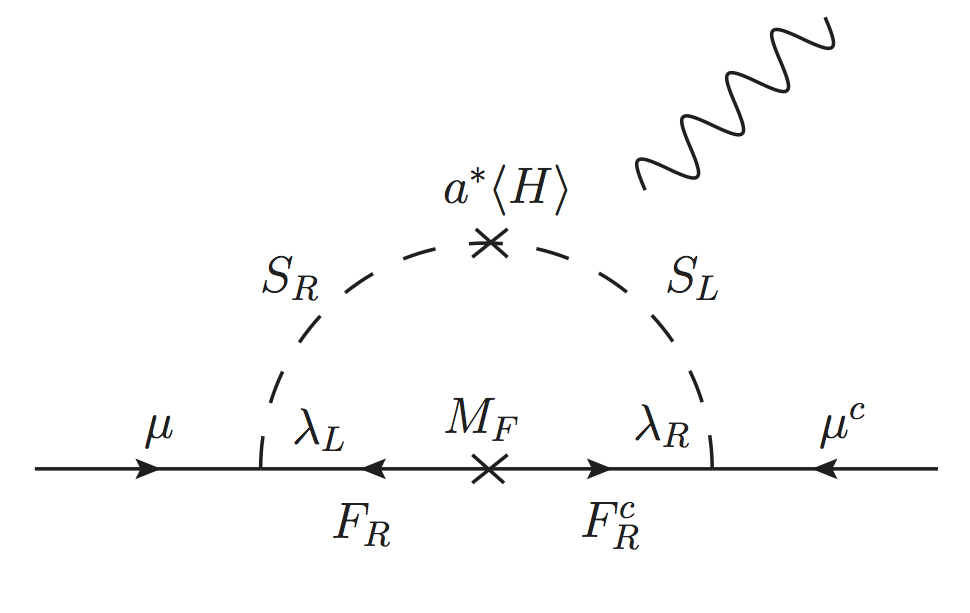}
\caption{\label{diagrams:g-2} 
Diagrams in the first row show one-loop contributions to $(g-2)_\mu$ for Class I (LL and RR) models, where Higgs insertions are on the external muon lines. The second row shows contributions for  Class II (FLR and SLR) models, where Higgs insertions occur on the internal fermion or scalar lines. }
\end{figure}

The contributions to $(g-2)_\mu$ from Class I models require a chirality flip from a Higgs vev insertion on the external muon line, see the first row of diagrams in Fig. \ref{diagrams:g-2}. In contrast, Class II models also receive contributions from Higgs vev insertions on the heavy internal fermion or scalar line (second row of diagrams in Fig. \ref{diagrams:g-2}), which can be parametrically enhanced by $1/y_\mu$.

In Tables \ref{tab:LLRR} and \ref{tab:LR} we show the possible $SU(2)_L\times U(1)_Y$ quantum number assignments for LL and RR models, and for FLR and SLR models, respectively (for further details see Appendix \ref{sec:QuantumNumbers}). In the Tables we restrict the NP fields, both fermions and scalars, to be at most in a triplet representation of $SU(2)_L$. For FLR and SLR models this restriction is made for simplicity, while for LL and RR we will demonstrate below that larger representations are not interesting for our purposes. Fields that contain a neutral state, i.e. a DM candidate, are denoted by a $\star$ in the superscript. There are in total 10 LL models, 9 RR models, 10 FLR models, and 10 SLR models. However only very few of the Class I (LL and RR) models are phenomenologically viable, as we show in the next two sections. 
The models that can successfully account for DM and $(g-2)_\mu$
are shown in boldface in Tables \ref{tab:LLRR} and \ref{tab:LR}.

\begin{table}[t]
\centering
\begin{tabular}{c||ccccccccccc}
 ${\rm \bf LL~}$ & & & & & & &  & & &  \\
 \hline
 $F_R$ & \Red{$1_0^{\star}$} & \Ylw{$1_1$} & \Red{$2_{-\frac{1}{2}}^\star$} & \Red{$2_{-\frac{1}{2}}^\star$} & \Ylw{$2_{\frac{1}{2}}^\star$} & \Red{$2_{\frac{1}{2}}^\star$} & \Red{$2_{\frac{3}{2}}$} & \Red{$3_{-1}^\star$} & \Red{$3_{0}^\star$} & \Red{$3_{1}^\star$} \\
 $S_R$ &  \Red{$2_{\frac{1}{2}}^\star$} &  \Ylw{$2_{- \frac{1}{2}}^\star$} &  \Red{$1_{1}$} &  \Red{$3_{1}^\star$} & \Ylw{$1_{0}^\star$} &  \Red{$3_{0}^\star$} & \Red{$3_{-1}^\star$} &  \Red{$2_{\frac{3}{2}}$} & \Red{$2_{\frac{1}{2}}^\star$ } & \Red{$2_{- \frac{1}{2}}^\star$}  \\
\multicolumn{1}{c}{}   \\
${\rm \bf RR~}$ &&&&&&&&& \\
\hline
$F_L$  &  \Red{$1_0^\star$} & \Ylw{$1_{-1}$} & \Red{$2_{- \frac{1}{2}}^\star$} & \Red{$2_{\frac{1}{2}}^\star$} & \Red{$2_{- \frac{3}{2}}$} & \Red{$3_{-1}^\star$} & \Red{$3_{0}^\star$} & \Red{$3_{1}^\star$} & \Red{$3_{-2}$} & \\
 $S_L$ & \Red{$1_{-1}$} & \Ylw{$1_{0}^\star$} &  \Red{$2_{- \frac{1}{2}}^\star$ }& \Red{$2_{- \frac{3}{2}}$} & \Red{$2_{\frac{1}{2}}^\star$} & \Red{ $3_{0}^\star$ } & \Red{ $3_{-1}^\star$} & \Red{ $3_{-2}$} & \Red{ $3_{1}^\star$} &
 \end{tabular}
\caption{ \label{tab:LLRR} Class I (LL and RR) models up to $SU(2)_L$ triplets, fields with $\star$ contain a DM candidate. Viable models are in boldface, see text for details.}
\end{table}

\begin{table}[t]
\centering
\begin{tabular}{c||cccccccccc}
${\rm \bf FLR~}$ &&&&&&&&&& \\
\hline
$F_R$  &  \Grn{$1_0^\star$} & \Grn{$1_{1}$} & \Grn{$2_{- \frac{1}{2}}^\star$} & \Grn{$2_{- \frac{1}{2}}^\star$} & \Grn{$2_{\frac{1}{2}}^\star$} & \Grn{$2_{\frac{1}{2}}^\star$} & \Grn{$2_{ \frac{3}{2}}$} & \Grn{$3_{-1}^\star$} & \Grn{$3_{0}^\star$} & \Grn{$3_{1}^\star$}  \\
$F_L$  &  \Grn{$2_{-\frac{1}{2}}^\star$}  & \Grn{$2_{-\frac{3}{2}}$}  & \Grn{$1_0^\star$ }& \Grn{$3_0^\star$ }& \Grn{$1_{-1}$} & \Grn{$3_{-1}^\star$ }& \Grn{$3_{-2}$} & \Grn{$2_{\frac{1}{2}}^\star$}  & \Grn{$2_{- \frac{1}{2}}^\star$} & \Grn{$2_{- \frac{3}{2}}$}  \\
$S_R$ & \Grn{$2_{\frac{1}{2}}^\star$}  & \Grn{$2_{-\frac{1}{2}}^\star$}  & \Grn{ $1_1$ }& \Grn{$3_1^\star$} & \Grn{$1_0^\star$} &  \Grn{$3_{0}^\star$}& \Grn{  $3_{-1}^\star$} &  \Grn{$2_{\frac{3}{2}}$ } &  \Grn{$2_{\frac{1}{2}}^\star$} & \Grn{$2_{-\frac{1}{2}}^\star$} \\
\multicolumn{1}{c}{}   \\
$ {\rm \bf SLR~}$ &&&&&&&& \\
\hline
$S_L$  & \Grn{  $1_0^\star$ } & \Grn{ $1_{-1}$ } & \Grn{ $2_{- \frac{1}{2}}^\star$} & \Grn{ $2_{- \frac{1}{2}}^\star$} & \Grn{ $2_{ \frac{1}{2}}^\star$} & \Grn{ $2_{- \frac{3}{2}}$} & \Grn{ $2_{- \frac{3}{2}}$} & \Grn{ $3_0^\star$}  & \Grn{ $3_{- 1}^\star$} & \Grn{ $3_{- 2}$ }\\
$S_R$  & \Grn{  $2_{- \frac{1}{2}}^\star$ }& \Grn{  $2_{ \frac{1}{2}}^\star$} &\Grn{  $1_0^\star$ }& \Grn{ $3_0^\star$} & \Grn{ $3_{-1}^\star$} &\Grn{  $1_1$ }& \Grn{ $3_1^\star$ }& \Grn{  $2_{- \frac{1}{2}}^\star$ }& \Grn{ $2_{\frac{1}{2}}^\star$ }& \Grn{ $2_{ \frac{3}{2}}$ }\\
$F_R$ & \Grn{ $1_1$ } & \Grn{ $1_0^\star$}  &  \Grn{ $2_{ \frac{1}{2}}^\star$} & \Grn{ $2_{\frac{1}{2}}^\star$} &\Grn{  $2_{\frac{3}{2}}$} & \Grn{  $2_{- \frac{1}{2}}^\star$} & \Grn{  $2_{- \frac{1}{2}}^\star$ }& \Grn{  $3_1^\star$ } & \Grn{  $3_0^\star$} & \Grn{ $3_{-1}^\star$ }
\end{tabular}
\caption{\label{tab:LR} Class II (FLR and SLR) models up to $SU(2)_L$  triplets, fields with $\star$ contain a DM candidate. Viable models are in boldface, see text for details.}
\end{table}


\section{Phenomenology}
\label{sec:pheno}
\setcounter{equation}{0}
To be considered viable, a model should explain the muon $g-2$ anomaly and reproduce the observed DM relic density. Of course, the model also needs to satisfy all experimental constraints, in particular from DM searches in direct and indirect detection experiments, and from direct production of heavy particles at the LHC. In this section we first introduce the relevant observables and provide approximate results that will prove useful for later discussions. 

\subsection{Muon $g - 2$}
\label{sec:g-2}
The anomalous magnetic moment of the muon, $a_\mu = (g-2)_\mu /2$, is one of the most important tests of the SM and provides a powerful probe of new physics. The longstanding $\sim 3.5 \sigma$ discrepancy between the SM prediction and the experimental value~\cite{Bennett:2006fi,Bennett:2004pv, Bennett:2002jb, Brown:2001mga, Jegerlehner:2009ry, Hagiwara:2011af, Davier:2010nc,Blum:2013xva}
\be
\Delta a_\mu =a_\mu^{\mysmall \rm EXP}-a_\mu^{\mysmall \rm SM} = 2.87 \, (80) \times 10^{-9} \, , 
\label{eq:gmu}
\ee
has triggered many speculations about NP scenarios that give additional contributions to $a_\mu$, see
\cite{Lindner:2016bgg} for a recent review. The new Muon (g-2) Experiment, E989, at Fermilab \cite{Grange:2015fou} has started to collect data at the end of 2017 and is expected to reach the precision of the E821 experiment \cite{Bennett:2006fi} within this year. After several years of running E989 should decrease the experimental error by a factor $4$, thus revealing possible new physics effects with high confidence. 

General NP contributions to $\ell\to\ell^{\prime}\gamma$ are described by 
the effective Lagrangian
\be
\mathcal L = \frac{e m_{\ell}}{8\pi^2}
C_{\ell\ell^{\prime}} \left({\bar\ell}_{R}^\prime\sigma_{\mu\nu} \ell_{L}\right) F^{\mu\nu} +{\rm h.c.},\,
\qquad
\ell,\ell^{\prime} = e,\mu,\tau\,,
\label{eq:eff_lagr_LFV}
\ee
where $C_{\ell\ell'}$ is a Wilson coefficient with mass dimension (GeV)$^{-2}$ . 
This leads to the NP contribution to the anomalous magnetic moment, $\Delta a_{\ell}$,
\be
\Delta a_{\ell} = \frac{1}{2\pi^2} m^{2}_{\ell}~{\rm Re}(C_{\ell\ell})\,,
\label{eq:gm2}
\ee
and to flavor violating transitions, with branching ratios that are in the $m_\ell\gg m_{\ell'}$ limit  given by
\begin{align}
\frac{{\rm BR}(\ell\to\ell^{\prime}\gamma)}{{\rm BR}(\ell\to\ell^{\prime}\nu\bar{\nu}^\prime)} = 
\frac{3\alpha}{\pi G_F^2} \left( |C_{\ell\ell^\prime}|^2+|C_{\ell^\prime\ell}|^2 \right).
\end{align}
We will be mostly interested in $\Delta a_\mu$ and $\tau\to \mu\gamma$, $\mu\to e\gamma$ transitions. 

Now we focus on the models introduced in the previous section. 
Their contributions to $\Delta a_\mu$ are captured 
by the general $SU(3)_c \times U(1)_{\rm em}$  Lagrangian
\begin{align}\label{eq:general:Lagr}
{\cal L} & \supset \left[S^* \, \bar{\mu} \left( \lambda_2^{R} P_L + \lambda_2^{L} P_R \right) F + {\rm h.c.}\right]   - M_{F} \bar{F} F - m_\mu \bar{\mu} \mu  - M_S^2 S^* S \, , 
\end{align}
where $S$ is a heavy complex scalar with electric charge $Q_S$, while $F$ is a heavy vector-like fermion with charge $Q_F=Q_S-1$. 
The LL (RR) model is recovered by setting $\lambda_2^R=0$ $(\lambda_2^L=0)$\,---\,cf.~Eqs.~\eqref{eq:LHmuon} and \eqref{eq:RHmuon}\,---\,while the FLR model has both couplings non-vanishing, $\lambda_2^L\ne 0, \lambda_2^R\ne 0$, see Eq.~\eqref{eq:FLR}. The SLR model has two scalars, $S_L$  with  $\lambda_2^L\ne 0$ but no coupling to $P_R F$, and $S_R$ with  $\lambda_2^{R}\ne 0$ but no coupling to $P_L F$, see Eq.~\eqref{eq:SLR}.  The results below apply once one sums over both contributions from $S_L$ and $S_R$.

The general Lagrangian  in Eq.~\eqref{eq:general:Lagr} gives a contribution to $(g-2)_\mu$ that reads, in agreement with the literature~\cite{Jegerlehner:2009ry},
\beq
\begin{split} \label{eq:Deltaamu:generalLagr}
\Delta a_{\mu} =& - \frac{m_\mu^2}{8 \pi^2 M_S^2}  \left( |\lambda^{L}_{2}|^2 + |\lambda^{R}_{2}|^2 \right) \left[ Q_F f_{LL}^F (x)+ Q_S f_{LL}^S (x) \right]  \\
&  - \frac{m_\mu M_F}{8 \pi^2 M_S^2}  {\rm Re} \left(   \lambda^{R*}_{2}  \lambda^{L}_{2}  \right) \left[ Q_F f_{LR}^F (x) + Q_S  f_{LR}^S (x) \right]  \, . 
\end{split}
\eeq
Here $x = M_F^2/M_S^2$, and the loop functions are given by
\begin{align}
\label{eq:fF}
f^F_{LL} (x) & =  \frac{2+ 3x - 6 x^2 + x^3 + 6 x \log x}{12 (1-x)^4} \, , & 
f_{LR}^F (x) & = - \frac{3-4x+x^2+2 \log x}{2 (1-x)^3} \, ,
\\
\label{loopf}
f^S_{LL}(x) & = \frac{1-6x + 3 x^2 + 2 x^3 - 6 x^2 \log x}{12 (1-x)^4} \, , &
f_{LR}^S(x) &  =  \frac{1-x^2+2x \log x}{2 (1-x)^3} \, .
\end{align}
The contributions in the first line of Eq.~(\ref{eq:Deltaamu:generalLagr}) are from diagrams with mass insertions on the external muon line (the first row in Fig.~\ref{diagrams:g-2}), while the terms in the second line come from diagrams with a chirality flip on the internal line (the second row in Fig.~\ref{diagrams:g-2}).  
The latter contributions are parametrically enhanced by $\lambda_2^R \lambda^L_2 M_F/m_\mu \propto v/m_
\mu$ and thus dominate $\Delta a_\mu$ (note that the product $\lambda^L \lambda^R$ must always be proportional to $v$, since it breaks $SU(2)_L$). The contributions with the muon mass insertion are therefore only relevant if either $\lambda_2^L$ or $\lambda_2^R$ is suppressed  or absent, as it is the case for Class I models.  Note that the signs of these Class I contributions (first line in  Eq.~(\ref{eq:Deltaamu:generalLagr})) are given simply by the signs of $Q_{S,F}$ because the loop functions are positive
\begin{align}
0 & \le  f_{LL}^F \left( x \right)  \le \frac{1}{6} \, , & 0 & \le  f_{LL}^S \left( x \right)  \le \frac{1}{12}  \, .
\end{align}
As we will show in the next section, Class I models in many cases predict the sign of $\Delta a_\mu$, which immediately allows to  discard many models from Table~\ref{tab:LLRR}. 

The couplings of heavy states to muons, $\lambda_2^{L,R}$, also enter the one-loop lepton-flavor-violating transitions 
$\tau\to \mu\gamma$ and $\mu \to e\gamma$ (see Ref.~\cite{Calibbi:2017uvl} for a recent review), 
along with the equivalent couplings to electrons, $\lambda_1^{L,R}$, and taus, $\lambda_3^{L,R}$.
The contributions in Class I models are due to diagrams with chirality flips on the external lines (similar to the diagrams in the first row of Fig.~\ref{diagrams:g-2}), which give in the limit $m_\tau\gg m_\mu \gg m_e$
\begin{align}
{\rm BR}(\mu\to e \gamma) &= \frac{12\pi^3}{m_\mu^4} \frac{\alpha}{G_F^2} \left(\frac{\lambda^{L,R}_1}{\lambda^{L,R}_2}\right)^2 \times (\Delta a_{\mu})^2
\nonumber\\ 
& \approx 4.1 \times 10^{-13}  \left(\frac{\lambda^{L,R}_1/\lambda^{L,R}_2}{1.7\times 10^{-5}}\right)^2  \left(\frac{\Delta a_\mu}{2.9\times 10^{-9}} \right)^2, \\
{\rm BR}(\tau\to \mu \gamma)& = \frac{12\pi^3}{m_\mu^4} \frac{\alpha}{G_F^2} \left(\frac{\lambda^{L,R}_3}{\lambda^{L,R}_2}\right)^2 \times 
(\Delta a_{\mu})^2 \times {{\rm BR}(\tau\to \mu \nu \bar{\nu})} \nonumber \\
& \approx
4.2 \times 10^{-8}  \left(\frac{\lambda^{L,R}_3/\lambda^{L,R}_2}{1.3\times 10^{-2}}\right)^2  \left(\frac{\Delta a_\mu}{2.9\times 10^{-9}} \right)^2.
\end{align}
 The results for Class II models (similar to the diagrams in the second row of Fig.~\ref{diagrams:g-2}) are obtained by replacing 
$(\lambda^{L,R}_i/{\lambda^{L,R}_2})^2 \to (\lambda^{L}_i/{\lambda^{L}_2})^2 + (\lambda^{R}_i/{\lambda^{R}_2})^2$ in the above expressions. 
The central value of $\Delta a_\mu$, and the present experimental bounds ${\rm BR}(\mu\to e \gamma) < 4.2 \times 10^{-13}$~\cite{TheMEG:2016wtm} and ${\rm BR}(\tau\to \mu \gamma) < 4.4 \times 10^{-8}$~\cite{Aubert:2009ag} then imply 
\beq
\lambda^{L,R}_1/\lambda^{L,R}_2 \lesssim 1.7\times 10^{-5}, \qquad\qquad \lambda^{L,R}_3/\lambda^{L,R}_2 \lesssim 1.3\times 10^{-2}.
\eeq
The heavy states thus need to couple to muons much more strongly than to electrons and taus. Such a muon-philic flavor structure is certainly possible (e.g.~through the ``flavour-locking'' mechanism, see Ref.~\cite{Knapen:2015hia}), although it is non-generic. In the remainder of the paper we assume that the above bounds are fulfilled and concentrate on the couplings to the muon sector. 

\subsection{DM Relic Density}
\label{relicapp}
We assume that the DM particle $\chi$ with mass $m_\chi$ is a thermal relic, so that its relic density is primarily determined by its annihiliation to SM particles. 
Expanding the annihilation cross-section $\sigma$ in the relative velocity $v$,  
\begin{align}
\label{XSexp}
\sigma v = a_0 + a_1 v^2 + a_2  v^4 + \cdots \, , 
\end{align}
the relic density is approximately given by (using the results of Ref.~\cite{Gondolo:1990dk}, see also Ref.~\cite{Giacchino:2013bta,Toma:2013bka})
\begin{align}
\Omega_{\chi} h^2 \approx  9 \cdot 10^{-11} \frac{x_f }{\sqrt{g_*(T_f)}}\, \frac{ \GeV^{-2} }{a_0 + 3 a_1/x_f + (20  a_2 - 9 a_1) /x_f^2}.
\label{Omapp}
\end{align}
Here $g_*(T_f) \sim {\cal O} (90) $ counts the effective number of relativistic degrees of freedom at the freeze-out temperature $T_f$.  The ratio $x_f = m_\chi/T_f\sim {\mathcal O}(20)$ is determined through the transcendental equation,
\begin{align}
x_{f} = \log \left[ 3.8 \cdot 10^{9}  \frac{g \sqrt{x_{f}} }{\sqrt {g_* (T_f)}}\,  \frac{m_\chi}{1\,\GeV} \, \frac{ a_0 + 6 a_1/x_f + (60  a_2 - 27 a_1) /x_f^2}{1\,{\rm pb}} \right] \, ,
\end{align}
for a DM particle with $g$ degrees of freedom.
\begin{figure}[t!]
\begin{center}
\includegraphics[scale=0.5]{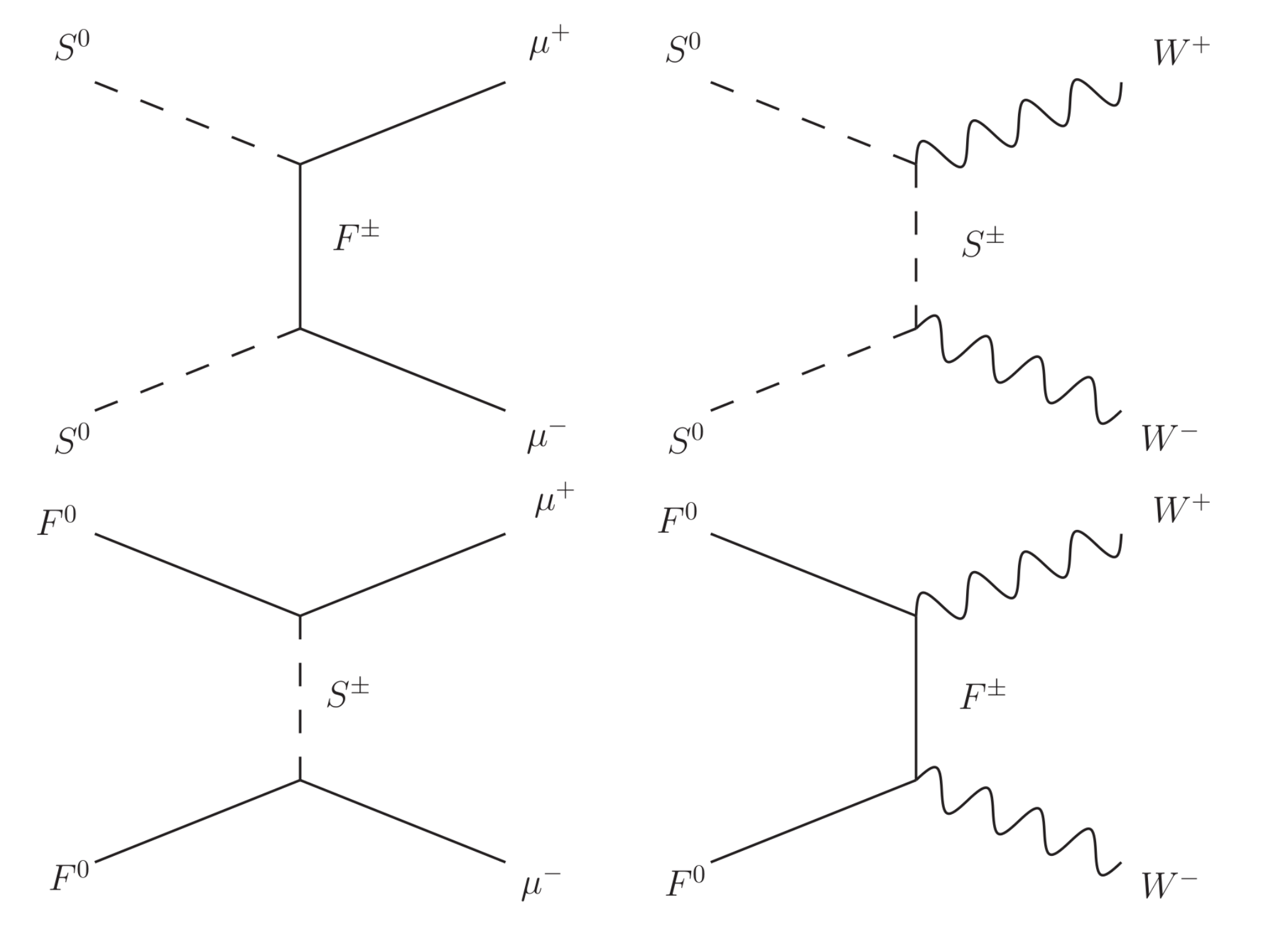}
\end{center} 
\caption{\label{fig:DMann} The two main DM annihilation processes for scalar DM (first row) and fermion DM (second row). }
\end{figure}

In both Class I and Class II models the two main annihilation channels of DM are the $t$-channel annihilation into muons and into gauge bosons, schematically depicted in Fig.~\ref{fig:DMann}.
In addition, one needs to take into account co-annhiliations if $M_F\simeq M_S$, or if the mass splittings in the DM $SU(2)_L$ multiplet are small. In this case  $\sigma v$ in Eq.~(\ref{XSexp}) has to be replaced with an effective annihilation cross-section. In the limit of negligible mass splittings it is given by~\cite{Griest:1990kh}
\begin{align}
\sigma v \to ( \sigma v )_{\rm eff} = \frac{1}{n^2} \sum_{ij} \sigma_{ij} v \, , 
\end{align}  
where $n$ is the dimension of the DM $SU(2)_L$ representation, and $ \sigma_{ij} $ the annihiliation cross-sections for $X_i X_j \to {\rm SM} $, with $X_{i,j}$ the DM multiplet elements. 

The effective cross sections for a DM $SU(2)_L$ multiplet with dimension $n$ and hypercharge~$Y$ annihilating  to gauge bosons in the limit $m_\chi\gg m_W$
are~\cite{MDM} \begin{align}
\label{annMDM:scalar}
 ( \sigma v )_{\rm eff}^{\rm scalar}  & = \frac{1}{64 \pi g_X m_\chi^2 }\left[{g_2^4 \left(n^4 - 4 n^2 + 3 \right) + 16 Y^4 g_Y^4 + 8 Y^2 g_2^2 g_Y^2  \left( n^2 - 1 \right)}\right] \, , 
 \\
 \label{annMDM:fermion}
 \begin{split}
   ( \sigma v )_{\rm eff}^{\rm ferm.}  & = \frac{1}{128 \pi g_X m_\chi^2 }\left[g_2^4 ( 2 n^4 + 17 n^2 -19)
 + 4 Y^2 g_Y^4 ( 8 Y^2+ 41) \right.
 \\
 & \qquad\qquad\qquad\qquad\qquad\qquad\qquad\qquad\qquad\left.+ 16 Y^2 g_2^2 g_Y^2 ( n^2 - 1 )\right] \, , 
 \end{split}
\end{align}
where in Eq.~\eqref{annMDM:scalar} one has $g_X = 2n (n)$ for a complex (real) scalar,  while in Eq.~\eqref{annMDM:fermion} $g_X=4n (2n)$ for a Dirac (Majorana) fermion. 

Using the general Lagrangian in Eq.~\eqref{eq:general:Lagr}, the annihilation cross-sections into muons are
\begin{align}
\label{eq:sigmamu:Cscalar}
(\sigma v)^{\rm C-scalar} & = \frac{1}{4 \pi M_F^2} \frac{1}{(1+r_S^2)^2} \left[ \lambda_L^2 \lambda_R^2 + \frac{\lambda_L^4 + \lambda_R^4}{4} \left( \frac{m_\mu^2}{M_F^2} + \frac{v^2 r_S^2}{3} \right) \right] \, , 
\\
\label{eq:sigmamu:Rscalar}
(\sigma v)^{\rm R-scalar} & = \frac{1}{ \pi M_F^2} \frac{1}{(1+r_S^2)^2} \left[ \lambda_L^2 \lambda_R^2 + \frac{\lambda_L^4 + \lambda_R^4}{4} \left( \frac{m_\mu^2}{M_F^2} + \frac{v^4 r_S^6}{15 (1+r_S^2)^2} \right) \right] \, , 
\end{align}
for a complex scalar in Eq.~\eqref{eq:sigmamu:Cscalar} and a real scalar in Eq.~\eqref{eq:sigmamu:Rscalar}, respectively. Here $r_S = M_S/M_F < 1$, and we have set $\lambda_{L,R} = \lambda^{L,R}_2$ to shorten the notation, assumed to be real for simplicity. Moreover, we have kept only the dominant terms, including the $\lambda_L \lambda_R = 0$ limit. 
Similarly, the cross sections for heavy fermions annihilating to muons are 
\begin{align}
\label{eq:sigmamu:Dfermion}
(\sigma v)^{\rm D-ferm.} & = \frac{1}{32 \pi M_S^2} \frac{r_F^2}{(1+r_F^2)^2} \left(  \lambda_L^2 + \lambda_R^2 \right)^2 \, , 
\\
\label{eq:sigmamu:Mfermion}
(\sigma v)^{\rm M-ferm.} & = \frac{1}{ 8 \pi M_S^2} \frac{1}{(1+r_F^2)^2} \left[ r_F^2 \lambda_L^2 \lambda_R^2 + \frac{\lambda_L^4 + \lambda_R^4}{4} \left( \frac{m_\mu^2}{M_S^2} + \frac{2 v^2 r_F^2 (1 + r_F^4)}{3 (1+r_F^2)^2} \right) \right] \, .
\end{align}
For  Dirac fermion the annihilation cross section is given in Eq. \eqref{eq:sigmamu:Dfermion}, and for Majorana fermion in Eq. \eqref{eq:sigmamu:Dfermion}. In both cases $r_F = M_F/M_S < 1$.

\subsection{DM Direct Detection}
\label{sec:DD}
Direct detection experiments provide strong bounds on the available parameter space of DM models. In our setup the most important constraints come from gauge interactions. It is well-known that bounds on DM--nucleus scattering due to tree-level $Z$-boson exchange exclude models with weak-scale Dirac fermion or scalar DM multiplets that have non-zero hypercharge. However, such models can still be viable, if they are just slightly modified. This is the case if there is a small Majorana mass term splitting the Dirac fermion DM into two Majorana states, or, in the case of scalar DM, if there is a mass splitting between CP-even and CP-odd components, see e.g.~Refs.~\cite{MDM, Hall:1997ah, TuckerSmith:2004jv}.
The $Z$-boson exchange then only leads to inelastic DM--nucleus scattering, which is kinematically forbidden for mass splitting of ${\cal O}(100 \, {\rm keV})$. In order to keep our discussion as general as possible, we therefore do not immediately discard models where a DM candidate is embedded into an $SU(2)_L$ multiplet with non-zero hypercharge, since even a tiny splitting or mixing can remove the constraints from direct detection experiments.  

\subsection{Electroweak Precision Observables}
Since at least some of the new states in both Class I and Class II models need to carry electroweak charges, and also couple to muons, the $Z$-couplings to muons are 
corrected at 1-loop. The corrections to the $Z \mu\mu$ vertex  due to heavy fermions and scalars running in the loop  parametrically scale as $\lambda^2/(16\pi^2) v^2/M^2$, where $M$ is the mass  scale of the heavy fields, and thus quickly decouple for $M\to \infty$. In the interesting regions of parameter space the resulting deviations in the coupling of $Z$ to muons, $\Delta g_{L,R}$, are therefore sufficiently small, as we have checked explicitly using the expressions in Appendix~\ref{Zmumu}. They are well below the experimental precision on $\Delta g_{L,R} \sim 10^{-3}$~\cite{PDG} in all parameter space regions that are not already excluded by direct searches at colliders.

\subsection{LHC Phenomenology}
\begin{figure}[t]
\begin{center}
\includegraphics[scale=0.7]{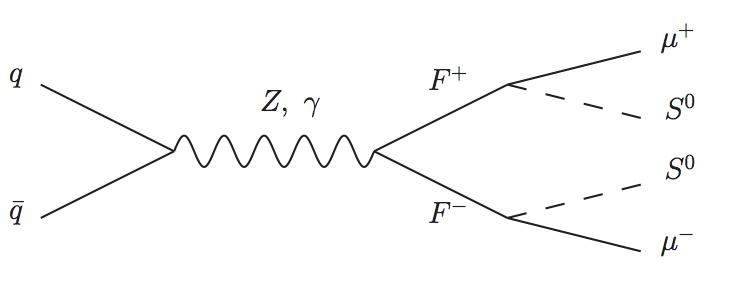}
\end{center} 
\caption{\label{fig:LHC} Drell-Yan production of a pair of heavy  charged fermions that subsequently decay to a scalar DM particle $S^0$ and a muon. Analogous diagram  for the case of fermion DM follows from replacing $F^+\to S^+$, $S^0\to F^0$, and 
resembles the production of supersymmetric smuons decaying into neutralinos.}
\end{figure}
The NP models that we consider must contain at least one new charged state, since a pair of new states must couple to the muon. 
The new charged particles, if sufficiently light, can be copiously produced in $pp$ collisions at the LHC through electroweak Drell-Yan production, i.e. through an $s$-channel $q\bar q \to \gamma^* /Z^*\to F\bar F$ (or $S S^*$) partonic process. Since they are odd under a conserved $Z_2$ symmetry, the NP particles then undergo a cascade of decays ending in the DM particle and SM states, in direct analogy to supersymmetric models with conserved R-parity. 

A decay channel that is always open for the lightest charged NP state is the decay to DM and a muon, through the very same couplings $\lambda^{L,R}_2$ that are
required by the $(g-2)_\mu$ diagrams. This leads to a signal topology of 2 opposite-charge muons and missing transverse energy (MET), see Fig.~\ref{fig:LHC}.
The LHC searches for events with $\mu^+\mu^-$ pairs in association with MET put severe constraints on the models without Higgs insertion, as we will discuss in Section \ref{sec:LL-RR-results}.

Other production modes besides Drell-Yan, as well as different decay modes, are possible for charged states that are part of $SU(2)_L$ multiplets. For instance,  associated production of a neutral and a singly-charged particle may be possible due to $s$-channel $W$-exchange.
In this case, in addition to the decay to muon and DM, the new charged states can also decay through emission of $W$ and $Z$ gauge bosons and, in the models with Higgs insertion, also Higgs bosons.

A detailed analysis of the LHC signatures will be performed in the next two Sections, for Class I models in Section \ref{sec:noHiggs} and for Class II models in Section \ref{sec:Higgs}.


\section{Models without Higgs Insertion}
\label{sec:noHiggs}
\setcounter{equation}{0}
We start our analysis with Class I models, i.e. the models without Higgs insertions. The defining feature of Class I models is that the diagrams contributing to $(g-2)_\mu$ have chirality flips only on the external lines, which implies that they are always proportional to the muon Yukawa coupling. Moreover, the sign of this contribution is fixed by the field content. 

The simple structure of the Class I models allows to spell out the Lagrangian as a general function of the $SU(2)_L$ quantum numbers, from which we will derive general expressions for the contribution to the $(g-2)_\mu$  and the DM annihilation cross-section. Using these analytical results we will identify viable models, which will turn out to be just the few models in which at least one of the new states is an $SU(2)_L$ singlet. 
Finally, we will perform a numerical analysis of these models including all constraints. 

\subsection{General Structure of LL Models}
\label{sec:general:LL}
The LL models contain two new Weyl fermions, $F_R \sim n^*_{(1/2 - Y)}$, $F_R^c \sim {n}_{(Y-1/2)}$ and a complex scalar, $S_R \sim (n\pm1)_{Y}$, that couple to the left-handed muon doublet, $\mu\sim 2_{-1/2}$, according to the Lagrangian in Eq.~\eqref{eq:LHmuon}.  The possible $S_R$ hypercharge assignments are dictated by requiring that there is at least one neutral state in the $F_R, F_R^c$, or $S_R$ multiplets, see Table~\ref{tab:LLRR}. From Eq.~\eqref{eq:LHmuon} one can derive the $SU(3)_c \times U(1)_{\rm em}$ Lagrangian  for LL models
\begin{align}
{\cal L}_{\rm LL} & =  {\cal L}_{\rm kin} + {\cal L}_{\rm gauge} +  \left( {\cal L}_{\rm yuk}^{n \pm1} + {\rm h.c.} \right), 
\end{align}
where the gauge boson couplings ${\cal L}_{\rm gauge}$ are spelled out in Appendix~\ref{appgauge}, and in terms of charge components of the new fields, 
\begin{align}
\label{eq:Lkin}
{\cal L}_{\rm kin}  & = \sum_{q_F} \Big(i \bar{F}_{q_F} \slashed{\partial} F_{q_F}  - M_F  \bar{F}_{q_F} F_{q_F}\Big)   +  \sum_{q_S} \Big(\partial_\mu S_{q_S}^* \partial^\mu S_{q_S} - M_S^2 S_{q_S }^*  S_{q_S }\Big) \, , 
\\
\begin{split}
\label{eq:L_yuk:LL}
{\cal L}_{\rm yuk}^{n \pm 1} & = \frac{\lambda_L }{\sqrt{{n - (1 \mp 1)/2}}} \sum_{q_F,q_S}    \left( \sqrt{\frac{n}{2}  \pm (q_F- Y+ 1) } \, \big(\bar{F}_{q_F} \mu_{L}\big)  S_{q_S}\delta_{q_S, q_F+1} \right.
\\
&\left.\qquad\qquad \mp    \sqrt{\frac{n}{2} \mp (q_F- Y) } \,  \big(\bar{F}_{q_F} \nu_{L}\big) S_{q_S }\delta_{q_S,q_F}  \right). 
\end{split}
\end{align}
where the Yukawa couplings are for $S_R\sim (n\pm1)_Y$, respectively. 
Here we also switched to a four-component notation, with the heavy Dirac fermion, $F_{q_F}$, defined as
\beq
F_{q_F} = 
\begin{pmatrix} 
F_{R, q_F}^c \\
F_{R, -q_F}^{ \dagger}
\end{pmatrix},
\eeq
and introduced the weak isospin components of the muon doublet $(\nu_L,\mu_L)$. The components of $S_R$ with charge $q_S$ are denoted as $S_{q_S}$. The label $q_F$ runs over the electric charges of the fermionic components, $q_F\in\{(Y-n/2),\ldots, (Y+n/2-1)\}$, while  $q_S\in\{Y-(n\pm1-1)/2,\ldots, Y+(n\pm 1-1)/2\}$ are the charges of scalar field components.  The prefactor in \eqref{eq:L_yuk:LL} ensures that $\lambda_L$ is the largest Yukawa coupling appearing in ${\cal L}_{\rm yuk}^{n \pm 1}$. 

\subsection{General Structure of RR Models}
\label{sec:general:RR}
The RR models contain two new Weyl fermions, $F_L \sim {n}^*_Y$, $F_L^c \sim n_{-Y}$ and a complex scalar, $S_L \sim  n_{-Y-1}$. The Lagrangian takes the form
\begin{align}
{\cal L}_{\rm RR} & =  {\cal L}_{\rm kin} + {\cal L}_{\rm gauge} + \big( {\cal L}_{\rm yuk} + {\rm h.c.}\big) \, , 
\end{align}
where the kinetic term takes the same form as in Eq.~\eqref{eq:Lkin}, the gauge boson couplings can be found in Appendix~\ref{appgauge}, while the Yukawa term is   
\beq
{\cal L}_{\rm yuk} =  \lambda_{R} \sum_{q_F, q_S} \bar{\mu}_{R} F_{q_F} S_{q_S} \delta_{q_F,-q_S-1}.
\eeq
The fermion charge runs over $q_F\in \{-(n-1)/2+Y, \ldots, (n-1)/2+Y\}$ and the scalar charge over $q_S\in\{-(n-1)/2-Y-1, \ldots, (n-1)/2-Y-1\}$. Above we simplified  the notation for the scalar, $S_L \to S$, and used Dirac fermion notation,
\begin{align}
F_{q_F} & = 
\begin{pmatrix} 
F_{L,q_F} \\
F_{L,-q_F}^{c \dagger}
\end{pmatrix}  \, , &
\mu_{R} & = 
\begin{pmatrix} 
0 \\
\mu^{c \dagger}
\end{pmatrix} \, .
\end{align}

\subsection{Models compatible with $(g - 2)_\mu$}
The contributions to $(g-2)_\mu$ for RR models are, using Eq.~\eqref{eq:Deltaamu:generalLagr},
\begin{align}
\Delta a_{\mu}^{\rm RR} &= - \frac{n \, m_\mu^2}{8 \pi^2 M_S^2}|\lambda_{R}|^2   \left[ f_{LL}^S + Y_{F_L} \left( f_{LL}^S + f_{LL}^F \right) \right]  \,,  
\label{g2RR}
\end{align}
where we have neglected loop-induced mass splittings within the $SU(2)_L$ multiplets, since they only amount to higher order corrections. The loop functions $f^S_{LL}, f^{F}_{LL}$ are given in Eqs.~\eqref{eq:fF}, \eqref{loopf}. These functions are both
positive definite, and satisfy the inequality $2 f_{LL}^S(x)\geq f_{LL}^F(x)\geq  f_{LL}^S(x)/2$. 
The RR models with $Y_{F_L} \ge -1/3$  can therefore be discarded since they give a negative contribution to $\Delta a_\mu$. 
The RR models with $Y_{F_L} \le -2/3$ are viable candidates to explain the $(g-2)_\mu$ anomaly as they always give a positive NP contribution to $\Delta a_\mu$, while the RR models with $Y_{F_L} = -1/2$ are viable only if $M_F < M_S$. 

In the LL models the NP contribution to $(g-2)_\mu$ is, for  $n_F =n$, $ n_S=n \pm 1$,
\begin{align}
\Delta a_{\mu}^{\rm LL} &=  - \frac{(n\pm1) \, m_\mu^2}{16 \pi^2 M_S^2} |\lambda_{L}|^2   \frac{n}{n- (1 \mp 1)/2} \left[ f_{LL}^S + \left( Y_{S} +   \frac{ \pm n - 4}{6} \right) \left( f_{LL}^S + f_{LL}^F \right) \right] \, . 
\label{g2LL}
\end{align}
The LL models  with $Y_S \ge  ({\mp n + 2})/{6}$ cannot explain the $(g-2)_\mu$ anomaly, since the predicted $\Delta a_{\mu}$ is always negative, irrespectively of the $M_F/M_S$ ratio. In contrast, the LL models with $Y_S \le  {\mp n}/{6}$ always give positive $\Delta a_{\mu}$ and are viable candidates for explaining $(g-2)_\mu$, as are the LL models with $Y_S = ({\mp n +1})/{6}$, but only if $M_F < M_S$. 

Taking the complete list of LL and RR models in Table \ref{tab:LLRR} and dropping the models with negative $\Delta a_\mu$ gives the field content of viable models in Table \ref{g2viable}. These already incorporate the occasional requirement $M_F < M_S$,  
which in those cases fixes the DM candidate to be a fermion. Note that, in order not to clutter the notation, we do not distinguish between $n$ and $n^*$ in the Tables.

\begin{table}[t]
\centering
\begin{tabular}{c||cccccc}
 $~{\rm LL}~$   & ~LL0~  &~LL1~ &  &  &  &  \\
 \hline
 $F_R$ & $1_1$ & $2_{\frac{1}{2}}^\star$ & $2_{\frac{3}{2}}$ &$3_{0}^\star$ &$3_{1}^\star$ \\
 $S_R$ &   $2_{- \frac{1}{2}}^\star$ &    $1_{0}^\star$ &  $3_{-1}^\star$ &   $2_{\frac{1}{2}}^\star$  & $2_{- \frac{1}{2}}^\star$  
 \end{tabular}
 \hspace{0.3cm}
 \begin{tabular}{c||cccccc}
 $~{\rm RR}~$ &~RR1~ & &&&& \\
\hline
$F_L$  &   $1_{-1}$ & $2_{- \frac{1}{2}}^\star$ & $2_{- \frac{3}{2}}$ & $3_{-1}^\star$ & $3_{-2}$ \\
 $S_L$ &  $1_{0}^\star$ &  $2_{- \frac{1}{2}}$ & $2_{\frac{1}{2}}^\star$ &  $3_{0}^\star$  &  $3_{1}^\star$ 
\end{tabular}
\caption{ \label{g2viable} LL and RR models that give a positive contribution to $(g-2)_\mu$. }
\end{table}

\subsection{Models compatible with $(g - 2)_\mu$ and DM relic density}
\label{sec:LLRR:relic:g-2}
Requiring that DM is a thermal relic excludes LL and RR models with large multiplicities of states, as they are too heavy to account for $(g-2)_\mu$. Indeed for $m_\chi \gg m_W$ and $n \gg 1$ 
the effective cross section for annihilation of dark multiplets into gauge bosons scales as $\sigma\propto n^3/m_\chi^2$, see Eqs.~\eqref{annMDM:scalar}, \eqref{annMDM:fermion}. In order to reproduce the measured relic density, $\Omega_\chi\propto 1/\langle \sigma v\rangle\propto m_\chi^2/n^3$, the mass of the DM candidate, and therefore the mass of the whole $n$-plet, needs to scale as $m_\chi \propto n^{3/2}$. On the other hand, the NP contribution to $(g-2)_\mu$ scales roughly as $\Delta a_\mu\propto n/m_\chi^2\propto n^{-2}$, implying an upper bound on $n$.
 Below we refine this argument for both RR and LL models, and show that only two models are left as potential candidates.
\paragraph{RR Models.}
We begin with the RR models that have $Y_{F_L} \le -2/3$ and a scalar DM candidate. In this case the NP contribution to $(g-2)_\mu$, Eq.~(\ref{g2RR}), is maximized for $M_S \lesssim M_F$.  This gives $f_{LL}^F \approx f_{LL}^S \approx 1/24$ and an upper bound 
\begin{align}
\Delta a_{\mu}^{\rm RR} & \le   \frac{n \, m_\mu^2}{192 \pi^2 M_S^2}|\lambda_{R}|^2   \left( 2 |Y_{F_L}| -1 \right)   \, . 
\end{align}
This translates to an upper bound on $m_\chi= M_S$, 
\begin{align}
m_\chi \le \big(0.26 \TeV\big) \times  \frac{|\lambda_{R}|}{\sqrt{4 \pi}}  \times  \sqrt{n (2 |Y_{F_L}|-1)}  \, ,
\end{align}
when one requires that the NP contribution to $(g-2)_\mu$ is within $2 \sigma$ of the measured central value, i.e., to have at least a NP shift of $\Delta a_\mu = 1.1 \times 10^{-9}$. This implies a lower bound on the annihilation cross section for $\chi\chi^\dagger \to WW, ZZ$, and thus, through use of Eqs.~(\ref{Omapp})~\eqref{annMDM:scalar} an upper bound on  
the relic density, 
\begin{align}
\label{eq:bound:Omega}
\Omega h^2 \le  \frac{0.10 \, n^2 \left( 2 |Y_{F_L}| - 1 \right)}
{1.8 \left(n^4 -4 n^2 +3\right) + 2.7 \, Y_{S_L}^4 + 4.3 Y_{S_L}^2 \left(n^2- 1\right)}\left(\frac{|\lambda_{R}|^2}{4 \pi} \right) \left( \frac{x_f}{30} \right) \left({\frac{50}{g_*}}\right)^{1/2}. 
\end{align} 
In the above inequality we assumed that the annihilation to gauge bosons is kinematically allowed, and also neglected mass splittings in dark multiplet.  Setting aside these caveats, the bound in Eq.~\eqref{eq:bound:Omega} also applies, if annihilations to muons are sizeable or non-perturbative corrections are taken into account, since an additional annihilation channel or Sommerfeld-enhancement factors would only reduce the relic density. 
For $n \ge 2$  one obtains $\Omega h^2 \le 0.10 $, which is below the required value of $0.12$ and already corresponds to the maximal possible value. The discrepancy becomes progressively worse for larger $n$, with the upper bound scaling as $\propto 1/n^2$. Therefore, out of the RR models in Table \ref{g2viable} with scalar DM only the  
 $n=1$ model, denoted henceforth as RR1, is potentially viable. The RR models with fermionic DM obey an even more stringent bound on the relic density, $\Omega h^2 \le 0.03 $. Therefore, RR1 is the only RR model that may simultaneously account for $(g-2)_\mu$ and give the correct relic abundance.

\paragraph{LL Models.}
 We now turn to the LL models, starting with scalar DM models that have  $Y_{S_R}  \le \mp n/{6} $. The NP contribution to $(g-2)_\mu$ is maximized for $M_S \lesssim M_F$, in which case $f_{LL}^S\approx f_{LL}^F=1/24$.  Eq.~(\ref{g2LL}) then translates to the following upper bound, 
\begin{align}
\Delta a_{\mu}^{LL} &\leq  \frac{(n\pm1) \, m_\mu^2}{384 \pi^2 M_S^2} |\lambda_{L}|^2    \left[ \frac{1}{3}  \mp    \frac{   n }{3} -2 Y_{S_R} \right] \frac{n}{n - (1\mp1)/2}\, .
\end{align}
This, in turn, implies an upper bound on $m_\chi$, when requiring that the NP contribution brings the prediction for $(g-2)_\mu$ within 2$\sigma$ of the measured value, 
\begin{align}
m_\chi \le \big(0.18 \TeV\big)\times  \frac{|\lambda_L|}{\sqrt{4 \pi}}\times \sqrt{ n_S  \left[ \frac{2}{3}  \mp  \frac{   n_S }{3} -2 Y_{S_R} \right]}  \sqrt{\frac{n_S \mp 1}{n_S -1/2  \mp1/2}}\,,
\label{eq:gm2-bound}
\end{align}
where $n_S=n\pm1$.
This translates to the following upper bound on the relic density,
\begin{align}
\Omega h^2 \le  \frac{0.01 \, n_S^2 \left(2 \mp n_S - 6 Y_{S_R}  \right)}{n_S^4 + n_S^2 \left( 2.5 Y_{S_R}^2 - 4 \right) + 1.5 Y_{S_R}^4 - 2.5 Y_{S_R}^2 + 3 } \frac{n_S \mp 1}{n_S -1/2  \mp1/2} \left( \frac{|\lambda_L|^2}{4 \pi} \right) \left( \frac{x_f}{30} \right) \left( {\frac{50}{g_*}}\right)^{1/2}. 
\end{align}
For the LL models in Table \ref{g2viable} with  $n_S \ge 2$ one has  $\Omega h^2 \le 0.08 $, which leaves the $n_S=1$ model, denoted by LL1, as the only viable option. All the LL models with fermionic DM give $\Omega h^2 \le 0.03$ and are thus disfavored. 

To summarize, the above analysis leaves only LL1 and RR1 models as the candidate models that could explain, without tuning,  the observed $(g-2)_\mu$ anomaly and give the correct DM relic density. In the remainder of the section we perform a detailed phenomenological analysis of these two models. Further details, including explicit expressions for the Lagrangians, are collected in Appendix~\ref{LLRRmodels}. As we have stressed, the above conclusions are valid only if DM is heavier than the $Z$-boson. Lighter DM is strongly constrained by LHC searches. Moreover, models with light DM would also typically lead to overabundant relic density, unless one has an efficient annihilation channel, for example through resonant Higgs exchange. Such models can then still potentially explain both $(g-2)_\mu$ and DM, but require significant tuning. 
We will illustrate this below with one example, the $n=1$ LL model, which we denote by LL0 in Table~\ref{g2viable}. The corresponding Lagrangian can be found in Appendix~\ref{LLRRmodels}.
\subsection{Numerical Results}
\label{sec:LL-RR-results}
Before we discuss the final results, we briefly recall the structure of the two potentially viable models, LL1 and RR1.  The LL1 (RR1) model contains a doublet (singlet)  heavy vectorlike lepton and a singlet scalar  that is the DM candidate in both models, i.e., $M_S < M_F$. Indeed there is no viable DM candidate if $M_S>M_F$, since in that case in the RR1 model the lightest $Z_2$-odd particle would be a charged fermion, while in the LL1 model the stable particle would be the neutral component of a fermion doublet. The latter is excluded by direct detection experiments because of the vector coupling of DM to $Z$ induced by $Y_F\neq0$, unless the model is extended by adding another field that mixes with the doublet, as discussed in Section~\ref{sec:DD}. Even then the relic density of the neutral fermion is lower than the observed value for $M_F \lesssim 1.1$ TeV, so that LL1 with $M_S>M_F$ cannot simultaneously explain DM and $\Delta a_\mu$. 

\begin{figure}[t!]
\begin{center}
\includegraphics[width=0.4\textwidth]{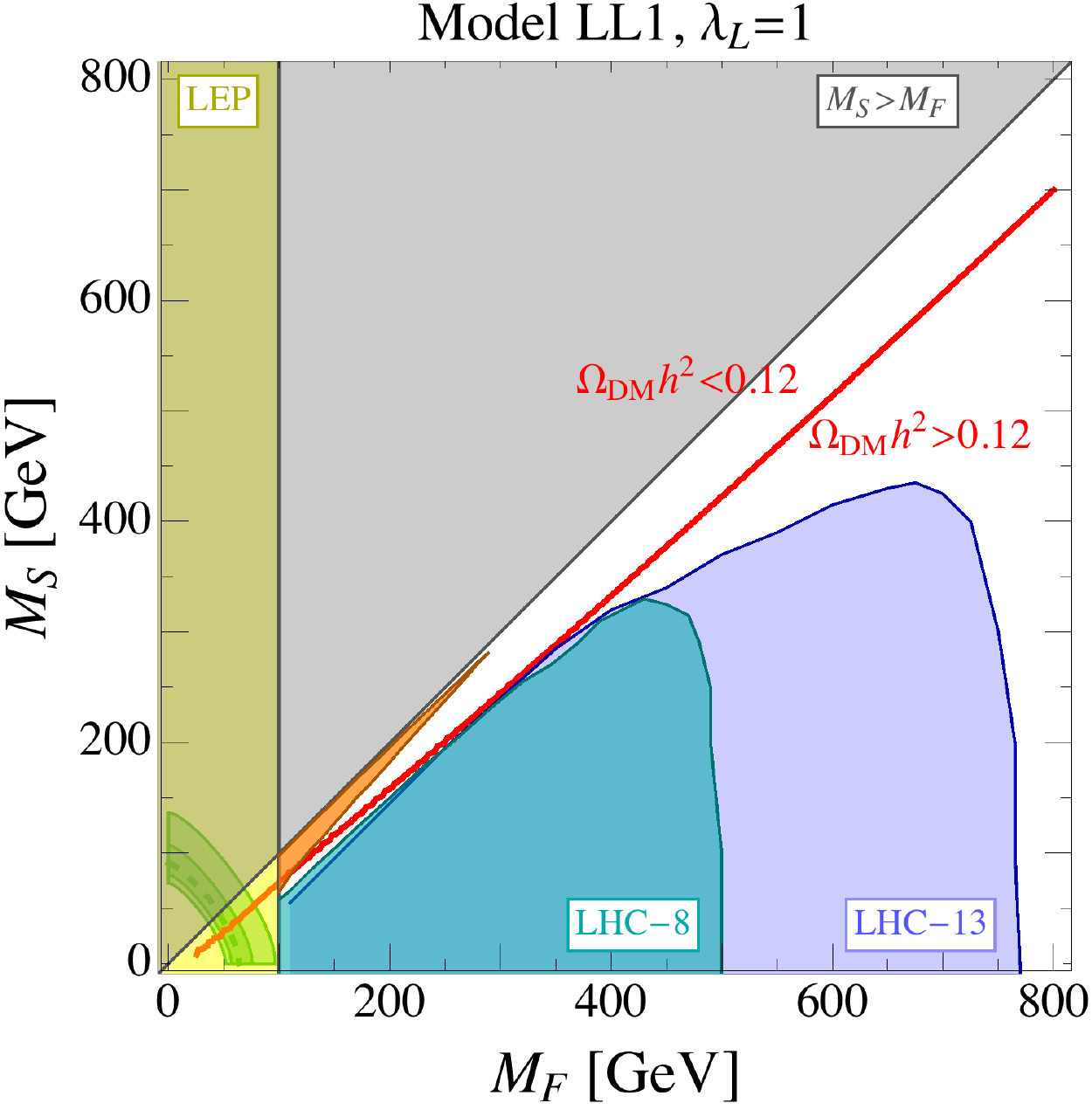}
\hspace{0.3cm}
\includegraphics[width=0.4\textwidth]{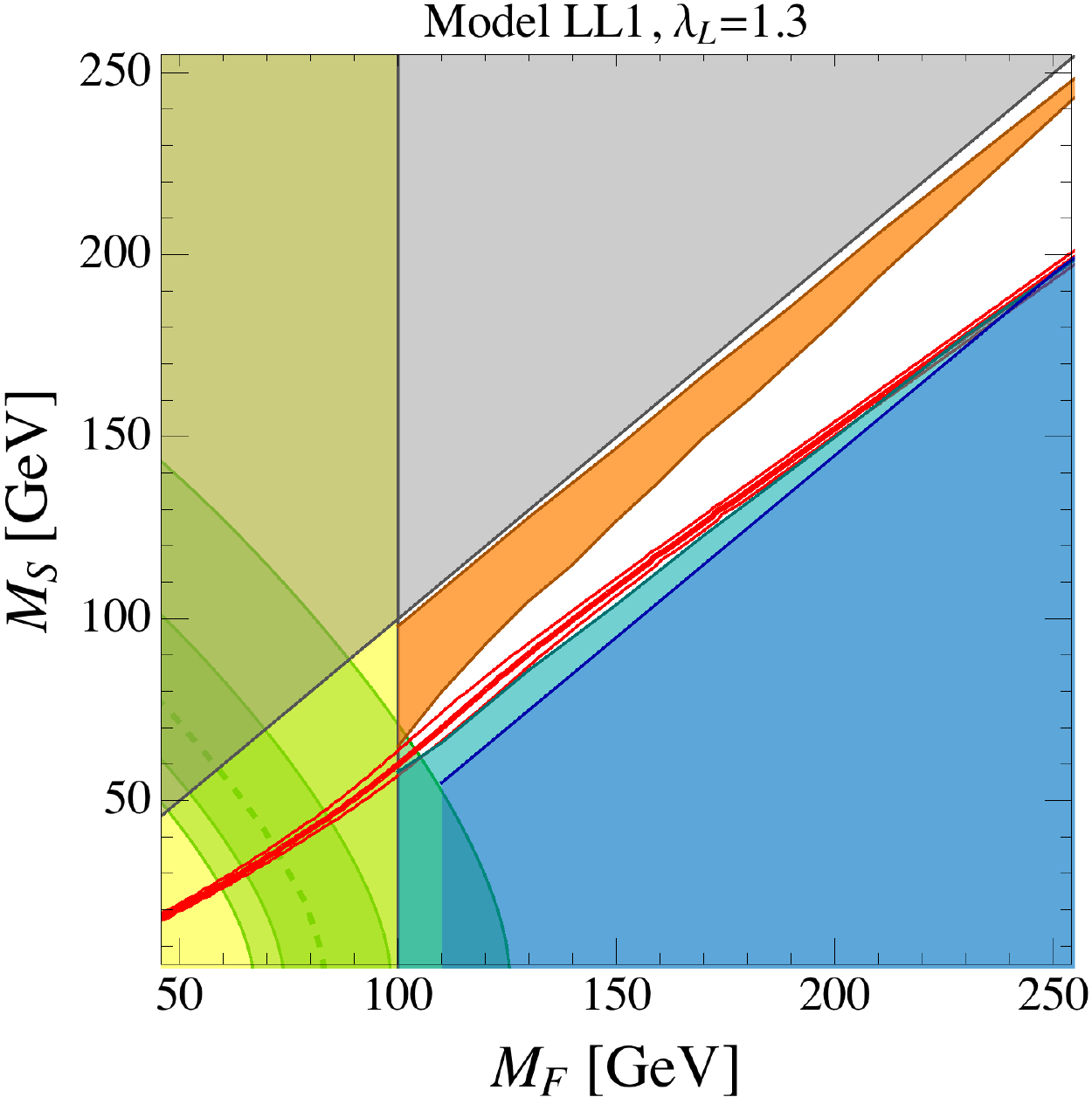}\\
\vspace{0.3cm}
\includegraphics[width=0.4\textwidth]{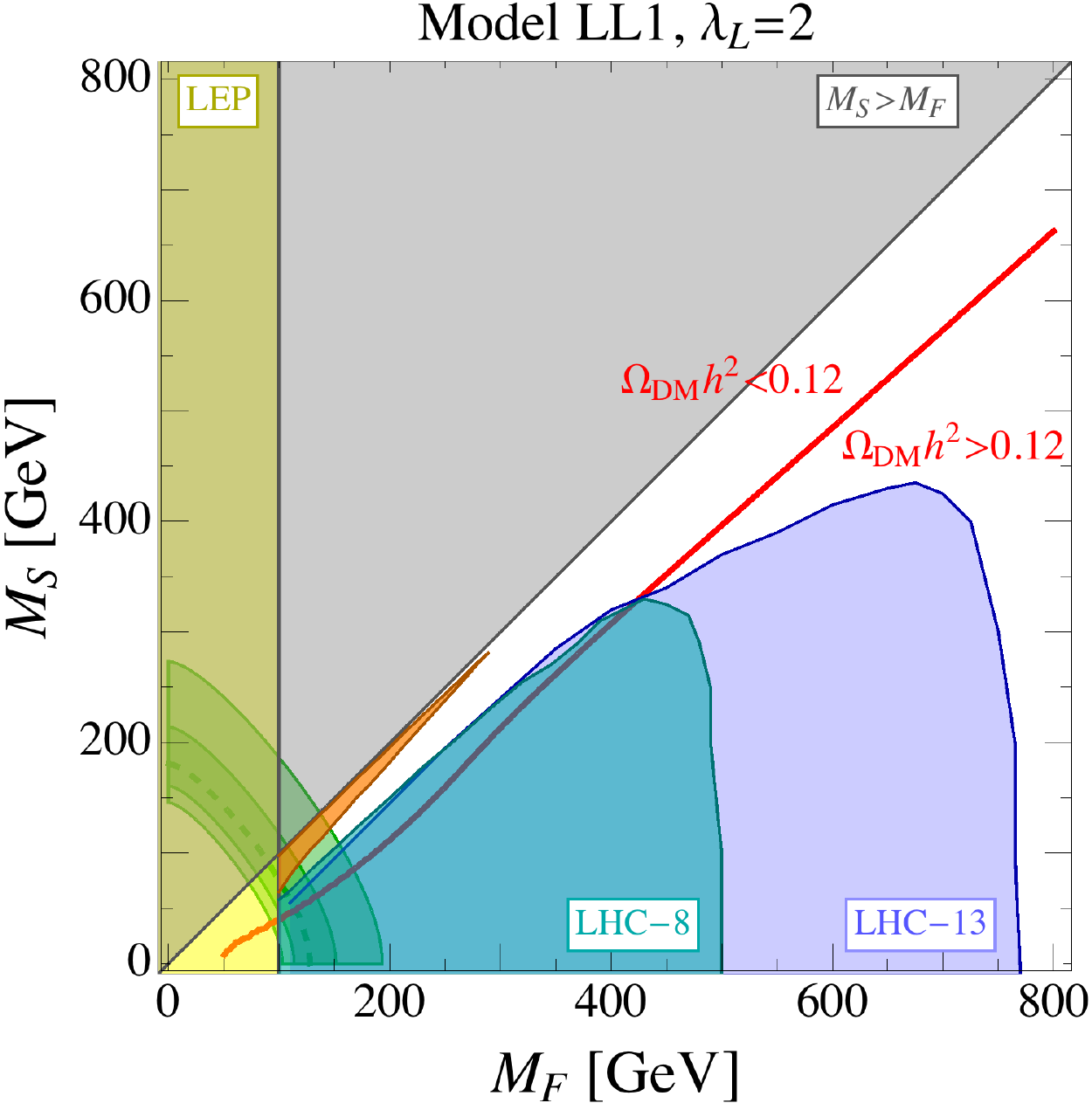}
\hspace{0.3cm}
\includegraphics[width=0.4\textwidth]{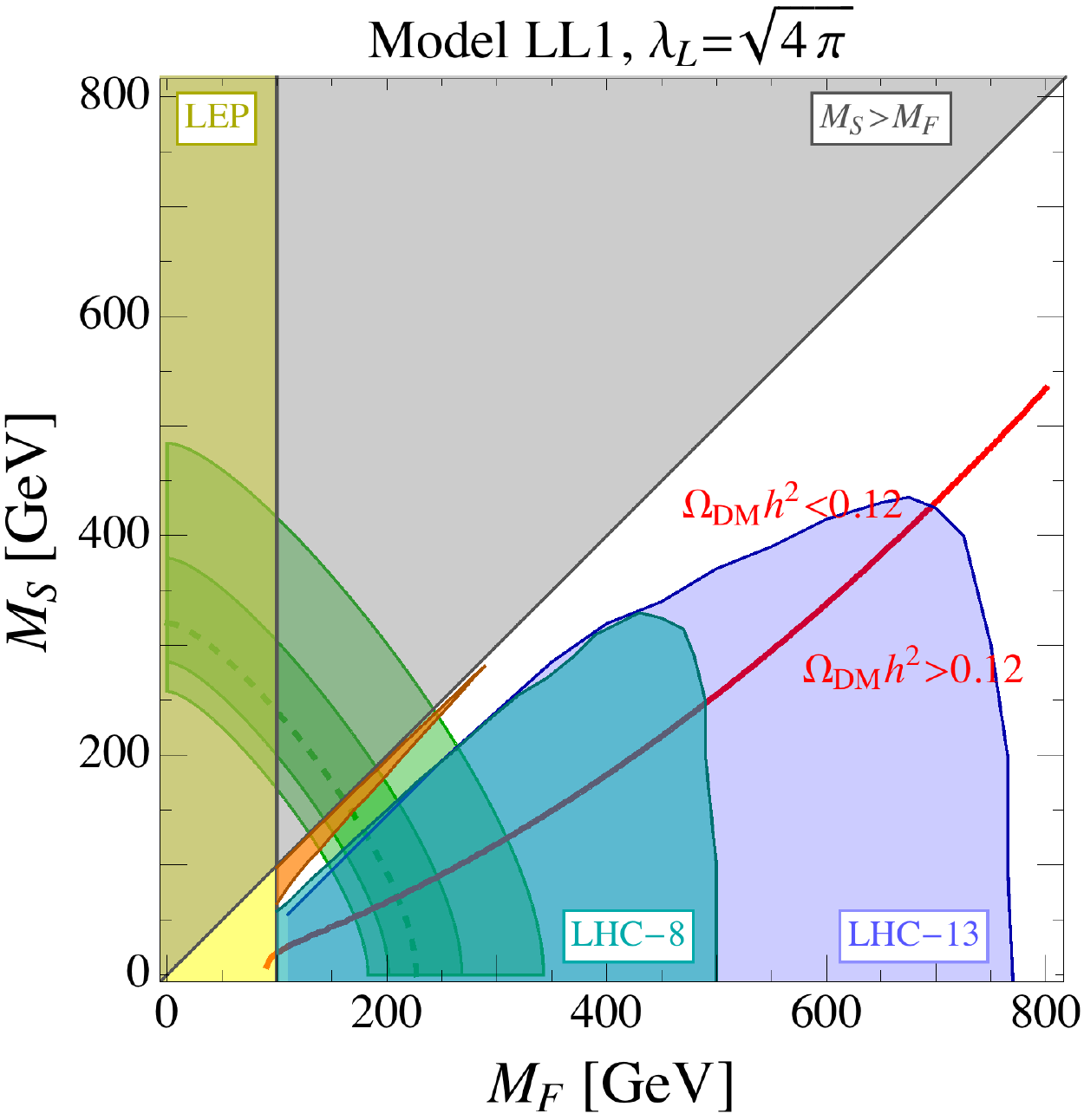}
\caption{Results for the LL1 model in the $M_F - M_S$ plane for increasing values of the coupling to leptons. The gray-shaded regions are excluded because the DM candidate needs to be scalar (see main text), hence $M_S < M_F$. In the dark (light) green region the total contribution to $(g-2)_\mu$  is compatible with the experimental value at $1 \,(2) \sigma$, and the red line indicate where the DM relic density is $\Omega h^2 = 0.12$ (in the upper right plot the red band corresponds to the conservative range $0.10 < \Omega h^2 < 0.14$). 
The yellow region is excluded by searches for heavy charged fermions at LEP \cite{Heister:2002mn,Abdallah:2003xe}, the cyan region (denoted as LHC-8) is excluded by $\sqrt{s}=$ 8 TeV LHC searches  \cite{Aad:2014vma,TheATLAScollaboration:2013hha}, the blue region (LHC-13) by $\sqrt{s}=$ 13 TeV searches \cite{ATLAS:2016uwq,ATLAS:2017uun}, and the orange area by the CMS soft leptons search \cite{Sirunyan:2018iwl}.
\label{fig:LL1}}
\end{center} 
\end{figure}

\begin{figure}[t!]
\begin{center}
\includegraphics[width=0.4\textwidth]{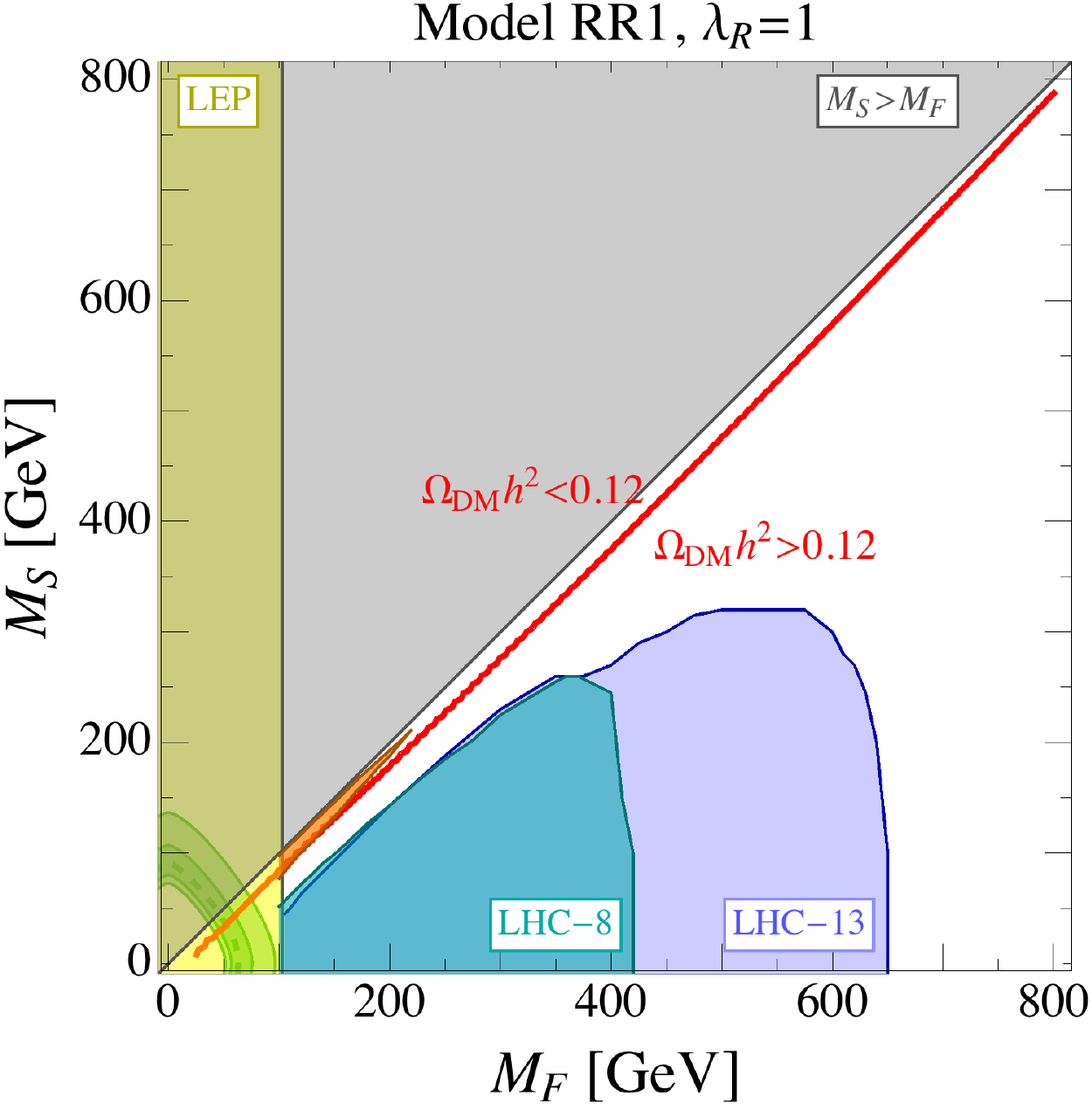}
\hspace{0.3cm}
\includegraphics[width=0.4\textwidth]{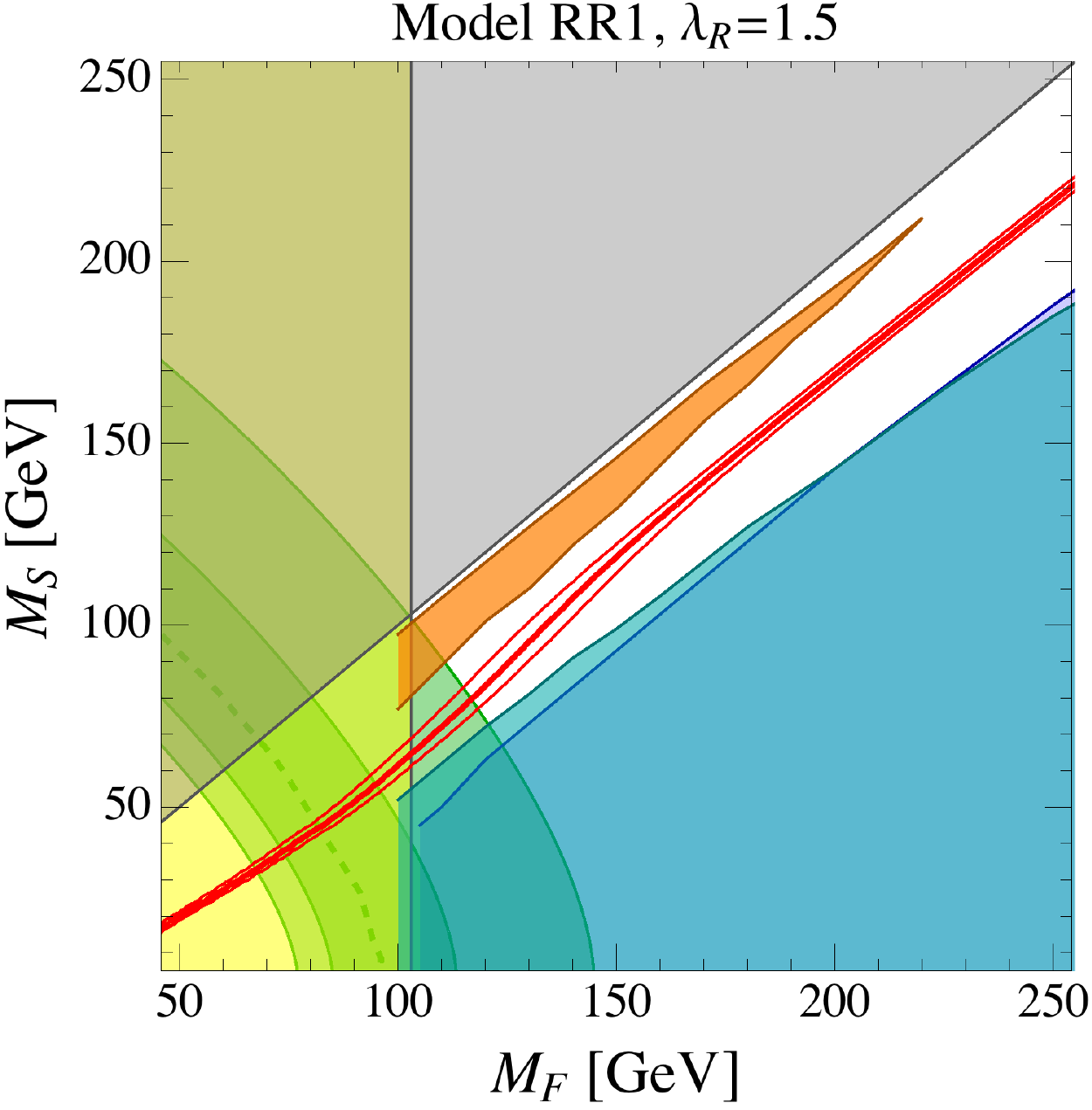}\\
\vspace{0.3cm}
\includegraphics[width=0.4\textwidth]{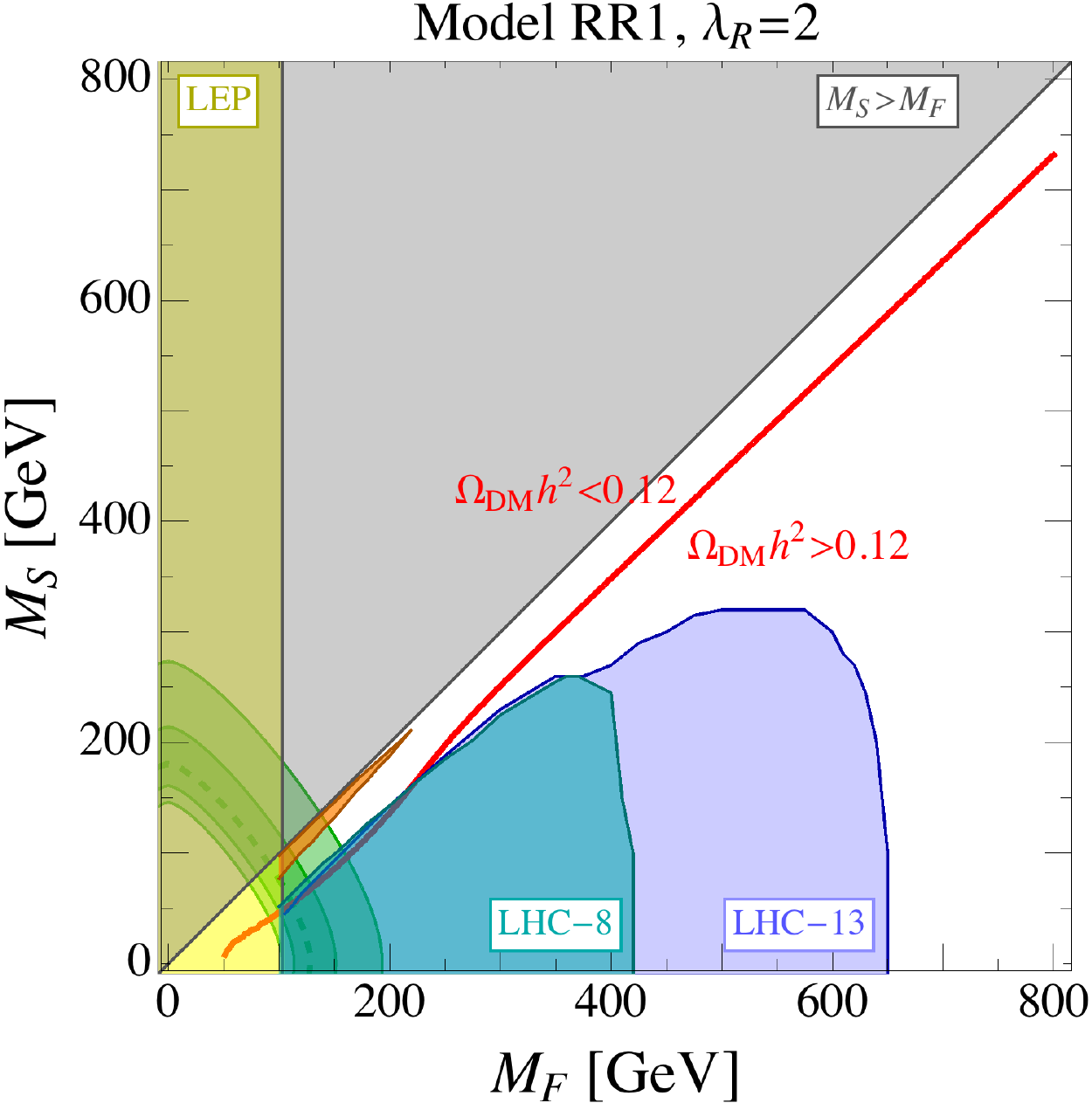}
\hspace{0.3cm}
\includegraphics[width=0.4\textwidth]{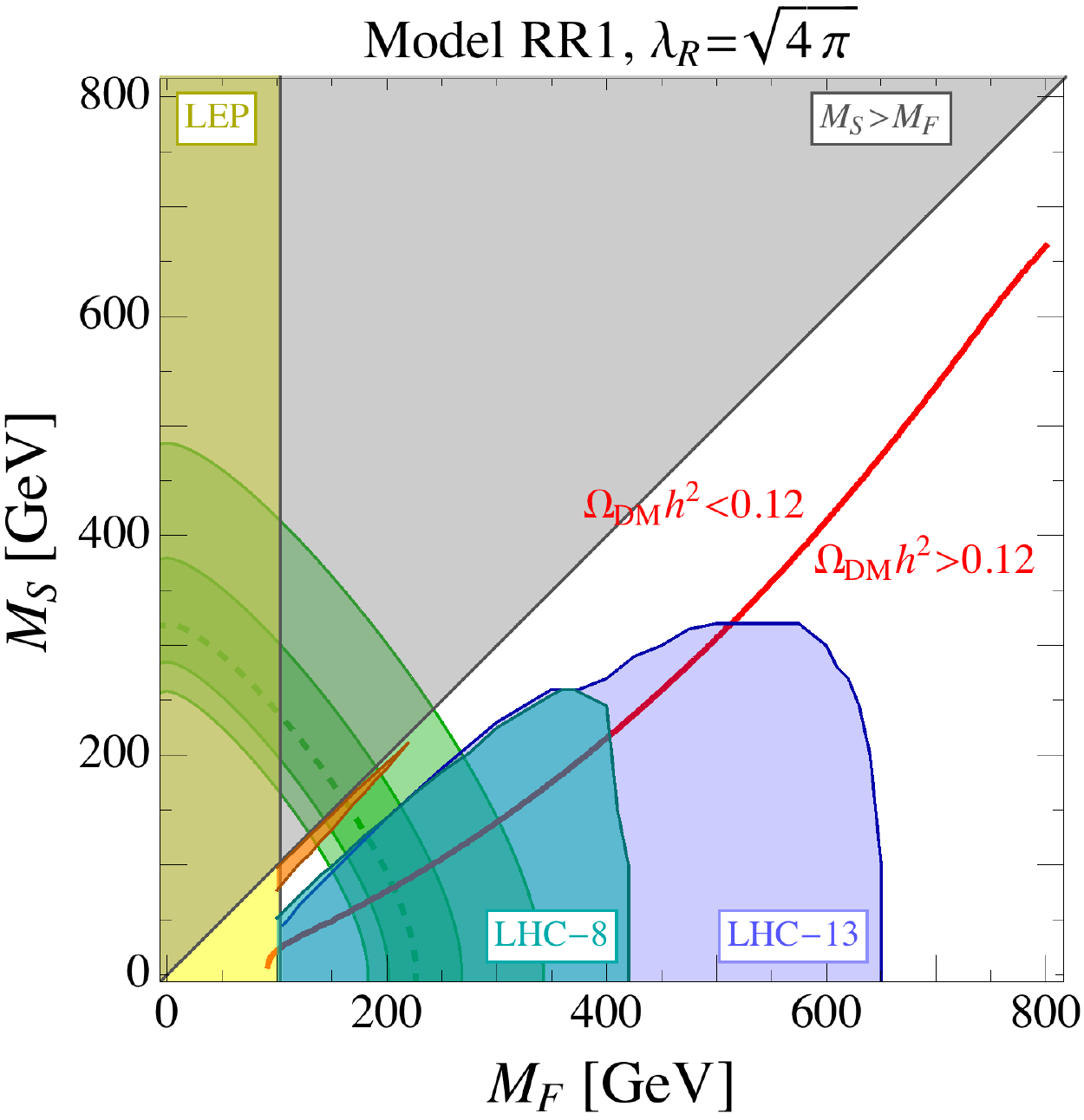}
\caption{\small Same as Fig.~\ref{fig:LL1} for the RR1 model. \label{fig:RR1} } 
\end{center}
\end{figure}

The allowed regions in the $M_F$--$M_S$ plane for LL1 and RR1 models are shown in Figs.~\ref{fig:LL1} and \ref{fig:RR1}, respectively,  fixing the couplings to muons, $\lambda_{L,R}$, to several representative values. 
In the dark (light) green regions  in Figs.~\ref{fig:LL1} and \ref{fig:RR1} the predicted $(g-2)_\mu$  is compatible with the experimental value in Eq.~(\ref{eq:gmu})
within $1\, (2) \sigma$, i.e.~$\Delta a_\mu\in [2.07\, (1.27),\,3.67\, (4.47)] \times 10^{-9}$, while the current experimental central value, $\Delta a_\mu=2.87 \times 10^{-9}$, is reached on the green dashed line.

The red line in Figs.~\ref{fig:LL1} and \ref{fig:RR1} indicates where the abundance of the neutral scalar matches the observed DM relic density, $\Omega_{\rm DM}h^2 \approx 0.12$ \cite{Ade:2015xua,Abbott:2017wau}. This is realized by means of DM annihilating to $\mu^+\mu^-$ via a $t$-channel fermion exchange, as shown in Fig.~\ref{fig:DMann}, and/or fermion-scalar coannihilation modes.  Since DM is a singlet there are no annihilations to gauge bosons. We computed the relic density numerically using {\tt micrOMEGAs}~\cite{Belanger:2014vza,Belanger:2013oya}. The results are in excellent agreement with the approximate expressions in Section~\ref{relicapp}, apart from regions of parameter space where coannihilations are important, since these were not covered in Section~\ref{relicapp}. Above (below) the red line DM annihilation rate is too large (small), giving a relic density that is below (above) the observed DM abundance. 
To keep the analysis minimal, we do not switch on the Higgs-portal coupling,  $\kappa S^2 |H|^2$, cf.~Appendix~\ref{LLRRmodels}. This coupling would open the possibility of achieving the correct relic density for $M_S\approx m_h/2$ through DM annihilation via the Higgs resonance \cite{Burgess:2000yq}, but would otherwise not change the conclusions of the analysis below.
\begin{figure}[t!]
\begin{center}
\includegraphics[width=0.4\textwidth]{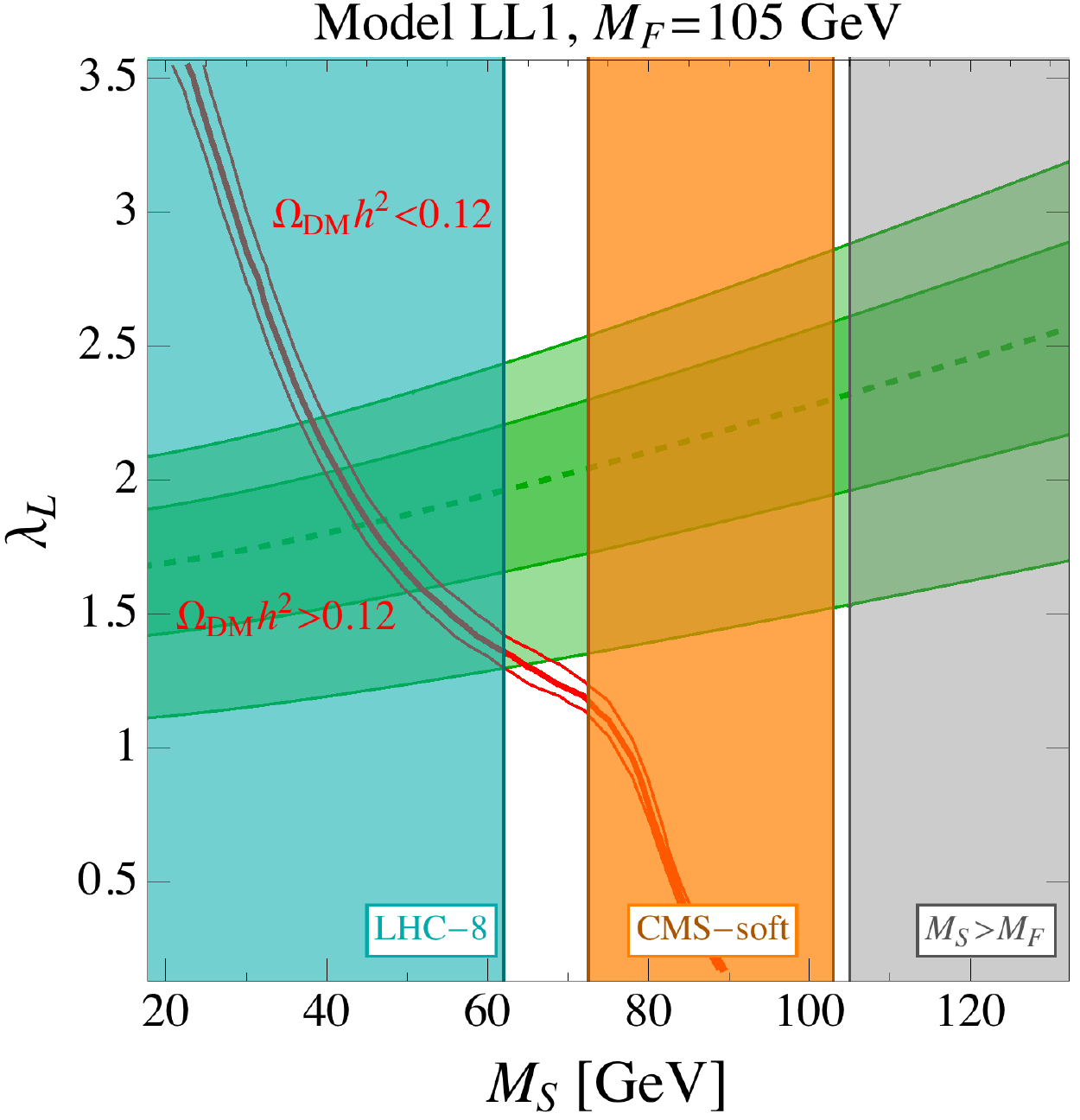}
\hspace{0.3cm}
\includegraphics[width=0.4\textwidth]{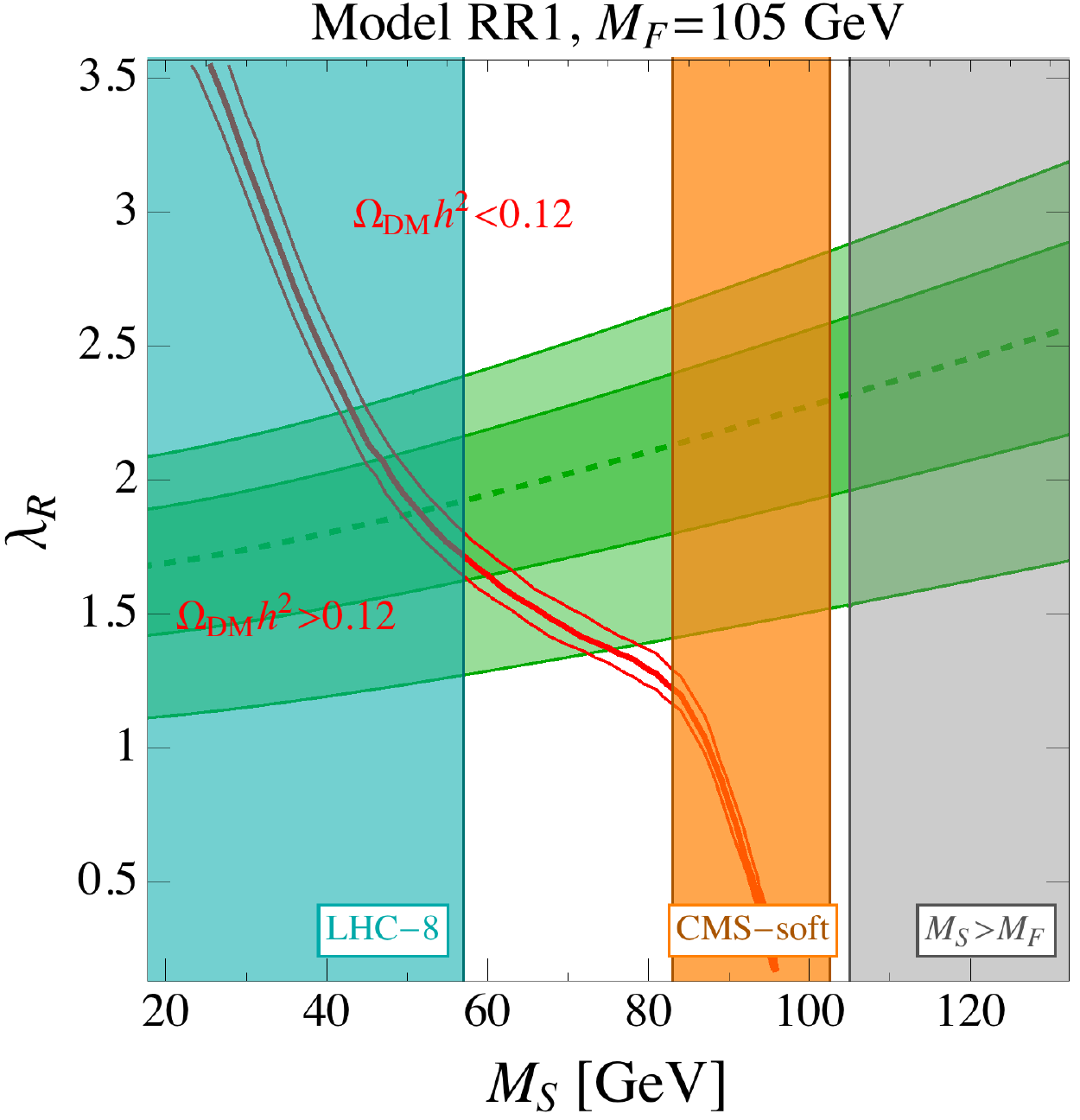}\\
\vspace{0.3cm}
\includegraphics[width=0.4\textwidth]{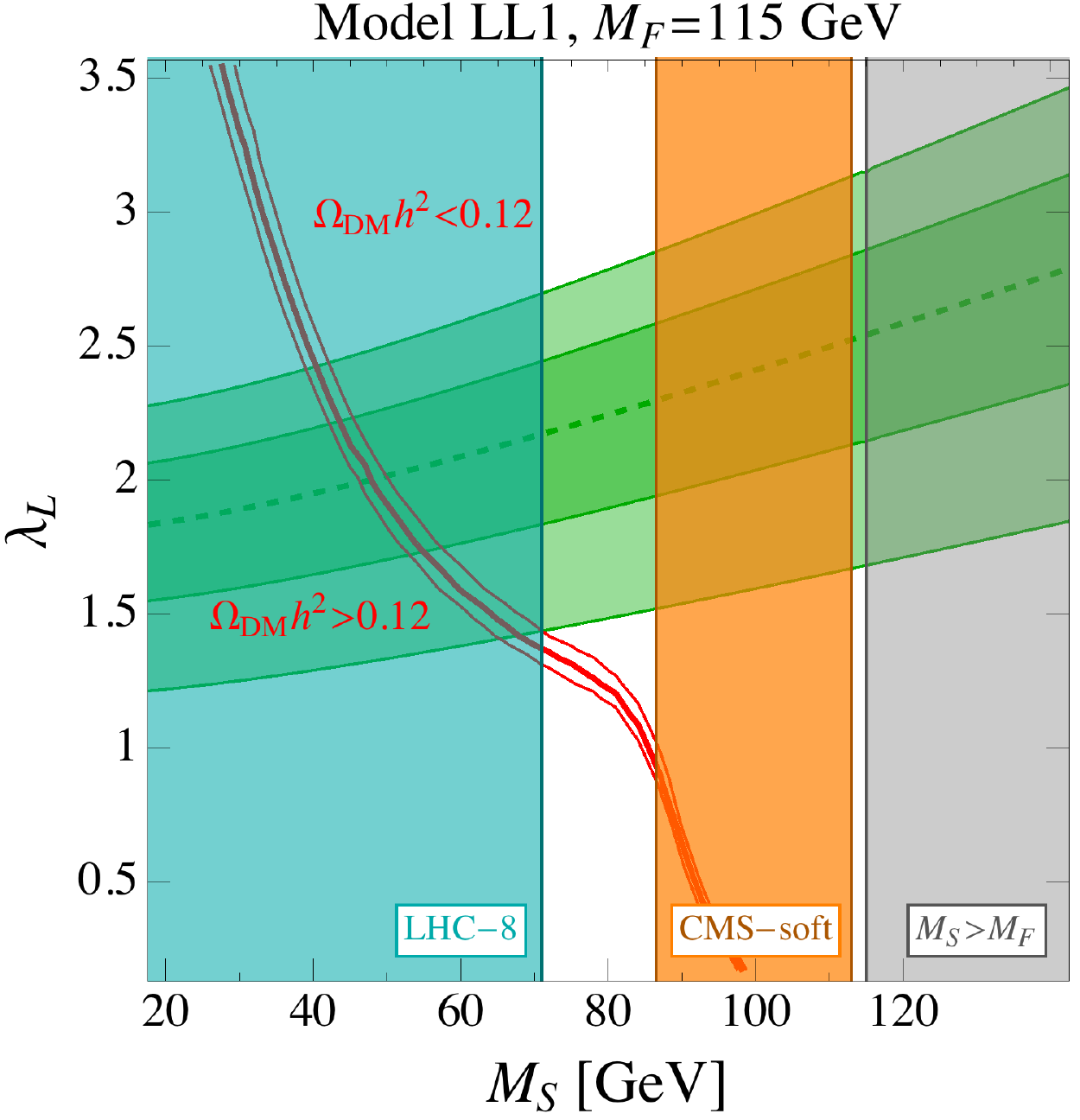}
\hspace{0.3cm}
\includegraphics[width=0.4\textwidth]{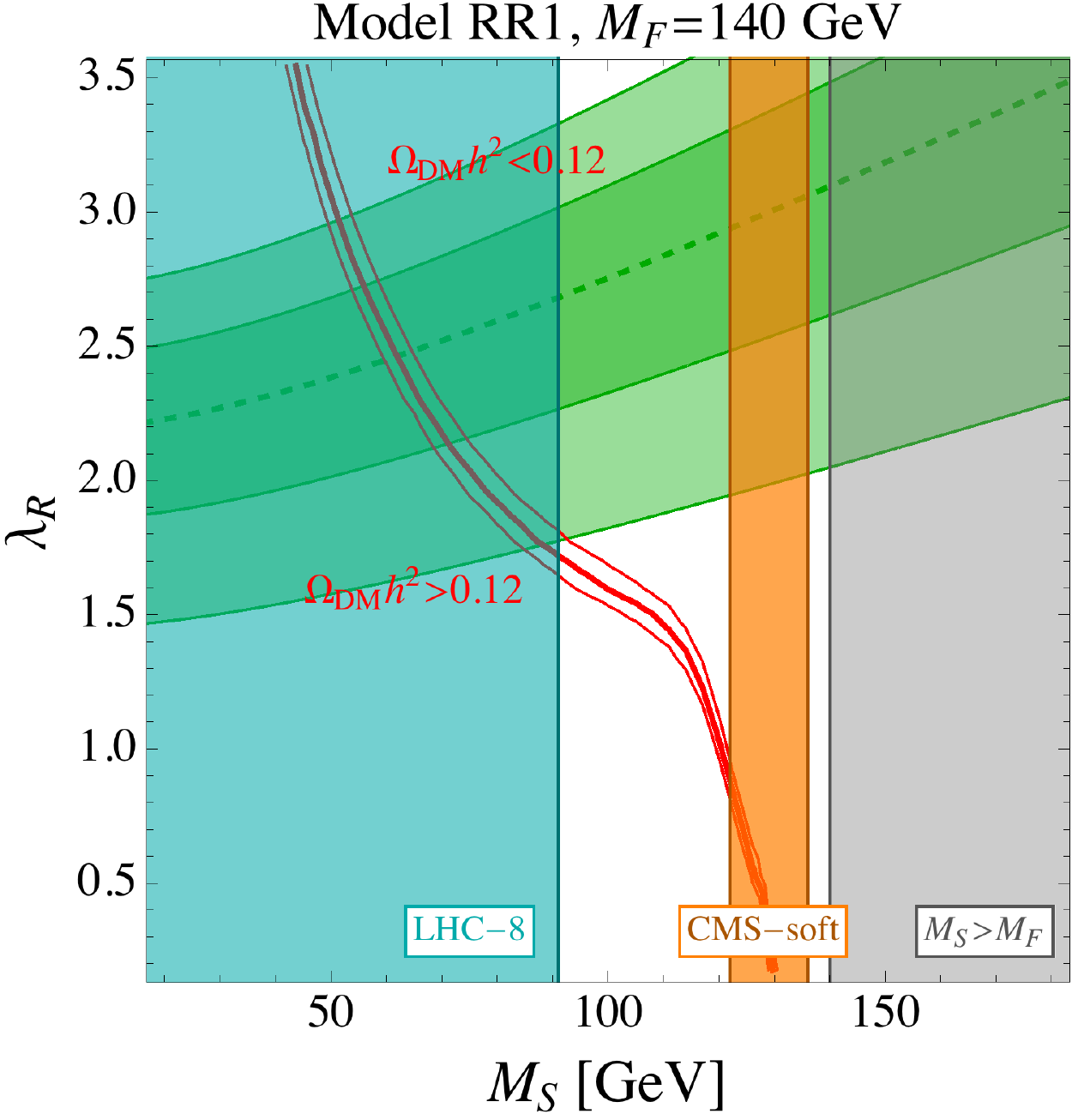}
\caption{Results for the LL1 (first column) and RR1 (second column) model in the $M_S - \lambda_{L,R}$ plane for different choices of $M_F$. Colored regions as in the previous figures. \label{fig:MS-lam}}
\end{center} 
\end{figure}

The overlap of green regions and red lines in Figs.~\ref{fig:LL1} and \ref{fig:RR1} confirms that within the two simple models LL1 and RR1 one can simultaneously obtain the correct relic density and fit $(g-2)_\mu$, as suggested by the analytical results above. However, in order to compensate for the large suppression of NP contributions due to the external Higgs insertions giving $\Delta a_\mu\propto m_\mu$, a sufficiently large contribution to $(g-2)_\mu$ requires rather light new fermions, $M_F\lesssim 350$ GeV, so that
direct searches for new charged fermions at colliders are relevant. 
In Figs.~\ref{fig:LL1} and \ref{fig:RR1} the region excluded by LEP experiments \cite{Heister:2002mn,Abdallah:2003xe}, $M_F\lesssim 100$ GeV, is indicated in yellow. 
Our estimate of the regions excluded by the LHC is based on recasting the searches for events with an energetic $\ell^+\ell^-$ pair ($\ell=e,\,\mu$) 
and missing transverse momentum performed by ATLAS at $\sqrt{s}=$ 8 TeV \cite{Aad:2014vma,TheATLAScollaboration:2013hha} (shown in cyan) and at $\sqrt{s}=$ 13 TeV \cite{ATLAS:2016uwq,ATLAS:2017uun} (in blue), 
as well as a 13 TeV CMS search for soft lepton pairs and missing transverse momentum \cite{Sirunyan:2018iwl} (in orange).
In our models, these signatures follow from Drell-Yan production of the heavy charged leptons, $pp \to F^+ F^-$, followed by the decays $F^{\pm} \to \mu^\pm S$, see  Fig.~\ref{fig:LHC}. In the case of the soft lepton search\,---\,which is sensitive to the region of small $M_F-M_S$ mass splitting\,---\,one energetic jet from initial state radiation is also required.
We have used {\tt Madgraph}+{\tt Pythia} \cite{Alwall:2014hca,Sjostrand:2014zea} to simulate the events and to compute the $F^+ F^-$ production cross sections, and {\tt CheckMate} \cite{Dercks:2016npn} to run the {\tt Delphes} \cite{deFavereau:2013fsa} detector simulation and compare the number of events obtained in a given signal region with the limits provided by the 8 TeV ATLAS searches \cite{Aad:2014vma,TheATLAScollaboration:2013hha} (already included in the  
{\tt CheckMate} framework), and with the limits of the 13 TeV searches \cite{ATLAS:2016uwq,ATLAS:2017uun,Sirunyan:2018iwl}, the last two of which we have implemented using the tools described in Ref.~\cite{Kim:2015wza}.

As one can see from the figures, the LHC bounds almost completely exclude the parameter region where the correct relic density and the solution to the $(g-2)_\mu$ anomaly overlap.
This is a consequence of the stringent LHC constraints, reaching up to 650 (750) GeV for a singlet (doublet) vectorlike lepton. These limits are considerably stronger than the corresponding limits for sleptons shown in the original analyses (see e.g.~Ref.~\cite{ATLAS:2017uun}),  
because the production cross section for a vector-like lepton is about an order of magnitude larger than for a scalar of the same mass. These constraints render a simultaneous explanation of DM and $(g-2)_\mu$  viable only  in a small corner of the parameter space with small $M_F-M_S$ splitting
and a limited range of $\lambda_{L,R}$ values. 
The upper-right plots in Figs.~\ref{fig:LL1} and \ref{fig:RR1} show instances of possible parameter choices, where the red line overlaps with the green region only. Specifically, we have found that LL1 and RR1 can simultaneously address DM and $(g-2)_\mu$ at the $2 \sigma$ level, without being in conflict with direct searches, if 
\begin{align*}
{\rm LL1}:& \quad 60~{\rm GeV}\lesssim M_S \lesssim 70~{\rm GeV},  & 100~{\rm GeV}\lesssim M_F \lesssim 115~{\rm GeV}, &\quad\quad1.2\lesssim\lambda_L\lesssim 1.4\,; \\
{\rm RR1}:& \quad 55~{\rm GeV}\lesssim M_S \lesssim 90~{\rm GeV}, & 100~{\rm GeV}\lesssim M_F \lesssim 140~{\rm GeV}, &\quad\quad1.3\lesssim\lambda_R\lesssim 1.8\,.
\end{align*}
This is best illustrated in Fig.~\ref{fig:MS-lam} for different choices of $M_F$ (close to the LEP limit and to the maximum attainable values). 
These plots also highlight the phenomenological differences between the two models. For the same value of $M_F$, the LHC excludes a broader range of parameter space in the LL1 model. This is a consequence of the larger production cross section of the heavy charged fermion in LL1 (which is part of a $SU(2)_L$ doublet) compared to the charged fermion in RR1 (which is a $SU(2)_L$ singlet). Similarly, for the same value of $M_F$, $M_S$ and $\lambda$, the DM annihilation cross section is larger in LL1 than in RR1, again because of the $SU(2)_L$ multiplicity of the fields. In particular, the LL1 scalar can also annihilate into neutrinos. 
As a consequence, for a given $M_S$, 
the correct relic abundance is obtained for smaller (larger) values of the coupling (of $M_F$) in LL1 than in RR1.
The net result of these two features is that the viable parameter region are smaller in LL1 compared to RR1, as is manifest from the plots in the first row of Fig.~\ref{fig:MS-lam} and the parameter regions above. 

\begin{figure}[t!]
\begin{center}
\includegraphics[width=0.45\textwidth]{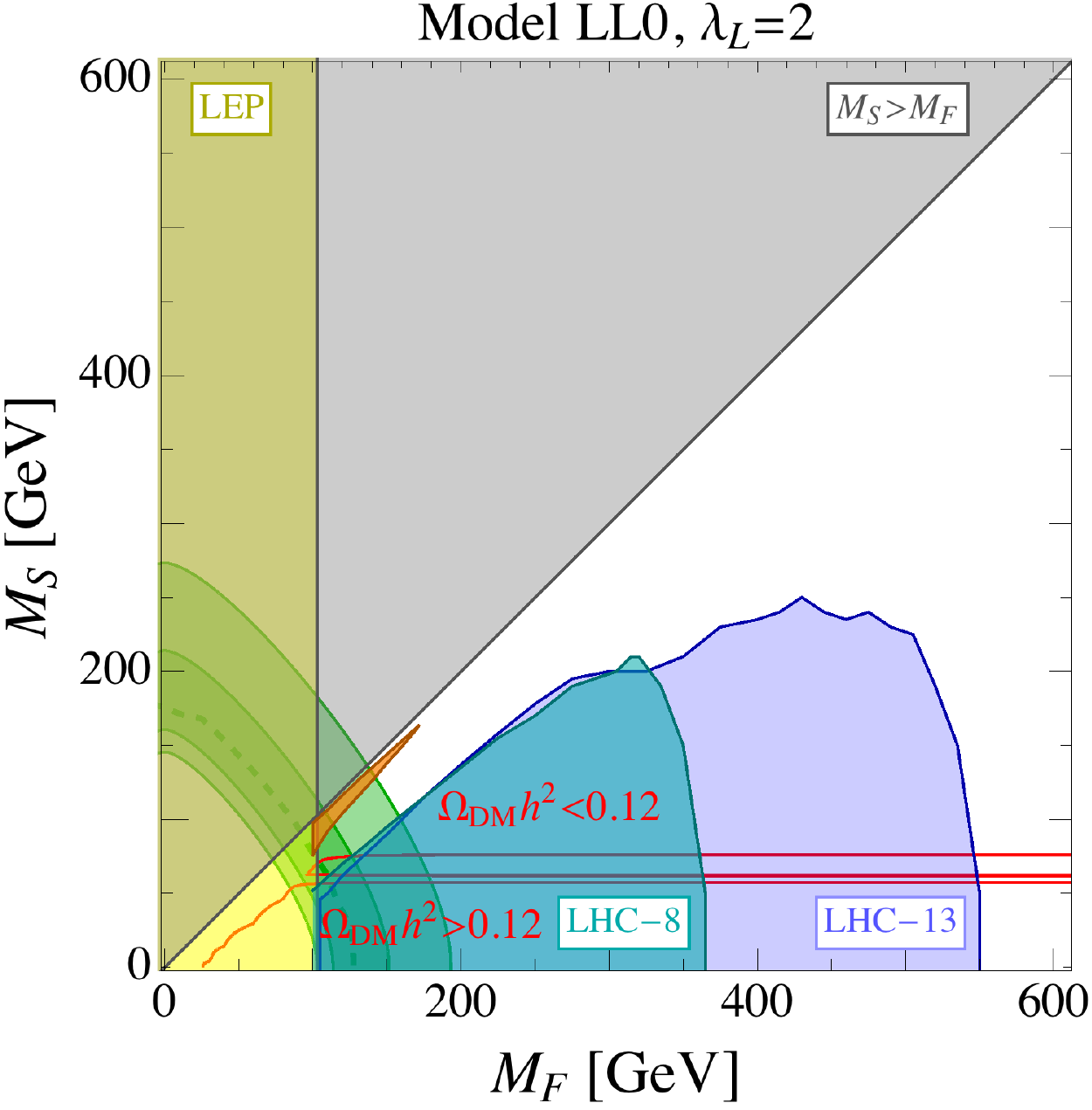}
\caption{Results for the LL0 model (inert doublet DM plus a vectorlike lepton singlet) in the $M_F - M_S$ plane for $\lambda_L=2$. Colored regions as in the previous figures. \label{fig:LL0}}
\end{center} 
\end{figure}
\begin{figure}[t!]
\begin{center}
\includegraphics[width=0.4\textwidth]{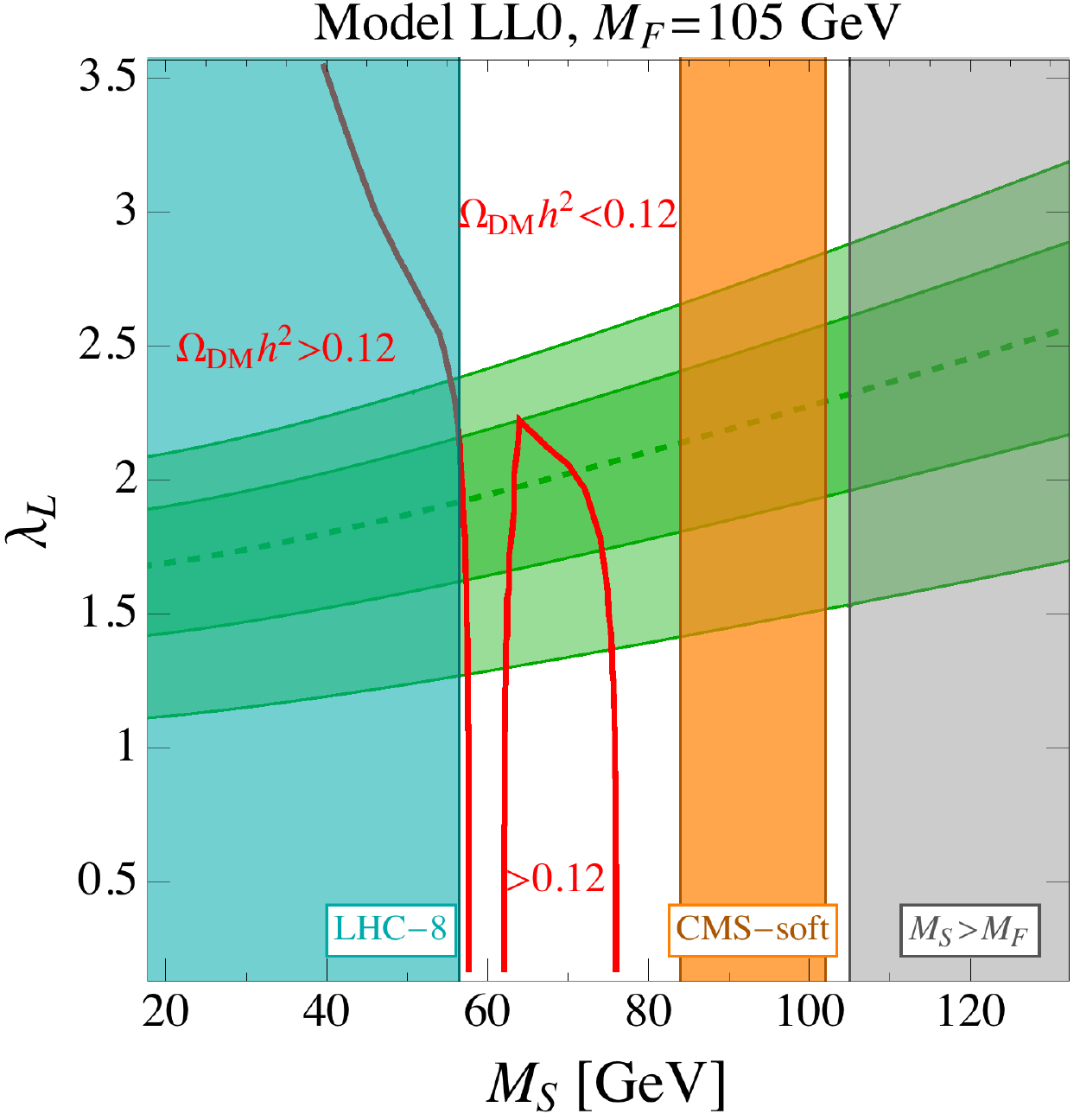}\hspace{0.3cm}
\includegraphics[width=0.4\textwidth]{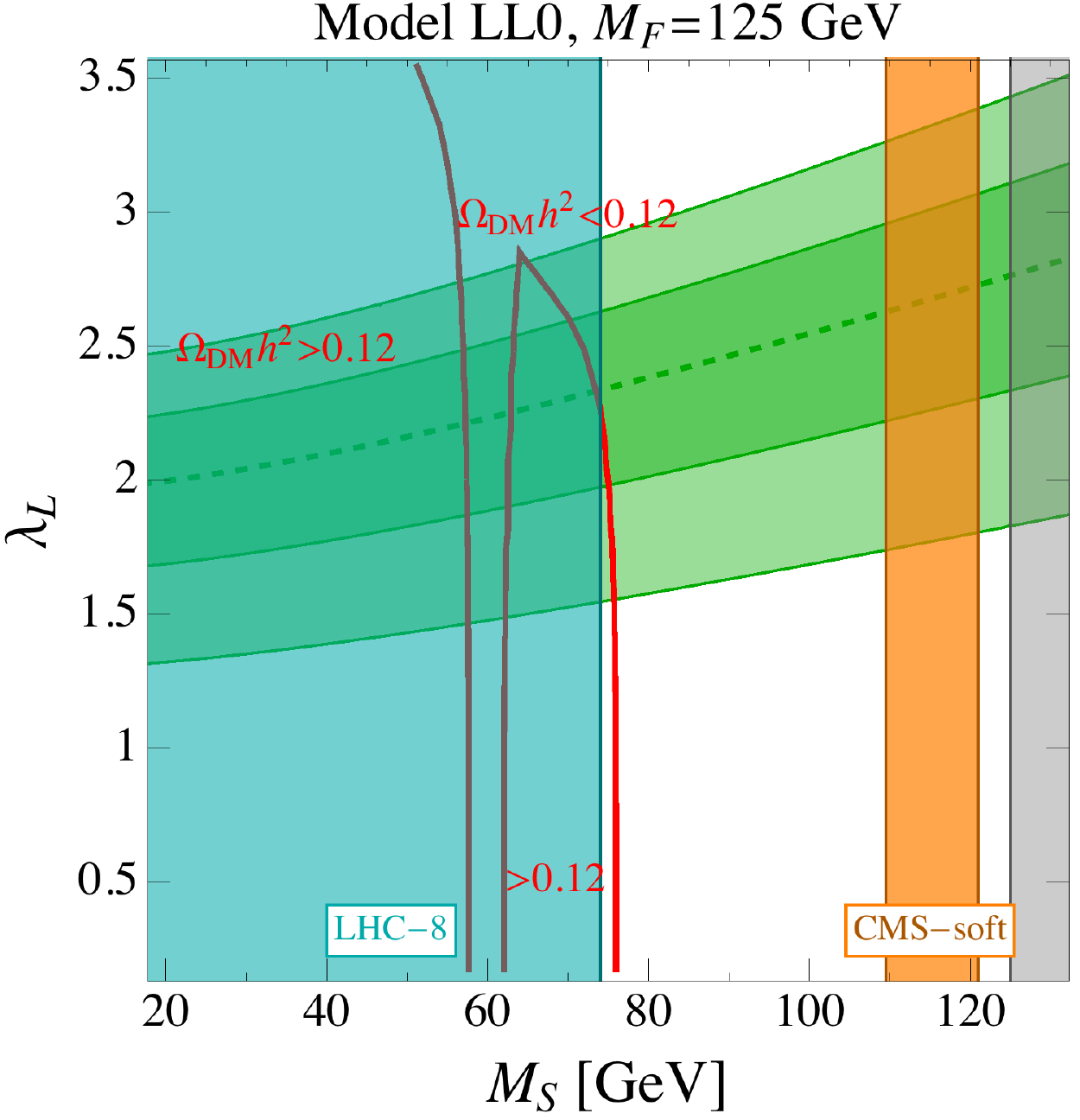}
\caption{Results for the LL0 model (inert doublet DM plus a vectorlike lepton singlet) in the $M_S - \lambda_L$ plane for specific choices of the vectorlike fermion mass $M_F$. Colored regions as in the Fig. \ref{fig:MS-lam}. \label{fig:LL0-2}}
\end{center} 
\end{figure}

We conclude the section with a brief discussion of the LL0 model. This is an example of an inert doublet DM model \cite{Ma:2006km,Barbieri:2006dq,LopezHonorez:2006gr}, where  DM is part of a scalar $SU(2)_L$ doublet. The LL0 model contains in addition the fermion partner\,---\,a singlet vectorlike lepton, see Table~\ref{g2viable}. Since DM is part of a weak doublet, it can efficiently annihilate into $W^+ W^-$, provided that $m_\chi > M_W$. In this case the relic density is too small for light DM masses,  $m_\chi \sim {\mathcal O}(100{\rm ~GeV})$, which is the region where the $(g-2)_\mu$ anomaly can be explained. Conversely, for $m_\chi \gtrsim 600$ GeV one obtains correct relic abundance, while the contribution to  $(g-2)_\mu$ is negligibly small, cf.~Eq.~(\ref{eq:gm2-bound}). 
This leaves only the case of very light DM masses, $m_\chi < M_W$, so that the  annihilation to on-shell $W^+ W^-$ is kinematically forbidden.
Lowering DM mass below the $W^+W^-$ threshold, the annihilation mode with one off-shell $W$ rapidly becomes less and less efficient, and the relic density 
$\Omega h^2 = 0.12$ is obtained for $m_\chi\approx 80$~GeV, see Figs.~\ref{fig:LL0} and \ref{fig:LL0-2}.  For even lower DM mass, DM is typically overabundant, with the exception of $m_\chi\approx m_h/2$. Then the annihilation through Higgs resonance is possible, and  $\Omega h^2 = 0.12$ can be obtained for perturbative quartic couplings between inert and Higgs doublets. As a consequence,  it is possible to find an overlap with the region favoured by $(g-2)_\mu$, 
see  Figs.~\ref{fig:LL0} and \ref{fig:LL0-2}, but only for tuned values of DM mass, and, given the regions excluded by the LHC,
a very limited range of the vectorlike fermion mass, 105 GeV$\lesssim M_F \lesssim$ 125 GeV, in the small $M_F-M_S$ region.
Furthermore, a non-trivial choice of the quartic couplings has to be made, in order to split the CP-even and CP-odd parts of the neutral component of the doublet, otherwise efficient co-annihilations would again make DM under-abundant. This also  gives the charged component a mass above the LEP bound of approximately 100 GeV. 

In summary, the simplest models  with two extra fields can marginally account for DM and $(g-2)_\mu$ simultaneously, but only for limited choices of the DM mass and the coupling to muons. This is due to the constraints from LHC searches for $\mu^+\mu^-$ + MET events, which are particularly stringent given the rather light charged fermions required to address $(g-2)_\mu$. Still, a small region in the parameter space is left open, squeezed between the exclusion limits from ATLAS searches with hard leptons and CMS searches with soft leptons, cf. Figs.~\ref{fig:LL1}, \ref{fig:RR1} and \ref{fig:MS-lam}. It would be desirable to study whether these regions can be excluded with future data. 

\section{Models with Higgs Insertion}
\label{sec:Higgs}
\setcounter{equation}{0}
We now turn to the discussion of models with three additional fields, which can have a qualitatively different phenomenology, if there are direct couplings of new fields with the SM Higgs\footnote{For a recent systematic discussion of models of DM coupling to the Higgs, see Ref.~\cite{Lopez-Honorez:2017ora}.}.
 These Class II models allow for a Higgs insertion in an internal line of the $(g-2)_\mu$ penguin diagram, cf. Fig \ref{diagrams:g-2}, giving  a parametric  enhancement $\propto 1/y_\mu$ over the penguin diagrams in Class I models. The NP fields can therefore be heavier by roughly
a factor  $\propto 1/\sqrt{y_\mu} \approx 15 $ compared to LL1 and RR1 models, resulting in a typical mass range of a few TeV.  This avoids many of the stringent constraints that are relevant for Class I models.

Because of the presence of Higgs couplings to new fields and the resulting mass mixing, the structure of Class II models after EWSB is more involved. Moreover, the presence of two couplings to both LH and RH muons allows for either sign of NP contribution to $(g-2)_\mu$, in contrast to Class I models, cf. Eq.~\eqref{eq:Deltaamu:generalLagr}. For this reason we do not attempt to perform an analytical discussion of Class II models, 
but rather turn directly to the numerical analysis after presenting the general structure of the models before EWSB.  

\subsection{General Discussion of FLR and SLR Models} 
The FLR and SLR models, Eqs.~\eqref{eq:FLR} and \eqref{eq:SLR}, are natural extensions of the LL and RR models obtained by either combining the fermionic or scalar content of two such models. The FLR models couple the NP fermions to the Higgs. They contain a single complex scalar, $S_R\sim n_Y$, as well as two pairs of vector-like fermions, $F_L \sim n_{Y-1}$, $F_L^c \sim n_{1-Y}^*$, and $F_R \sim (n \pm 1)_{1/2 - Y}$, $F_R^c \sim {(n \pm 1)^*}_{- 1/2 + Y}$, which couple directly to the Higgs through a Yukawa interaction, $y_F H F_L F_R$, see Eq. \eqref{eq:FLR}. In contrast, in SLR models the couplings to the Higgs are through the NP scalars, $S_R \sim n_{Y}$, $S_L \sim (n\pm1)_{-1/2-Y}$, allowing for the interaction term $a H S_L S_R$, see Eq.~\eqref{eq:SLR}. The SLR models also include a single pair of vector-like fermions, $F_R \sim (n \pm 1)_{1/2 - Y}$, $F_R^c \sim {(n \pm 1)^*}_{- 1/2 + Y}$.

As for Class I models, also in Class II models the sizes of $SU(2)_L$ representations, $n$ or $n\pm1$, are bounded from above, if they are to explain simultaneously the $(g-2)_\mu$ anomaly and the DM relic density.  However, there are two important differences.  Unlike in Class I models, the  NP contribution to $(g-2)_\mu$ is not fixed simply by the quantum numbers of the NP fields and their mass hierarchies, but depends also on the new Higgs couplings, $y_F$ and $a$, see Eqs.~\eqref{eq:FLR}, \eqref{eq:SLR}. Second, since $\Delta a_\mu$ is enhanced by $1/y_\mu$ compared to Class I models, the bounds on allowed $SU(2)_L$ representations are much weaker, allowing for values as large as $n\sim {\mathcal O}(20)$. This prevents us from discussing the full set of Class II models. Table~\ref{selecLR} contains possible models that contain the smallest representations of SM gauge group, up to triplets of $SU(2)_L$, excluding only models where DM candidates with $Y \ne 0$ can not mix with a self-conjugate  particle (Majorana fermion or real scalar), cf.~the discussion in Section \ref{sec:DD}.
In the remainder of the section we perform phenomenological analyses for the two simplest  models, FLR1 and SLR1, which are representative cases for the whole set of Class II models.  

\begin{table}[t]
\centering
\begin{tabular}{c||cccccc}
FLR & FLR1&&  &&  & \\
\hline
$F_R$  &  $1_0^\star$ & $2_{- \frac{1}{2}}^\star$ & $2_{- \frac{1}{2}}^\star$& $2_{\frac{1}{2}}$ & $2_{\frac{1}{2}}$ &$3_{0}^\star$ \\
$F_L$  &  $2_{-\frac{1}{2}}^\star$ & $1_0^\star$ & $3_0^\star$ & $1_{-1}$ & $3_{-1}$   &$2_{- \frac{1}{2}}^\star$   \\
$S_R$ &$2_{\frac{1}{2}}$  &   $1_1$ &$3_1$ & $1_0^\star$ &  $3_{0}^\star$ &    $2_{\frac{1}{2}}$   \\
\multicolumn{1}{c}{}   \\
SLR & SLR1& &&&&\\
\hline
$S_L$  &  $1_0^\star$  & $1_{-1}$  & $2_{- \frac{1}{2}}^\star$ & $2_{- \frac{1}{2}}^\star$  & $3_0^\star$  &$3_{- 1}$   \\
$S_R$  &  $2_{- \frac{1}{2}}^\star$ & $2_{ \frac{1}{2}}$ & $1_0^\star$ & $3_0^\star$& $2_{- \frac{1}{2}}^\star$ &$2_{\frac{1}{2}}$  \\
$F_R$ &$1_1$  & $1_0^\star$  &  $2_{ \frac{1}{2}}$ &$2_{\frac{1}{2}}$  &   $3_1$  &  $3_0^\star$   
\end{tabular}
\caption{ \label{selecLR} Viable FLR and SLR models with smallest $SU(2)_L$ representations, excluding models where DM candidates with $Y \ne 0$ cannot mix with a self-conjugate particle. 
}
\end{table}

\subsection{Numerical results for the FLR1 model}
The FLR1 model is the simplest example of a Class II model where the NP fermion fields couple to the Higgs. Beside the SM fields, it contains also a heavy (Majorana) fermion that is an electroweak singlet, $F_R\sim 1_0$, a vector-like pair of heavy fermions that are weak doublets, $F_L\sim 2_{-1/2}$, $F_L^c \sim 2_{1/2}^*$, and a vector-like scalar doublet, $S_R\sim 2_{1/2}$. In this subsection we therefore use a more suggestive notation
\beq
F_S\equiv F_R\sim 1_0, \qquad F_D\equiv F_L\sim 2_{-1/2},  \qquad F_D^c \equiv F_L^c \sim 2_{1/2}^*, \qquad S\equiv S_R\sim 2_{1/2},
\eeq
in terms of which the interaction and mass part of the Lagrangian is, 
\begin{align}\label{eq:FLR1}
{\cal L}_{\rm FLR1} & \supset \left( \lambda_{1H} H F_D F_S + \lambda_{2H} \tilde H F_D^c F_S +\lambda_{1} \mu F_S S + \lambda_{2} \mu^c F_D S^*  +{\rm h.c.} \right) \nonumber \\
& - \left( M_{F_D} F_D F_D^c + \frac{M_{F_S}}{2} F_S F_S +{\rm h.c.} \right)  - M_{S}^2 S^* S \, .
\end{align}
In Appendix \ref{sec:FLR} we also give component-wise form of the Lagrangian. 
The Yukawa-like interactions, proportional to $\lambda_{1H}$ and $\lambda_{2H}$, induce mixing between $F_S$ and $F_D$, after the Higgs acquires its vev.  The DM candidate is a Weyl fermion that is the lightest admixture of $F_S$ and the two Weyl fermions forming the neutral components of $F_D$ and $F_D^c$,  while the DM mass is approximately equal to the smallest of the two fermion mass parameters, $M_{F_S}, M_{F_D}$. 
\begin{figure}[t!]
\begin{center}
\includegraphics[width=0.4\textwidth]{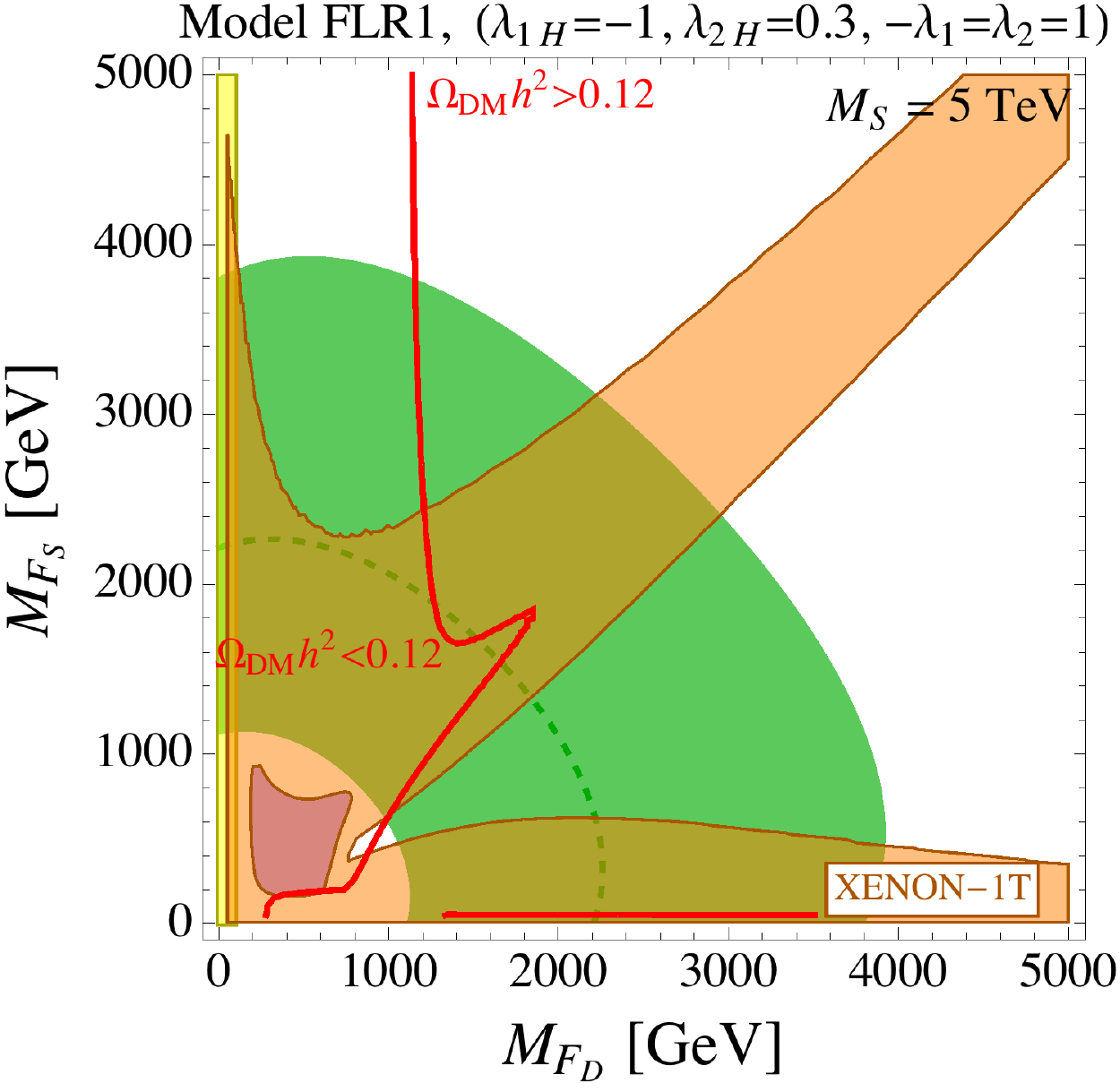}
\hspace{0.3cm}
\includegraphics[width=0.4\textwidth]{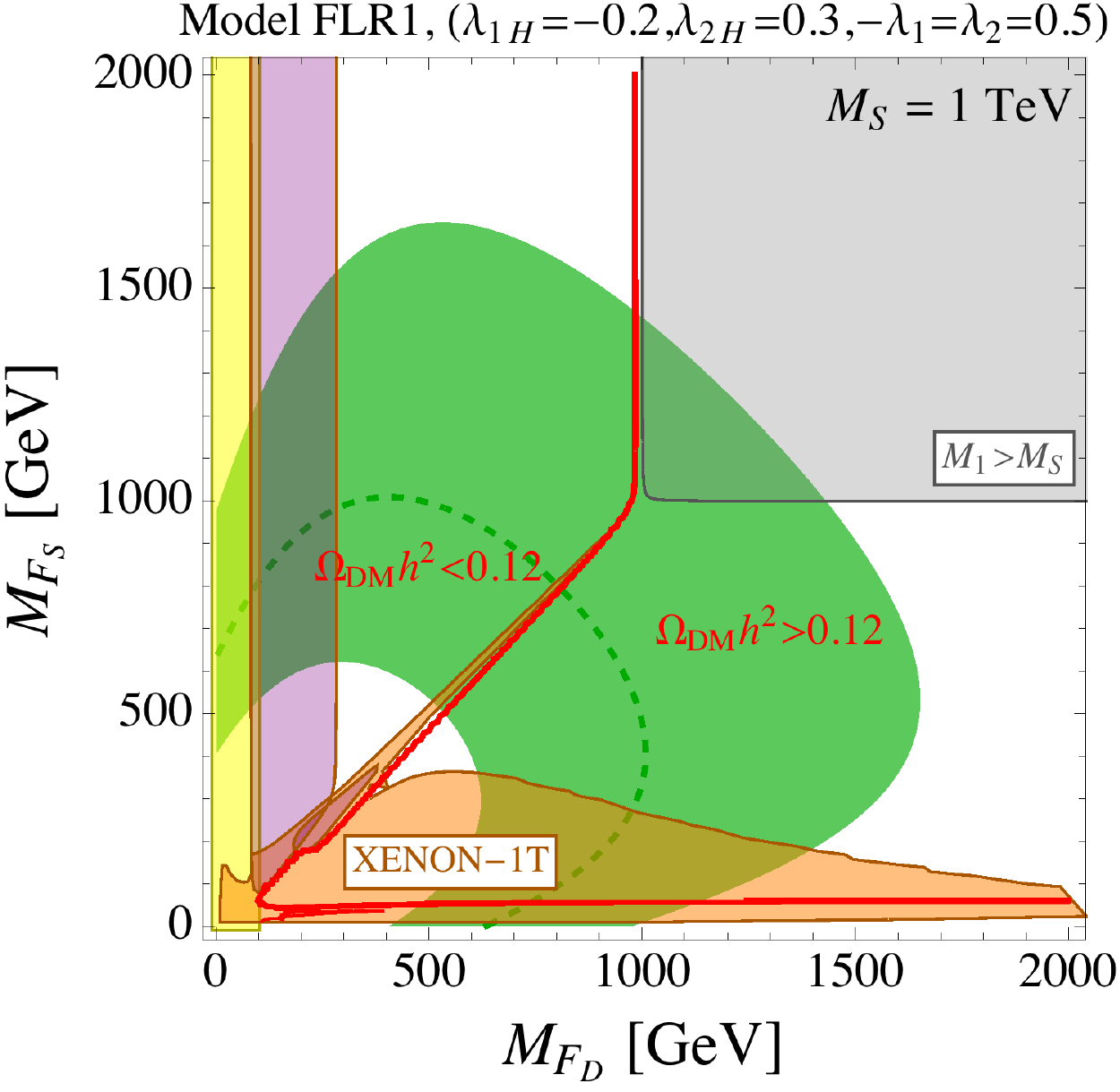}
\caption{Results for the FLR1 model in the doublet-singlet mass plane $M_{F_D} - M_{F_S}$. The two choices of the couplings $\lambda_i$ and of the scalar doublet mass $M_S$ are as indicated. In the dark green region the total contribution to $(g-2)_\mu$  is compatible with the experimental value within $1\sigma$. The red line indicates where the DM relic density equals $\Omega_{\rm DM}h^2 = 0.12$. The pale-orange region is excluded by direct detection \cite{Aprile:2017iyp}.  The light-purple areas are excluded by indirect detection \cite{Calibbi:2015nha}.
\label{LR1}}
\end{center} 
\end{figure}

Fig.~\ref{LR1} shows two illustrative choices of FLR1 parameters, taking $\lambda_i\sim {\mathcal O}(1)$, with signs chosen such that there is a positive contribution to $(g-2)_{\mu}$. In the dark green region $(g-2)_{\mu}$ matches the experimental measurement within 1$\sigma$, with dashed line denoting the central value.  One can see that a sufficiently large contribution to $(g-2)_\mu$ is obtained even for NP particles in  the multi-TeV range. In the left (right) panel the scalar has a mass of $M_S=5\,(1)$ TeV, while the required fermion mass is in the $1$ to $4$ TeV ($0.5$ to $1.5$ TeV) range.
This is an order of magnitude heavier than what was found for Class I models, and a direct consequence of the relative $\sim v/m_\mu$  enhancement of the contribution to $\Delta a_\mu$.

In the left panel of Fig. \ref{LR1} the scalar is heavier than the fermions and can be to good approximation ignored in the  DM phenomenology (this then resembles closely the phenomenology of the so-called Singlet-Doublet DM model \cite{Mahbubani:2005pt,Cohen:2011ec,DEramo:2007anh,Calibbi:2015nha}). We calculate the relic abundance with  {\tt micrOMEGAs}, indicating $\Omega_{\rm DM} h^2 = 0.12$ by a red line as before. One can recognize three distinct regimes where the correct relic density is reproduced: {\bf i)} For $M_{F_D} \ll M_{F_S}$, DM is predominantly the neutral component of $F_D$. The main annihilation channel is $W^+W^-$, due to the $t$-channel charged fermion exchange. The correct relic abundance is achieved for a DM mass of $m_\chi \approx M_{F_D}\approx 1.1$ TeV, independent of the singlet mass, $M_{F_S}$.
This is analogous to the familiar case of pure Higgsino DM in Supersymmetry. 
{\bf ii)} For $M_{F_S} < M_{F_D}$, DM tends to be mainly singlet, but with a sizable doublet contribution, which is needed to avoid DM overabundance. This is the reason why around 1\,TeV the red line runs close to the diagonal, $M_{F_S} = M_{F_D}$. {\bf iii)} For even lower $M_{F_S}$ masses a resonant annihilation trough Higgs or $Z$ is possible, giving a low mass red line independent of $M_{F_D}$. 

The right panel of Fig. \ref{LR1} shows a similar behavior for a light singlet. However, when the DM mass approaches the scalar mass, $M_S =1$ TeV, the annihilations to muons through the $t$-channel scalar exchange start to dominate (co-annihilations are also important in this region), and the red line become $M_{F_S}$ independent. In the gray-shaded region the scalar doublet is the DM candidate. This is either excluded by direct detection\,---\,unless the electroweak-breaking effects split the CP-even and CP-odd components as discussed in Sec.~\ref{sec:DD}\,---\,or one has an overabundant DM (which is the case for scalar doublets with $M_S \gtrsim 550$ GeV).

The pale-orange region in Fig.~\ref{LR1} is excluded by the latest bound from the direct detection experiment 
XENON1T \cite{Aprile:2017iyp}. The spin-independent DM--nucleon scattering cross section is mainly due to tree-level Higgs exchange. This vanishes both in the limit of a pure singlet DM and of pure doublet DM, cf.~Eq.~(\ref{eq:LR1-Lh}). Furthermore,
the direct detection constraints are weakened  for $\lambda_{H1}/\lambda_{H2} <0$, in which case a partial cancellation in the coupling of DM to $h$ occurs. Incidentally, this condition is also compatible with a positive sign of the contribution to $(g-2)_{\mu}$. The light-purple areas are excluded by indirect detection as explained in Ref.~\cite{Calibbi:2015nha}.

The NP particles of the FLR1 model can manifest at the LHC through different production modes and decays. 
The mass of the scalar doublet can be constrained by searches for pair production of the charged scalar followed by a decay to the muon and DM: $pp\to S_+S_-$, $S_\pm \to \mu^\pm F_{0 1}$ (cf.~the notation in \ref{sec:FLR}). Rescaling the production cross section, we can estimate that the limit obtained in Ref.~\cite{Sirunyan:2018iwl} for the case of slepton production translates into a bound on $M_S\gtrsim 400$ GeV for a DM mass $m_\chi \lesssim$ 200 GeV.
Another sensitive mode is the production of the charged and neutral components of the fermion doublets, $F_D$ and $F_D^c$, followed by the decays of these particles to a lighter fermion singlet and SM Higgs or gauge bosons. This mode resembles the familiar case of production of Higgsino-like charginos and neutralinos decaying into bino or gravitino.
Considering a combination of neutralino and chargino searches recently published by CMS \cite{Sirunyan:2018ubx}, 
we find that the strongest constraint is set by the following mode: $pp\to F_\pm F_{02/03}$ with $F_\pm \to W^\pm F_{01}$, $F_{02/03} \to Z F_{01}$, and the gauge bosons decaying leptonically. Taking into account the difference in the production cross sections and branching ratios between our case and the models studied in Ref.~\cite{Sirunyan:2018ubx}, we assess a limit on the fermion doublet mass at about $500$ GeV for a relatively light (mainly-singlet) DM, $M_{F_S}\lesssim$ 200 GeV. As we can see, these limits do not impact much on the parameter space displayed in Fig.~\ref{LR1}, where the bounds from direct detection are much more prominent.

In conclusion, the FLR1 model can easily accommodate both the correct DM relic density and the $(g-2)_{\mu}$ anomaly, evading the bounds from direct and indirect DM searches. Since DM and other NP particles can be rather heavy, of ${\mathcal O}(1)$ TeV or more, the constraints from LHC searches are easily evaded.

\subsection{Numerical results for the SLR1 model}
The SLR1 models is one of the simplest representatives of Class II models, where new scalars couple to the Higgs, see Eq.~\eqref{eq:SLR}.  The SLR1 model contains, in addition to the SM fields, a heavy (real) singlet scalar, $S_L\sim 1_0$, a heavy doublet scalar, $S_R\sim 2_{-1/2}$, and a charged weak singlet fermion, $F_R\sim 1_1$. In this section we use a more suggestive notation, 
\beq
S\equiv S_L\sim 1_0,\qquad D\equiv S_R\sim 2_{-1/2}, \qquad F\equiv F_R\sim 1_1,
\eeq
with the interaction Lagrangian
\begin{align}\label{eq:SLR1}
{\cal L}_{\rm SLR1} & \supset \left( a_H H S D + \lambda_{1} D \mu F +\lambda_{2} S \mu^c  F^c + {\rm h.c.} \right)  \nonumber \\
& - \left( M_{F} F F^c + {\rm h.c.} \right) - \frac{M_{S}^2}{2} S^2 - M_{D}^2 D^* D \, .
\end{align} 
Further details on the Lagrangian are given in~Appendix \ref{sec:FLR}.

\begin{figure}[t!]
\begin{center}
\includegraphics[width=0.4\textwidth]{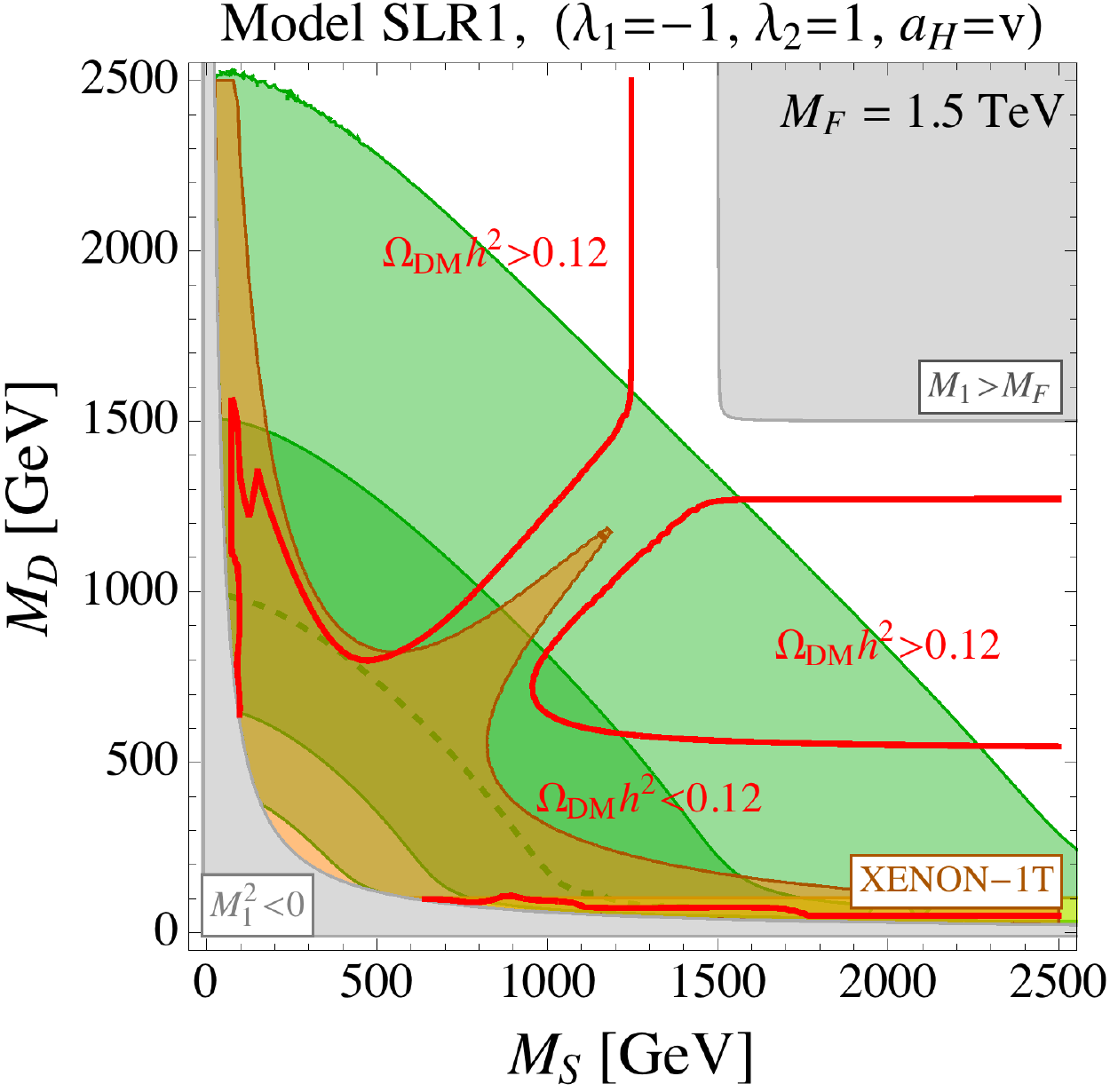}
\hspace{0.3cm}
\includegraphics[width=0.4\textwidth]{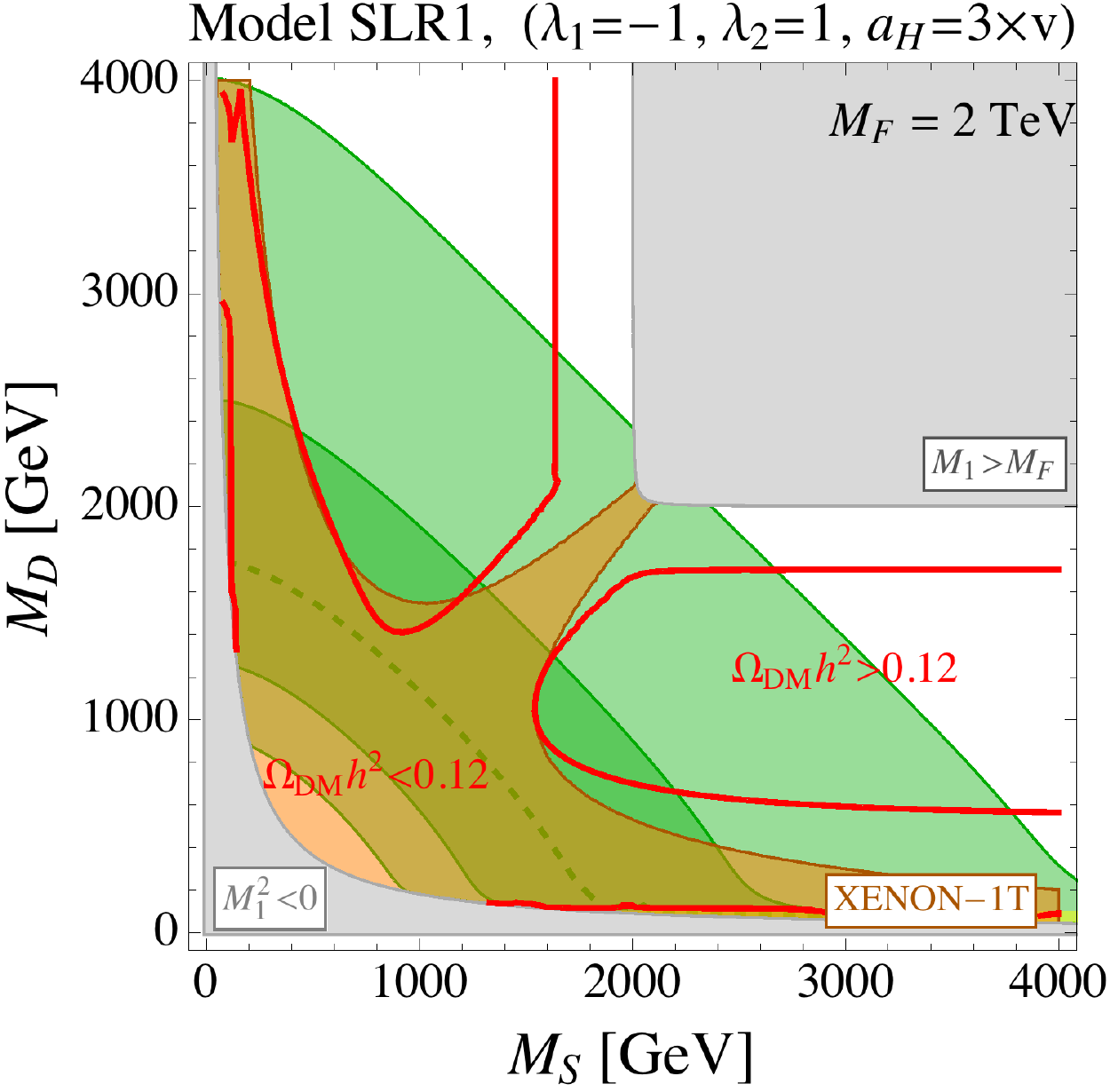}
\caption{Results for the SLR1 model in the scalar singlet-doublet mass plane $M_{S} - M_{D}$. The two choices for the values of the couplings and the vectorlike lepton mass $M_F$ are as indicated. The light gray areas are excluded by negative DM squared mass, while in the dark gray regions the heavy charged fermion is lighter than the scalars. The other colors are as in Fig.~\ref{LR1}.\label{fig:SLR1}}
\end{center}  
\end{figure}

In this model the DM candidate is a mixed singlet-doublet scalar, with the mixing arising from the trilinear $a_H H S D$ coupling once the Higgs acquires its vev. Fig.~\ref{fig:SLR1} shows two illustrative slices of the model's parameter space in the $M_S-M_D$ plane. In the left (right) panel we take for the trilinear coupling $a_H=v$ ($a_H = 3 \, v$), for the fermion mass $M_F = 1.5$ TeV (2 TeV), and $\lambda_1 = -1$, $\lambda_2 = 1$. The opposite signs insure that the contribution to $(g-2)_\mu$ is positive. The light gray areas on the bottom-left corners of the panels are excluded by a negative scalar squared mass, while in the top-right gray areas the charged fermion is lighter than the lightest scalar.
The light (dark) green areas correspond to a contribution to $(g-2)_\mu$ that fits the experimental value at $1\sigma \,(2 \sigma)$, while the red line corresponds to $\Omega_{\rm DM} h^2 = 0.12$. The pale-orange regions are excluded by XENON1T \cite{Aprile:2017iyp}. 

In the  $M_S < M_D$ region DM is mostly an electroweak singlet and is typically overabundant, since efficient annihilation is only induced through mixing with the doublet. The correct relic density is thus obtained either in the large mixing regime $M_S \approx M_D$, or when the DM mass
is close to the fermion mass, $M_S \approx M_F$, in which case $t$-channel annihilation to muons and coannihilations become effective. In contrast, for $M_S > M_D$ DM mostly behaves as a scalar doublet. In this case, efficient annihilation to $W^+W^-$ induces  correct relic abundance for $M_D \approx 550$ GeV, and under-abundant DM for $M_D \lesssim 550$ GeV. Above this mass, the relic density is too large, unless  in a regime of small mass splitting, either with the scalar singlet of with the vector-like lepton, so that new annihilation channels and co-annihilations become important.
 
From the two numerical examples we see that also in the SLR1 model
a large positive contribution to $(g-2)_\mu$ can be easily compatible with DM masses in the multi-TeV range. As in the FLR1 model, we expect that also for the SLR1 model NP searches at the LHC are at present not able to constrain the parameter space of  Fig.~\ref{fig:SLR1} at a level that can compete with the bounds from XENON1T. In fact, searches for a charged fermion decaying to the muon and DM give a constraint similar to those we obtained for the RR1 and LL0 models, cf.~Figs.~\ref{fig:RR1} and \ref{fig:LL0}. Production of the states of the scalar doublet, decaying to gauge bosons and DM, leads instead to the same topology that we discussed for the FLR1 model, $pp\to S_\pm S_2 \to W^\pm Z$ + MET. However, the production cross section for such scalars is about one order of magnitude smaller than for a fermion doublet of the same mass. Hence, we expect that searches as in Ref.~\cite{Sirunyan:2018ubx} are only sensitive to doublet masses up to 200 GeV for light singlet DM, $M_S < 100$ GeV.

\section{Summary and Conclusions}
 \label{ref:conclusions}

 In this paper we have performed a systematic study of models with minimal field content that can simultaneously address the muon $g-2$ anomaly and account for the observed DM relic density. We have first classified all such models in Tables~\ref{tab:LLRR} and \ref{tab:LR} and grouped them into two classes. ``Class I models'' involve only two additional fields so that the new physics contribution to $(g-2)_\mu$  scales as $\Delta a_\mu \propto m_\mu^2/M^2$ (where $M$ is the typical scale of the new fields). ``Class II models'' give a parametrically enhanced new physics contribution, $\Delta a_\mu \propto m_\mu v/M^2$, at the price of having three additional fields. Two of these fields couple to the Higgs which is responsible for the $v/m_\mu$ enhancement. 
 
 The structure of Class I models is simple enough to write down their Lagrangians as a general function of the dimension $n$ of their $SU(2)$ representations. This allows to obtain simple analytic expressions for $\Delta a_\mu$ and the relic density as a function of $n$. Reproducing the correct relic density results in an upper bound on the DM mass, $m_\chi \propto n^{3/2}$,  while the contribution to $(g-2)_\mu$ scales as $\Delta a_\mu \propto n/m_\chi^2 \propto n^{-2}$. This implies an upper bound on $n$, restricting the Class I models to only two viable ones. Taking into account constraints from collider searches strongly restricts the parameter space of the two models. After all the constraints are taken into account  only tuned regions remain in the parameter space, for which both the relic density and $(g-2)_\mu$ can be simultaneously reproduced, see Figs.~\ref{fig:LL1},~\ref{fig:RR1},~\ref{fig:LL0}. In these regions the model parameters  are essentially fixed (see Figs.~\ref{fig:MS-lam},~\ref{fig:LL0-2}), and the new particles have masses of $\ord{100}$ GeV. It would be interesting to study whether the remaining parameter space of these  models can be completely covered by upcoming data.
It is unlikely that searches for energetic leptons plus MET at the LHC can further test the small mass splitting region that we are interested in. In fact, comparing 8 and 13 TeV searches, we have seen that the bound given by these searches on small mass splittings did not benefit from increased centre-of-mass energy and luminosity. On the other hand, searches targeting soft leptons, as in Ref.~\cite{Sirunyan:2018iwl}, appear to be still statistically limited and could then provide a discovery channel (or more stringent constraints) for the Class I models at the LHC. On the other hand, the proposed future linear or circular $e^+e^-$ colliders would easily test the whole parameter space of interest, given that the new heavy particles must have masses considerably below 150 GeV. 
 
Class II models have a more involved structure due to mass mixing. We have analyzed only the two simplest models which we expect to be representative for the whole class. Thanks to the $v/m_\mu$ enhancement of new physics contributions to $(g-2)_\mu$, these models can successfully explain the relic density and resolve the muon $g-2$ anomaly with new particles in the multi-TeV range, and thus easily evade constraints from collider and DM searches, see Figs.~\ref{LR1},~\ref{fig:SLR1}. 
Future bounds from direct detection experiments appear to be the most suitable way to further probe this kind of models. 
A more detailed phenomenological analysis of these models and possibly other Class II scenarios would be desirable in the future.

\section*{Acknowledgements}
\noindent We are grateful to Paride Paradisi for discussions that have stimulated this work and collaboration at initial stages. 
We also thank Daniel Dercks for the patient help with {\tt CheckMate}, and Manuel Perez-Victoria, Paolo Panci, Alex Pomarol, and Susanne Westhoff for helpful discussions. JZ acknowledges support in part by the DOE grant DE-SC0011784.

\appendix
\numberwithin{equation}{section}

\section{Electroweak Representations}
\label{sec:QuantumNumbers}
In this appendix we explain how the possible electroweak representations for the models presented in Section \ref{Setup} were obtained. 
We start with Class I models of LL type, so that the new states couple only the left-handed muon, see Eq. \eqref{eq:LHmuon}. 
\paragraph{LL Models.}
Fixing  the quantum numbers of the fermion as $F_R\sim \left(n_F\right)_{Y_F}$ determines the quantum numbers of the scalar to be
\beq
Y_S = \frac{1}{2} - Y_F, \qquad\qquad  n_S = (n_F -1){\rm ~or~}(n_F +1) \, .
\eeq
For given $n_F$ there are only two possibilities for $Y_F$, such that the model has a DM candidate:
\begin{itemize}
\item $F$ contains a neutral state for $Y_F = - (n_F -1)/{2}, -(n_F -1)/{2} + 1, \hdots, (n_F -1)/{2}$. This gives $n_F$ possible assignments for $Y_F$. For each of these possiblities there are two possible choices of $n_S$ (with the exception of $n_F=1$ where only $n_S=2$ is possible). 
\item $F$ contains no neutral state but $S$ does. In this case there is only one possible charge assignment, $Y_F = (n_F + 1)/{2}$ and $n_S = n_F + 1$.
\end{itemize}
In all other cases neither $F_R$ nor $S_R$ contains a neutral state. Thus for  $n_F > 1$ there are exactly $2 n_F +1$ viable NP models differing in the choices for $Y_F, Y_S$ and $n_S$  (for $n_F = 1$ there are $n_F+1=2$ viable NP models).  These charge assignments are listed in Table \ref{tab:LLRR} for $n_F\leq 3$, $n_S\leq 3$.  
The fields that contain a neutral state are indicated by a $\star$.

\paragraph{RR Models.}
For the Class I models of RR type  the new fields couple only to the RH muon, see Eq. \eqref{eq:RHmuon}. Fixing the representation for the new fermion to be $F_L\sim  \left(n_F\right)_{Y_F}$, determines uniquely the scalar quantum numbers to be
\beq
Y_S  = -1 - Y_F, \qquad\qquad n_S = n_F  \, .
\eeq
Given $n_F$, there are two possibilities for $Y_F$, such that there is a DM candidate:
\begin{itemize}
\item  $F$ contains a neutral state for $Y_F = - (n_F -1)/{2}, - (n_F -1)/{2} + 1, \ldots, ({n_F -1})/{2}$, giving $n_F$ possible $Y_F$ assignments.
\item $F$ contains no neutral state but $S$ does, which happens for  $Y_F = -(n_F + 1)/{2}$.
\end{itemize}
For given $n_F \ge 1$ there are therefore $ n_F +1$ potentially viable models. They are listed in Table \ref{tab:LLRR} for $SU(2)_L$ representations up to triplets.  We indicate the fields that contain a neutral state by $\star$.
\paragraph{FLR and SLR Models.}
The electroweak charge assignments for the Class II models of FLR type, where the Higgs couples to the fermions, Eq.~\eqref{eq:FLR}, can be read off from the previous two cases of LL and RR models listed in Table \ref{tab:LLRR}. There is a single scalar that is identified with $S_R\sim S_L^*$, so that the scalar loop in the lower-left diagram in Fig.~\ref{diagrams:g-2} can be closed. The viable representations for the two fermions $F_{L,R}$ are thus obtained from the RR and LL models in Table \ref{tab:LLRR} that have the same $n_S$ but opposite $Y_S$. 

For the Class II models of the SLR type,  Eq.~\eqref{eq:SLR}, one needs two scalars, $S_L$ and $S_R$, whose quantum numbers combine to the one of a conjugated Higgs, $2_{-1/2}$. All such combinations of $S_L$ and $S_R$  are listed in Table~\ref{tab:LR}. For each of these cases one can check that the quantum numbers allow to identify the single fermion with $F_R\sim F_L^c$ in Table \ref{tab:LLRR}, so that the fermion loop in the lower-right diagram in Fig.~\ref{diagrams:g-2} can be closed.

\section{Lagrangians for explicit models}
\setcounter{equation}{0}
In this appendix we give the Lagrangians for the models for which we performed detailed phenomenological analyses in the main text, the representatives of Class I models\,---\,the LL0, LL1 and RR1 models, and the representatives of Class II models\,---\,the FLR1 and SLR1 models. We use four-component Dirac spinor notation, so that the  lepton Dirac spinors are
\beq
e_{Ri} \mathrel{\overset{\makebox[0pt]{\mbox{\normalfont\tiny\sffamily Weyl}}}{=}} \begin{pmatrix} 
0 \\
e_i^{c \dagger}
\end{pmatrix}, 
\quad{\rm and~~}
L_i \mathrel{\overset{\makebox[0pt]{\mbox{\normalfont\tiny\sffamily Weyl}}}{=}} \begin{pmatrix} 
\ell_i  \\
0
\end{pmatrix} \,, 
{\rm ~where~~}
\ell_i \mathrel{\overset{\makebox[0pt]{\mbox{\normalfont\tiny\sffamily SU(2)}}}{=}} \begin{pmatrix} 
\nu_{Li} \\
e_{Li}
\end{pmatrix}.
\eeq

\subsection{Models without Higgs insertions (LL and RR models)}
\label{LLRRmodels}
We first give the field content and Lagrangians for the Class I models, LL0, LL1, and RR1, which can be obtained from the general expressions in Sections~\ref{sec:general:LL} and \ref{sec:general:RR}.

\paragraph{\bf  \underline{Model LL0:}} The field content is $ F_R \sim  1_{1}, F_R^c \sim 1_{-1}, S_R \sim 2_{-1/2}$, see Table \ref{tab:LLRR}. The two Weyl fermions form a negatively charged Dirac fermion, $F_-\sim  1_{-1}$, while for the complex scalar we use the short-hand notation $S\equiv S_R$. The latter has two charge components, $S_-$ and $S_0$, so that we have the following composition in terms of Dirac and charge components, 
\beq
F_-  \mathrel{\overset{\makebox[0pt]{\mbox{\normalfont\tiny\sffamily Weyl}}}{=}}
  \begin{pmatrix} 
F_R^c  \\
F_R^{\dagger}
\end{pmatrix}, 
\qquad\qquad 
S\, \, \mathrel{\overset{\makebox[0pt]{\mbox{\normalfont\tiny\sffamily SU(2)}}}{=}}  \begin{pmatrix} 
S_0  \\
S_-
\end{pmatrix}.
\eeq
The Lagrangian is 
\beq
\begin{split}
{\cal L}_{\rm LL0}=&   \bar F_-( i\slashed \partial - M_F )F_-  +|\partial_\mu S_-|^2 - M_S^2 |S_-|^2 +|\partial_\mu S_0|^2 - M_S^2 |S_0|^2 
\\
&+ \lambda_{L} \bar{F}_-   \left( \mu_{L} S_0 -  \nu_{\mu L} S_- \right) +  \lambda_{L}^{*}    \left( \bar{\mu}_{L} S_0^* -  \bar{\nu}_{\mu L} S_-^* \right) F_- + {\cal L}_{\rm gauge} + {\cal L}_{\rm scalar} \, , 
\end{split}
\eeq
with
\begin{align}
\begin{split}
{\cal L}_{\rm gauge} & =    |e| A_\mu  \Big( \bar{F}_- \gamma^\mu F_- + i S^*_- \partialLR S_- \Big) +  \frac{i g}{\sqrt2} \Big( W_\mu^+ S_0^* \partialLR S_- + W_\mu^- S^*_- \partialLR S_0  \Big)
\\
& + \frac{g}{c_W} Z_\mu \Big[  s_W^2   \bar{F}_- \gamma^\mu    F_- + i \Big( -\frac{1}{2} + s_{ W}^2 \Big) \Big( S^*_- \partialLR S_-  \Big)  + \frac{i}{2} S_0^* \partialLR S_0 \Big] + {\cal L}_{SSVV}
 \, , 
 \end{split}
 \\
{\cal L}_{\rm scalar} &=   \kappa |S|^2 |H|^2 + \xi |S|^4 = \frac{\kappa}{2} |S_0|^2   \left( h^2 + 2 v h + v^2 \right)+ \xi |S_0|^4 +\cdots \, , 
\end{align}
where $v = 246 \GeV$ and we do not write out explicitly the couplings of scalars with two gauge bosons, collected in ${\cal L}_{SSVV}$ (see Appendix~\ref{appgauge}). The dark matter candidate is $S_0$, and therefore $M_S < M_F $.\\[-4mm]

\paragraph{\bf  \underline{Model LL1:}} The field content for this model, as given in Table \ref{tab:LLRR}, is $ F_R \sim 2^*_{1/2}$, $F_R^c \sim 2_{-1/2}$, $S_R \sim 1_0$. The two Weyl fermions are combined into a  
 Dirac fermion $F\sim  2_{-\frac{1}{2}}$. In terms of Weyl spinors and $SU(2)_L$ components this field decomposes as 
 \beq
 F  \mathrel{\overset{\makebox[0pt]{\mbox{\normalfont\tiny\sffamily Weyl}}}{=}}
  \begin{pmatrix} 
F_R^c  \\
F_R^\dagger
\end{pmatrix}, \qquad\qquad F \, \, \mathrel{\overset{\makebox[0pt]{\mbox{\normalfont\tiny\sffamily SU(2)}}}{=}}  \begin{pmatrix} 
F_0  \\
F_-
\end{pmatrix}. 
\eeq
To emphasize that the real scalar has charge zero, we use the notation $S_0 \equiv S_R$.  The 
Lagrangian of LL1 model is thus
\beq
\begin{split}
{\cal L}_{\rm LL1}  =&  \bar{F}_- \big(i \slashed \partial - M_F\big) F_- + \bar{F}_0 \big(i \slashed \partial - M_F\big) F_0 + \frac{1}{2} \big( \partial_\mu S_0 \partial^\mu S_0 - M_S^2 S_0^2 \big)
\\
& + \lambda_{L} 
\left(\bar{F}_- \mu_{L} + \bar{F}_0 \nu_{\mu L}\right)S_0 +  \lambda_{L}^{*}   \left( \bar{\mu}_{L} F_-  + \bar{\nu}_{\mu L} F_0 \right) S_0     + {\cal L}_{\rm gauge} + {\cal L}_{\rm scalar} \, , 
\end{split}
\eeq
with
\begin{align}
\begin{split}
{\cal L}_{\rm gauge} & =    |e| A_\mu  \bar{F}_- \gamma^\mu F_- + \frac{g}{c_W} Z_\mu \Big[ \Big( - \frac{1}{2} + s_W^2  \Big) \bar{F}_- \gamma^\mu    F_- +\frac{1}{2 }  \bar{F}_0 \gamma^\mu   F_0 \Big]  \\
& + \frac{g}{\sqrt2} \left( W_\mu^+ \bar{F}_0 \gamma^\mu F_- + W_\mu^- \bar{F}_- \gamma^\mu F_0 \right) \, , 
\end{split}
\\
{\cal L}_{\rm scalar} &=   \kappa S_0^2 |H|^2 + \xi S_0^4 = \frac{\kappa}{2} S_0^2   \left( h^2 + 2 v h + v^2 \right)+ \xi S_0^4 \, . 
\end{align}
 The dark matter candidate is $S_0$, so that we take $M_S < M_F $.\\[-4mm]

\paragraph{\bf  \underline{Model RR1:}} The field content for this model, as given in Table \ref{tab:LLRR}, is $F_L \sim 1_{-1}$, $F_L^c \sim 1_{1}$, $S_L \sim 1_0$. The two Weyl fermions are combined into a Dirac fermion $F_-\sim 1_{-1}$, 
\beq
F_-  \mathrel{\overset{\makebox[0pt]{\mbox{\normalfont\tiny\sffamily Weyl}}}{=}}
  \begin{pmatrix} 
F_L  \\
F_L^{c \dagger}
\end{pmatrix},
\eeq
while we denote the neutral real scalar as $S_0 \equiv S_L$.  The Lagrangian for the RR1 model is
\beq
\begin{split}
{\cal L}_{\rm RR1} = & \bar{F}_- \big( i \slashed \partial - M_F\big) F_- + \frac{1}{2}\big(\partial_\mu S_0  \partial^\mu S_0 - M_S^2 S_0^2\big)\\
&+ \big( \lambda_{R}  \bar{\mu}_{R} F_- S_0 +  {\rm h.c.}\big)   + {\cal L}_{\rm gauge} + {\cal L}_{\rm scalar} \, , 
\end{split}
\eeq
where
\begin{align}
{\cal L}_{\rm gauge} & =    |e| A_\mu  \bar{F}_- \gamma^\mu F_- + \frac{g s_W^2}{c_W} Z_\mu    \bar{F}_- \gamma^\mu    F_-  \, , \\
{\cal L}_{\rm scalar} &=   \kappa S_0^2 |H|^2 + \xi S_0^4 = \frac{\kappa}{2} S_0^2   \left( h^2 + 2 v h + v^2 \right)+ \xi S_0^4 \, .
\end{align}
The DM candidate is $S_0$, so that we take $M_S < M_F $. \\
\subsection{Models with Higgs insertion (FLR and SLR models)}
\label{sec:FLR}
In this subsection we present the Lagrangians for two examples of Class II models, the FLR1 model, where the Higgs couples to the new fermions, and the SLR1 model, where the Higgs couples to the new scalars. 
 
\paragraph{\bf  \underline{Model FLR1:}}  The field content of the model is $F_S\equiv F_R \sim  1_0, F_D\equiv F_L \sim 2_{-1/2}, F_D^c\equiv F_L^c \sim 2^*_{1/2}, S\equiv S_R = 2_{1/2}$, cf. Table \ref{tab:LR} and Eq. \eqref{eq:FLR1}. The scalar doublet has two components, 
the charged complex scalar $S_+$, and a neutral complex scalar $S_0$. For the fermions we use the 4-component notation. The charged components of $F_L$ and $F_L^c$ combine into a Dirac fermion 
\beq
F_{-}   \mathrel{\overset{\makebox[0pt]{\mbox{\normalfont\tiny\sffamily Weyl}}}{=}}
  \begin{pmatrix} F_{L-} \\ F_{L+}^{c \dagger} \end{pmatrix} \, , 
\eeq
while the neutral components of $F_R, F_L$, and $F_L^c$ mix into 3 Majorana fermions,
\begin{align}
F_{0i} &  \mathrel{\overset{\makebox[0pt]{\mbox{\normalfont\tiny\sffamily Weyl}}}{=}}
  \begin{pmatrix} F_i \\ F_i^\dagger \end{pmatrix} \, , \quad i=1,2,3 \, .
  \end{align}
  In terms of these mass eigenstates the Lagrangian is given by
  \begin{align}
{\cal L}_{\rm FLR1} & \supset {\cal L}_{\rm mass} + {\cal L}_S + {\cal L}_h +   {\cal L}_{\rm gauge} +  {\cal L}_{\rm scalar} \, , 
\end{align}
with
\begin{align}
{\cal L}_{\rm mass} & = -\frac{1}{2} M_i \bar{F}_{0i} F_{0i} - M_{F_D} \bar{F}_- F_- - M_S^2 \left( |S_+|^2 + |S_0|^2 \right) \, , 
\label{eq:LR1-Lm}
\\
\begin{split}
{\cal L}_S  & = \lambda_{1} V_{1j}  \left( S_0\bar{F}_{0j} \nu_{\mu L} - S_+ \bar{F}_{0j} \mu_L \right) + \lambda_{2} S_0^* \big(\bar{\mu} P_L F_-\big) +
 \lambda_{2} V_{2j} S^*_+ \big( \bar{\mu} P_L F_{0j}\big) +{\rm h.c.} \, , \label{eq:LR1-LS}
\end{split}
\\
{\cal L}_h  & = - \frac{h}{\sqrt2} \left( \lambda_{1H} V_{2i} V_{1j} + \lambda_{2H} V_{3i} V_{1j} \right) \bar{F}_{0i} P_L F_{0j} + {\rm h.c.} \, , \label{eq:LR1-Lh} 
\\
\begin{split}
   {\cal L}_{\rm gauge}  & = \frac{g}{c_W} Z_\mu \Big[ \frac{1}{4} \big( V_{2i}^* V_{2j} - V_{3i}^* V_{3j} \big) \bar{F}_{0i} \gamma^\mu P_L F_{0j} 
  - \frac{1}{4} \big( V_{2i} V_{2j}^* - V_{3i} V_{3j}^* \big) \bar{F}_{0i} \gamma^\mu P_R F_{0j}  \\
   &  \qquad\qquad\quad-\Big(\frac{1}{2} - s_W^2 \Big) \bar{F}_- \gamma^\mu F_- \Big]   
    + |e| A_\mu \bar{F}_- \gamma^\mu F_- \\
    &\qquad\qquad\quad+\frac{g}{\sqrt{2}} \left[ W_\mu^+ \left( V_{2i}^* \bar{F}_{0i} \gamma^\mu P_L F_- + V_{3i} \bar{F}_{0i} \gamma^\mu P_R F_- \right) + {\rm h.c.} \right] \, .  \label{eq:LR1-Lg}
   \end{split}
\end{align}
We do not write out $   {\cal L}_{\rm scalar}$, which describes the scalar gauge interactions and the scalar--Higgs interactions, since they are not needed in our analysis, and also show only the mass part of the free-field kinetic terms. The mixing matrix $V$ diagonalizes the Majorana mass matrix, 
\begin{align}
V^T \begin{pmatrix} M_{F_S} & \frac{\lambda_{1H} v}{\sqrt2}  & \frac{\lambda_{2H} v}{\sqrt2} \\ \frac{\lambda_{1H} v}{\sqrt2} & 0 &  M_{F_D} \\ \frac{\lambda_{2H} v}{\sqrt2} & M_{F_D} & 0 \end{pmatrix} V = \begin{pmatrix} M_1 && \\ & M_2 & \\ & & M_3 \end{pmatrix} \, .
\end{align}
We take $M_1 \le M_2 \le M_3$, so that $F_{01}$ is the DM candidate. Note that this model resembles the (bino)--(Higgsino)--(left-handed slepton) sector of the MSSM. 

The NP contribution to $(g-2)_\mu$ is
\begin{align}
\Delta a_\mu =& \frac{m_\mu^2}{8 \pi^2 M_S^2 } |\lambda_{2}|^2 f^F_{LL} \left( \frac{M_{F_D}^2}{M_S^2}\right) +  \frac{m_\mu }{8 \pi^2 M_S^2 } \sum_{A=1,2,3} M_{A}  {\rm Re} \left( \lambda_{1} \lambda_{2} V_{1A} V_{2A} \right) f^S_{LR} \left( \frac{M_{A}^2}{M_S^2}\right)\nonumber \\
& - \frac{m_\mu^2}{8 \pi^2 M_S^2 } \sum_{A = 1,2,3}  \left(  |\lambda_{2}|^2 |V_{2A}|^2 + |\lambda_{1}|^2 |V_{1A}|^2 \right) f^S_{LL} \left( \frac{M_{A}^2}{M_S^2}\right) \, .
\end{align}
In the limit of approximately equal masses $M_{A}$ the loop function $f_{LR}^S$ is approximately constant, so that the relevant term simplifies to 
\begin{align}
 \sum_{A=1,2,3} M_{A}  {\rm Re} \left( \lambda_{1} \lambda_{2} V_{1A} V_{2A} \right) f^S_{LR} \left( \frac{M_{A}^2}{M_S^2}\right) & \approx  f^S_{LR} \left( \frac{M_{A}^2}{M_S^2}\right)  {\rm Re} \left[ \lambda_{1} \lambda_{2} \left( V M_{\rm diag} V^T \right)_{12}  \right] \nonumber \\
 & =  \frac{v}{\sqrt 2} f^S_{LR} \left( \frac{M_{A}^2}{M_S^2}\right)  {\rm Re} \left( \lambda_{1} \lambda_{2} \lambda_{1H} \right) \, , 
\end{align}
which corresponds to the leading diagram in the mass insertion approximation.\\[-4mm]

\paragraph{\bf  \underline{Model SLR1:}}  The field content is a vector-like pair of Weyl fermions, $F\equiv F_R \sim 1_{1}, F^c\equiv F_R^c \sim 1_{-1}$, a real singlet scalar $S\equiv S_L \sim 1_{0}$, and a complex scalar doublet  $D\equiv S_R \sim 2_{-1/2}$. The two Weyl fermions combine into a charged Dirac fermion, 
\begin{align}
F_{-}   \mathrel{\overset{\makebox[0pt]{\mbox{\normalfont\tiny\sffamily Weyl}}}{=}}
  \begin{pmatrix} F_{R}^c \\ F_{R}^{ \dagger} \end{pmatrix} \, . 
  \end{align}
The scalar sector contains a CP-odd neutral scalar $A_0$, a charged complex scalar $S_-$,  and two CP-even neutral scalars $S_\alpha, \alpha=1,2$ that are admixtures of the neutral CP-even components in $S_L, S_R $. 

In terms of the mass eigenstates the Lagrangian is given by
  \begin{align}
{\cal L}_{\rm SLR1} & \supset {\cal L}_{\rm mass}  +{\cal L}_S + {\cal L}_h +  {\cal L}_{\rm gauge}  \, , 
\end{align}\
with
\begin{align}
{\cal L}_{\rm mass} & = - M_F \bar{F}_{-} F_{-}  - \frac{1}{2} M_\alpha^2 S_\alpha^2  - \frac{1}{2} M_D^2 A_0^2 - M_D^2 |S_-|^2 \, , 
\\
{\cal L}_S  & = \frac{\lambda_{1}}{\sqrt2} \left( U_{2\alpha}  S_\alpha + i A_0 \right) \big(\bar{F}_- \mu_L\big) - \lambda_{1}  S_-\big(\bar{F}_- \nu_{\mu L}\big) + \lambda_{2} U_{1\alpha} S_\alpha \big(\bar{\mu}_R F_-\big)  +{\rm h.c.} \, , \\
{\cal L}_h  & = - \frac{a_H}{2} h  \, U_{1
\alpha} U_{2 \beta} S_\alpha S_\beta +  {\rm h.c.} \, , 
\\
\begin{split}
   {\cal L}_{\rm gauge}  & \supset \frac{g}{c_W} Z_\mu \Big[ i \big( - \tfrac{1}{2} + s_W^2 \big) \big(S_-^* \partialLR S_-\big)  + \tfrac{1}{2} U_{2 \alpha} \big( A_0 \partialLR S_\alpha\big) \Big] 
 + i |e| A_\mu \big(S_-^* \partialLR S_-\big)
     \\
    & + \frac{g^2}{8 c_W^2 } \left( i A_0 + U_{2 \alpha} S_\alpha \right) \left( - i A_0 + U_{2 \beta} S_\beta \right) \left( 2 c_W^2 W_\mu^- W_\mu^+ + Z_\mu Z^\mu \right)
   \\
 &  + S_-^* S_-   \Big[ \tfrac{1}{2} g^2 W^-_\mu W^+_\mu + \Big( e A_\mu  + \frac{g}{c_W} \Big( -\tfrac{1}{2} + s_W^2 \Big) Z_\mu \Big)^2 \Big]\,
   \\
   & +\frac{g}{2} W_\mu^+
    \Big[i U_{2\alpha} \big(S_\alpha  \partialLR S_-\big) +  A_0   \partialLR S_-  +S_-  \left(U_{2 \alpha} S_\alpha -i A_0\right)  \Big( e A^\mu  + \frac{g }{c_W} s_W^2 Z^\mu\Big)  \Big] + {\rm h.c.}
 \end{split}
\end{align}
Above we do not write out explicitly the fermion-gauge couplings, as well as the free-field kinetic terms, apart from masses. We take $a_H$ to be real so that the mixing matrix $U$ is also real. It diagonalizes the mass matrix for the neutral scalars, 
\begin{align}
U^T \begin{pmatrix} M_S^2 & v a_H   \\ v a_H &  M_D^2 \end{pmatrix} U = \begin{pmatrix} M_1^2 && \\ & M_2^2  \end{pmatrix} \, .
\end{align}
By convention we take $M_1^2 \le M_2^2 $. Since $S_1$ is a DM candidate, we have $M_1 < M_F$.
The NP contributions to  $(g-2)_\mu$ are given by
\beq
\begin{split}
\Delta a_\mu = &  \frac{|\lambda_1|^2 m_\mu^2}{16 \pi^2 M_D^2 }  f^F_{LL} \left( \frac{M_{F}^2}{M_D^2}\right) + \frac{m_\mu M_F}{8 \sqrt2 \pi^2 } \sum_{\alpha=1,2}   \frac{\lambda_{1} \lambda_{2} U_{1\alpha} U_{2\alpha}  }{M_{S \alpha}^2}f^F_{LR} \left( \frac{M_{F}^2}{M_{S_\alpha}^2}\right)
 \\
& + \frac{m_\mu^2}{16 \pi^2 } \sum_{\alpha = 1,2}  \frac{  |\lambda_{1}|^2 |U_{2 \alpha}|^2 + 2 |\lambda_{2}|^2 |U_{1\alpha}|^2 }{M_{S_\alpha}^2} f^F_{LL} \left( \frac{M_{F}^2}{M_{S_\alpha}^2}\right)  
 \, .
 \label{SLRg2}
\end{split}
\eeq
Note that the off-diagonal entries of $U_{ij}$ are proportional to the weak scale, therefore all contributions in Eq.~\eqref{SLRg2} scale at least as $\propto v^2$. 

\section{Correction to the $Z\mu \mu$ Vertex}
\label{Zmumu}
Using the general Lagrangian in Eq.~\eqref{eq:general:Lagr} with explicit $Z$-couplings,
\begin{align}
{\cal L} & = \left[S^* \, \bar{\mu} \left( \lambda_2^{R} P_L + \lambda_2^{L} P_R \right) F + {\rm h.c.}\right]   - M_{F} \bar{F} F   - M_S^2 S^* S  \nonumber \\
& + Z_\mu \left[ \overline{\mu} \, \gamma^\mu \left( g^{Z \mu \mu}_L P_L + g^{Z\mu \mu}_R P_R \right) \mu  + \overline{F} \gamma^\mu \left( g^{ZFF}_L P_L + g^{ZFF}_R P_R \right) F  + i S^\dagger g^{ZSS} \overset{\leftrightarrow}{\partial^\mu} S \right] \, , 
\end{align}
one gets for the 1-loop contribution to the $Z\mu\mu$-couplings (in the limit $m_\mu \to 0$)
\begin{align}
\Delta g^{Z\mu \mu}_{L} = \frac{|\lambda^L_{2}|^2 }{16 \pi^2} \left[  g^{ZFF}_R I_a +  ( g^{ZFF}_L -g^{ZFF}_R) I_b  + g^{Z\mu \mu}_L I_c   \right] \, , 
\end{align}
with the Feynman integrals 
\begin{align}
 I_a (M_F, M_S)  & =  \int_0^1  \int_0^1 dx dy \left( 2 (1-x)^2 y (1-y) \frac{M_Z^2}{\tilde{\Delta}_F} + x \log \frac{\Delta_S}{\tilde{\Delta}_F} \right) \, , \\
 I_b (M_F, M_S)  & =  \int_0^1  \int_0^1 dx dy \left( (1-x)   \frac{M_F^2}{\tilde{\Delta}_F}  \right) \, , \\ 
 I_c (M_F, M_S)  & =  \int_0^1  \int_0^1 dx dy \left( x  \log \frac{\Delta}{\Delta_S}  \right) \, , 
\end{align}
and
\begin{align}
\Delta & = x M_S^2  + (1-x) M_F^2 \, , \\
\tilde{\Delta}_F & = x M_S^2  + (1-x) M_F^2 - (1-x)^2 y (1-y) M_Z^2 \, , \\
\Delta_S & = x M_S^2  + (1-x) M_F^2 - x^2 y (1-y) M_Z^2 \, ,
\end{align}
with $M_Z$ the $Z$ boson mass.
Moreover, one has
\begin{align}
\Delta g^{Z \mu \mu}_{R} & = \Delta g^{Z \mu \mu}_{L} (\lambda^L_2 \to \lambda^R_2, g_L^{Z \mu \mu} \to g_R^{Z \mu \mu},  g_R^{ZFF} \leftrightarrow g_L^{ZFF}  )\, .
\end{align}
The above results were written in a form that ensures efficient numerical evaluation even in the decoupling limit, $M_{S,F}\gg m_Z$, since in the $v \to 0$ limit both the Feynman integrals $I_a$ and $I_c$, as well as the prefactor of $I_b$, go to zero.

\section{ Gauge Boson Couplings of an SU(2) $n$-plet}
\label{appgauge}
The general gauge boson couplings of  a Dirac fermion, $F\sim (n_F)_{Y_F)}$, and a complex scalar $S\sim (n_S)_{Y_S}$ are given by
\begin{align}
{\cal L}_{\rm gauge} (n_F, Y_F, n_S, Y_S) & = \frac{g}{c_W} Z_\mu J^\mu_n - |e| A_\mu J^\mu_{\rm em} + \frac{g}{\sqrt 2} \left( W_\mu^+ J^\mu_+ +{\rm h.c.} \right) + {\cal L}_{SSVV}, \, 
\end{align}
with the currents 
\begin{align}
J^\mu_n & =  \sum_{T_3} \bar{F}_{T_3 + Y_F} \gamma^\mu \left( c_W^2 T_3 - s_W^2 Y_F \right) F_{T_3 + Y_F}  
 + i \sum_{T_3}  \left( c_W^2 T_3 - s_W^2 Y_S \right)  S^*_{T_3 + Y_S} \partialLR  S_{T_3 + Y_S }  \, 
 \\
J^\mu_{\rm em} & =  \sum_{T_3}  \bar{F}_{T_3 + Y_F} \gamma^\mu (T_3 + Y_F) F_{T_3 + Y_F} + i  \sum_{T_3}  S^*_{T_3 + Y_S} \partialLR (T_3 + Y_S)  S_{T_3 + Y_S}  \, , 
\\
\begin{split}
J^\mu_{\pm} & =   \frac{1}{2} \sum_{T_3} \sqrt{n_F^2- (1\mp 2 T_3 )^2} \left(\bar{F}_{T_3 + Y_F} \gamma^\mu F_{T_3 + Y_F  \mp 1} \right) 
\\
 &\quad + \frac{i}{2} \sum_{T_3}  \sqrt{n_S^2- (1\mp 2 T_3 )^2}   S^*_{T_3 + Y_S}\partialLR S_{T_3 + Y_S \mp 1} \, . 
 \end{split}
\end{align}
The sum is over $T_3\in [- (n_{F(S)}-1)/{2}, - (n_{F(S)}-1)/{2} + 1 , \cdots,  (n_{F(S)}-1)/{2}]$ for the fermion (scalar), and $X \partialLR Y \equiv X \partial^\mu Y - (\partial^\mu X) Y$.  The boson quartic Lagrangian is given by 
\begin{align}
{\cal L}_{SSVV} 
& = \frac{g^2}{c_W^2} Z_\mu Z^\mu \sum_{T_3} S^*_{T_3 + Y_S} \left( c_W^2 T_3 - s_W^2 Y_S \right)^2 S_{T_3 + Y_S}  \nonumber \\
& - \frac{2 g |e|}{c_W} Z_\mu A^\mu \sum_{T_3} S^*_{T_3 + Y_S} \left( c_W^2 T_3 - s_W^2 Y_S \right) \left( T_3 + Y_S \right)  S_{T_3 + Y_S}  \nonumber \\
& + e^2 A_\mu A^\mu \sum_{T_3} S^*_{T_3 + Y_S}  \left( T_3 + Y_S \right)^2  S_{T_3 + Y_S} 
\nonumber \\
& + \frac{g^2}{4} W_{\mu}^+ W^{-\mu} \sum_{T_3} S^*_{T_3 + Y_S}  \left( n_S^2 - 1 - 4 T_3^2 \right)  S_{T_3 + Y_S} 
 \\
& + \frac{ g^2}{\sqrt{2}  c_W} Z^\mu W_\mu^\pm \sum_{T_3} \left( c_W^2 ( T_3 \mp \frac{1}{2}) -  s_W^2 Y_S \right)  \sqrt{n_S^2- (1\mp 2 T_3 )^2}  S^*_{T_3 + Y_S} S_{T_3 + Y_S \mp 1}
\nonumber \\
& - \frac{|e| g}{ \sqrt{2}  }  A^\mu W_\mu^\pm \sum_{T_3} \left(  T_3 \mp \frac{1}{2} +  Y_S \right)  \sqrt{n_S^2- (1\mp 2 T_3 )^2}  S^*_{T_3 + Y_S} S_{T_3 + Y_S \mp 1} \nonumber \\
& + \frac{g^2}{8} W^{\mu \pm} W_\mu^\pm \sum_{T_3}  \sqrt{n_S^2- (1\mp 2 T_3 )^2} \sqrt{n_S^2- (3\mp 2 T_3)^2}  S^*_{T_3 + Y_S} S_{T_3 + Y_S \mp 2} . \nonumber
\end{align}
Note that the above results also apply to fields in complex conjugate representations of $SU(2)_L$, $F^\prime\sim ({n}^*_F)_{Y_F}$ and $S^\prime\sim ({n}^*_S)_{Y_S)}$, with $S_{T_3 +Y_S} \to S^\prime_{T_3 + Y_S}$, $F_{T_3 +Y_F} \to F^\prime_{T_3 + Y_F}$,  $W_\mu^\pm \to - W_\mu^\pm$ and all explicit factors of $T_3$ under the sums replaced as $T_3 \to -T_3$.

\bibliographystyle{JHEP} 
\bibliography{paper_ref}

\end{document}